\documentclass[manuscript,screen]{acmart}
\AtBeginDocument{%
  }

\setcopyright{acmcopyright}
\copyrightyear{2024}
\acmYear{2024}
\acmDOI{XXXXXXX.XXXXXXX}

\usepackage{bbm}
\usepackage{subfigure}
\usepackage{caption}
\usepackage{ulem}
\usepackage{makecell}

\usepackage{soul}
\usepackage{url}
\usepackage{booktabs}
\usepackage{bm}
\usepackage{bbm}
\usepackage{amsfonts}
\usepackage{enumitem}
\usepackage{verbatim}
\usepackage{threeparttable}
\usepackage{multirow}
\usepackage{color}
\usepackage{xcolor}
\usepackage{tablefootnote}
\usepackage{threeparttable}

\usepackage{tikz}
\usepackage[edges]{forest}

\usepackage{amsmath,amsfonts,bm, multirow}

\def\eqref#1{equation~\ref{#1}}
\def\1{\bm{1}}

\DeclareMathAlphabet{\mathsfit}{\encodingdefault}{\sfdefault}{m}{sl}
\SetMathAlphabet{\mathsfit}{bold}{\encodingdefault}{\sfdefault}{bx}{n}

\begin{document}

%%
%% The "title" command has an optional parameter,
%% allowing the author to define a "short title" to be used in page headers.
\title{A Comprehensive Survey on Retrieval Methods in Recommender Systems}
% \title{A Comprehensive Survey on the First-Stage Retrieval Methods for Recommendation}

\author{Junjie Huang}
\affiliation{%
  \institution{Shanghai Jiao Tong University}
  \country{China}}
\orcid{0000-0002-5637-0735}
\email{huangjunjie2019@sjtu.edu.cn}

\author{Jizheng Chen}
\affiliation{%
  \institution{Shanghai Jiao Tong University}
  \country{China}}
\orcid{0009-0003-3509-4537}
\email{humihuadechengzhi@sjtu.edu.cn}

\author{Jianghao Lin}
\affiliation{%
  \institution{Shanghai Jiao Tong University}
  \country{China}}
\orcid{0000-0002-8953-3203}
\email{chiangel@sjtu.edu.cn}

\author{Jiarui Qin}
\affiliation{%
  \institution{Shanghai Jiao Tong University}
  \country{China}}
\orcid{0000-0002-9064-885X}
\email{qjr1996@sjtu.edu.cn}

\author{Ziming Feng$^\dagger$}
\affiliation{%
  \institution{China Merchants Bank Credit Card Center}
  \country{China}}
\orcid{0000-0002-4558-1246}
\email{zimingfzm@cmbchina.com}

\author{Weinan Zhang$^\dagger$}
\affiliation{%
  \institution{Shanghai Jiao Tong University}
  \country{China}}
\orcid{0000-0002-0127-2425}
\email{wnzhang@sjtu.edu.cn}

\author{Yong Yu}
\affiliation{%
  \institution{Shanghai Jiao Tong University}
  \country{China}}
\orcid{0000-0003-0281-8271}
\email{yyu@sjtu.edu.cn}

\thanks{$^\dagger$Corresponding authors}

\renewcommand{\shortauthors}{J. Huang et al.}

%%
%% By default, the full list of authors will be used in the page
%% headers. Often, this list is too long, and will overlap
%% other information printed in the page headers. This command allows
%% the author to define a more concise list
%% of authors' names for this purpose.
% \renewcommand{\shortauthors}{J. Huang et al.}

\begin{abstract}
In an era dominated by information overload, effective recommender systems are essential for managing the deluge of data across digital platforms. Multi-stage cascade ranking systems are widely used in the industry, with retrieval and ranking being two typical stages. Retrieval methods sift through vast candidates to filter out irrelevant items, while ranking methods prioritize these candidates to present the most relevant items to users. Unlike studies focusing on the ranking stage, this survey explores the critical yet often overlooked retrieval stage of recommender systems. To achieve precise and efficient personalized retrieval, we summarize existing work in three key areas: improving similarity computation between user and item, enhancing indexing mechanisms for efficient retrieval, and optimizing training methods of retrieval. We also provide a comprehensive set of benchmarking experiments on three public datasets. Furthermore, we highlight current industrial applications through a case study on retrieval practices at a specific company, covering the entire retrieval process and online serving, along with practical implications and challenges. By detailing the retrieval stage, which is fundamental for effective recommendation, this survey aims to bridge the existing knowledge gap and serve as a cornerstone for researchers interested in optimizing this critical component of cascade recommender systems.
\end{abstract}

%%
%% The code below is generated by the tool at http://dl.acm.org/ccs.cfm.
%% Please copy and paste the code instead of the example below.
%%



\maketitle

\section{INTRODUCTION}\label{sec:introduction}
\begin{figure}[!htbp]
\centering
\includegraphics[width=1\linewidth]{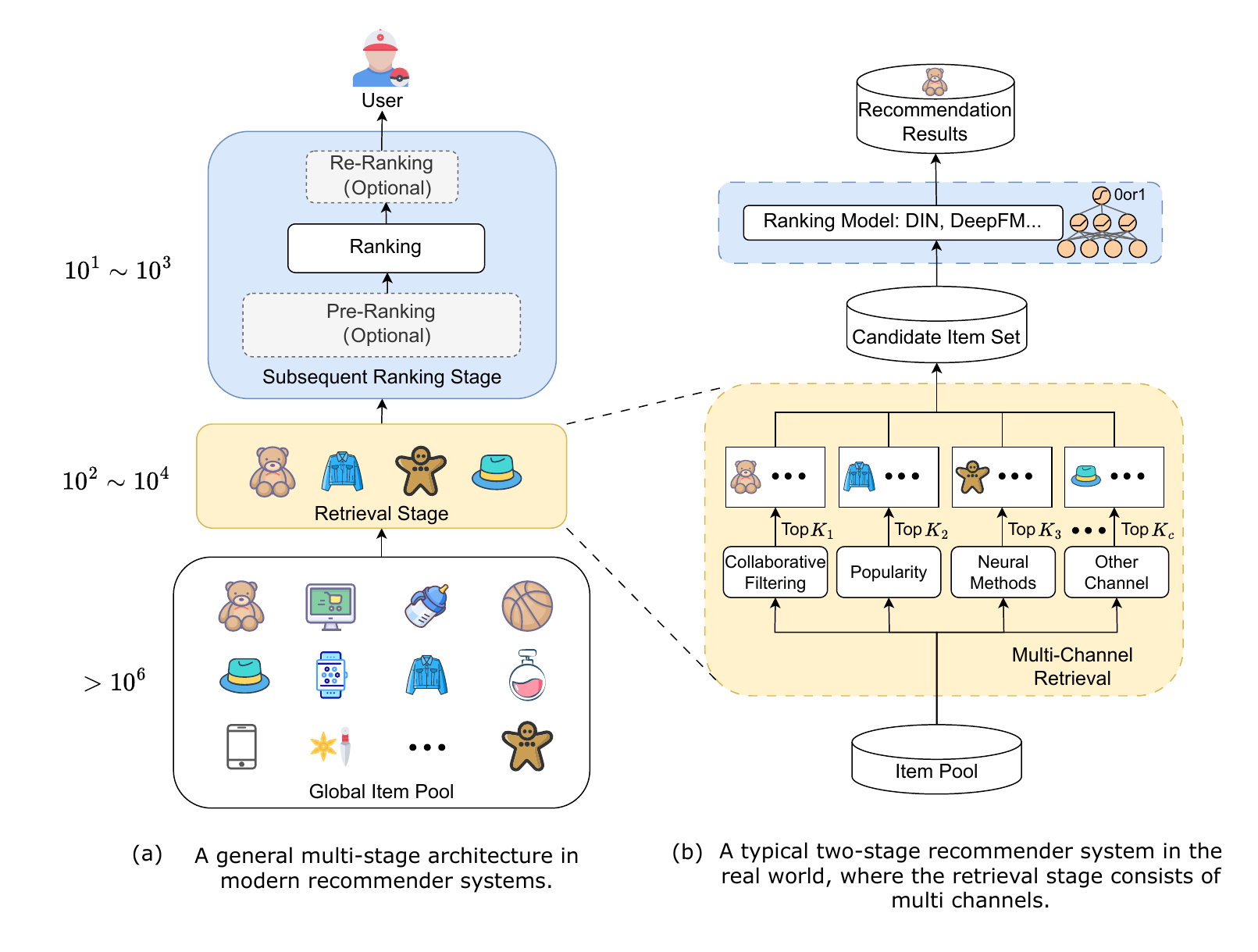}
\caption{The multi-stage architecture in modern recommender systems and the illustration of multi-channel retrieval. The latter will be detailed further in Section ~\ref{sec:mc}.}
\vspace{-3mm}
\label{fig:cascade}
\end{figure}

% 介绍rs的目的以及引出multi-stage rs
We are being bombarded with a vast amount of information due to the growing popularity of the Internet and the development of User Generated Content (UGC)~\cite{krumm2008user} in recent years.
To save users from information overload, recommender systems have been widely applied in today's short video~\cite{liu2019user}, news~\cite{wang2018dkn} and e-commerce~\cite{chen2019behavior} platforms. 
While complicated models~\cite{pi2020search, qin2021retrieval,lin2023map,wang2023flip} often offer higher accuracy, their poor efficiency makes online deployment challenging because of latency restrictions~\cite{pi2019practice}. On the other hand, simple models~\cite{huang2013learning, rendle2010factorization} have capacity limitations, but they could evaluate a great number of items efficiently because of their low time complexity. Therefore, striking a balance between efficacy and efficiency becomes crucial in order to quickly filter out information that users are interested in. As is shown in Figure~\ref{fig:cascade} (a), one widely used solution in the industry is multi-stage cascade ranking systems~\cite{wang2011cascade}.
The system includes a retriever and a variety of subsequent rankers.
In the very first stage of the cascade system, referred to as the retrieval stage in this paper (also called matching stage or recall stage in some literature~\cite{qin2022rankflow, zhu2022bars}), a retriever is typically used to quickly eliminate irrelevant items from a large pool of candidates, whereas rankers in the later stages aim to accurately rank the items. Each stage selects the top-$K$ items it receives and feeds them to the next stage. 
% The user will ultimately see the items that were selected by the ranking stage.
As shown in Figure~\ref{fig:cascade} (a), rankers in multi-stage cascade ranking systems are arranged in the shape of a funnel, narrowing from bottom to top. The retrieval and ranking stage are two typical stages, while pre-ranking~\cite{wang2020cold} and re-ranking~\cite{xi2023bird} stages are relatively optional, and the number of rankers in the system may vary depending on different scenarios. Additionally, on the left side of Figure~\ref{fig:cascade} (a), we display the approximate output scale of each stage, noting that the range of this scale is specific to the particular platform and scenario.

% 接下来讲召回和排序两阶段的对比, 强调召回阶段的特点
Although both the retrieval and ranking stages aim to select the most relevant items, each stage has its own unique characteristics.
\begin{itemize}[topsep = 3pt,leftmargin =*]
    \item \textbf{Difference in candidate sets (i.e., inference spaces).}
    The retrieval stage needs to quickly filter through the entire item pool, which may contain millions of items; while the ranking stage only needs to score and order the items that have been selected by the retrieval methods, typically narrowing down to hundreds or thousands of items.
    \item \textbf{Difference in input features.}
    During the retrieval stage, due to time constraints and the need to filter through a large candidate set quickly, utilizing complex feature interactions is impractical for real-time online requirements. As a result, only limited, coarse-grained features of users and items are considered.
    In contrast, the ranking stage can utilize a diverse set of features by designing various feature interaction operators, such as product operators~\cite{qu2016product}, convolutional operators~\cite{li2019fi}, and attention operators~\cite{xiao2017attentional}. The ranking stage further enhances its capability by integrating multiple feature domains and providing detailed modeling of behavioral and interaction features. These intricate and meaningful patterns serve as valuable indicators for evaluating user preferences, allowing for a more precise and refined ranking of the items.
    \item \textbf{Difference in model architectures.}
    Typically, retrieval models employ a dual-tower architecture, where user and item representations are generated separately, and their matching scores are calculated using basic similarity metrics~\cite{huang2013learning}, such as inner product or cosine similarity. This dual-tower architecture allows for independent optimization of user and item representations before they interact, which is often referred to as \textit{late interaction}. On the other hand, ranking models often utilize a single-tower architecture, where user features, item features, and cross features are concatenated and processed through a deep neural network (DNN). This setup, known as \textit{early interaction}, facilitates multi-level interactions between features, thereby increasing the model's expressive power. The single-tower architecture processes these concatenated features deeply within the network, allowing for richer and more complex interactions compared to the more straightforward combination in dual-tower setups. The distinction between single and dual-tower architectures essentially lies in how and where the interactions between users and items are modeled, that is, late interaction in dual-tower setups via direct similarity metrics, versus early interaction in single-tower setups through deeper, concatenated feature processing.
    \item \textbf{Difference in optimization objectives.}
    The retrieval stage aims to achieve high recall efficiently, i.e., to quickly identify a subset containing as many relevant items as possible. In this stage, the focus is on ensuring that the relevant items are included in the set, without prioritizing their exact order. In contrast, the goal of the ranking stage is to refine the retrieved candidate set by accurately determining the relative importance of each item. Since only a small number of items are fed into the ranking model, these models can utilize more complex architectures to enhance precision, ensuring that the most relevant items are positioned at the top of the list. Thus, the retrieval stage prioritizes the inclusion of relevant items (i.e., recall), while the ranking stage prioritizes precise ordering to maximize the relevance of the top-ranked items.
\end{itemize}

% 召回模型发展脉络
% 以及有ranking survey，但没有retrieval的现状
% In the retrieval phase of real-world systems, efficiency is critically important due to the vast number of items involved. Typically, the retrieval models begin by generating representations for users and items and then calculate matching scores between them using simple similarity metrics such as inner products or cosine similarity.
In the past few decades, there has been substantial progress in the retrieval stage.
The traditional retrieval methods are based on the concept of collaborative filtering (CF)~\cite{su2009survey}, which include neighborhood based~\cite{sarwar2001item} and matrix factorization (MF) based methods~\cite{koren2009matrix, manotumruksa2017deep}. However, these methods primarily focus on learning embeddings for user and item IDs, without considering other important features such as user profiles, item attributes, or contextual information. Additionally, using just shallow structures in these models is insufficient to capture complex patterns and interactions, resulting in suboptimal representations. 
Consequently, in recent years, there has been a shift towards embedding based retrieval (EBR), which treats the retrieval problem as a nearest neighbor search in the vector space. EBR is not a specific algorithm but a broad family of algorithms, including Item2Vec~\cite{barkan2016item2vec}, YouTube DNN~\cite{covington2016deep}, EGES~\cite{wang2018billion} of Alibaba, and Pinsage~\cite{pinsage} of Pinterest, e.t.c. These models, which vary from two-tower to graph-embedding structures, aim to learn comprehensive representations for users and items.
Given the extensive research conducted, it is timely to assess the current landscape, learn from existing methods, and identify directions for future research. While there are comprehensive surveys on the ranking stage~\cite{zhang2021deep, wu2022survey, he2023survey}, the retrieval stage lacks systematic reviews. This oversight may stem from the greater emphasis traditionally placed on the ranking stage, which, as the final stage in recommender systems, interacts directly with users and has a more immediate impact on business objectives. 
However, the retrieval stage is also crucial as it sets the stage for all subsequent processes. Without effective retrieval, even the most advanced follow-up stages cannot perform optimally.

\subsection{A Taxonomy and Major Content}
% 强调我们是第一篇，和其他survey区别，涵盖的工作范围
To our knowledge, this is the first survey that comprehensively summarizes retrieval-related efforts in recommender systems. While there have been several surveys on recommender systems, none have specifically focused on the retrieval stage.
Batmaz et al.~\cite{batmaz2019review} provide a thorough review of deep learning based recommendation methods, primarily focusing on autoencoders, recurrent neural networks (RNNs), and convolutional neural networks (CNNs) for recommendations. Similarly, Zhang et al.~\cite{zhang2019deep} offer a detailed overview of current efforts in deep learning based recommender systems and discuss some unresolved issues. Xu et al.~\cite{xu2018deep} propose a unified framework for search and recommendation and mention some retrieval models, but their discussion is scattered and lacks a focus on the retrieval stage. In contrast, our work provides a comprehensive overview of retrieval methods in recommender systems under a unified framework.

% We review numerous works published in major conferences and journals. 
\begin{figure}[!t]
\centering
\includegraphics[width=1\linewidth]{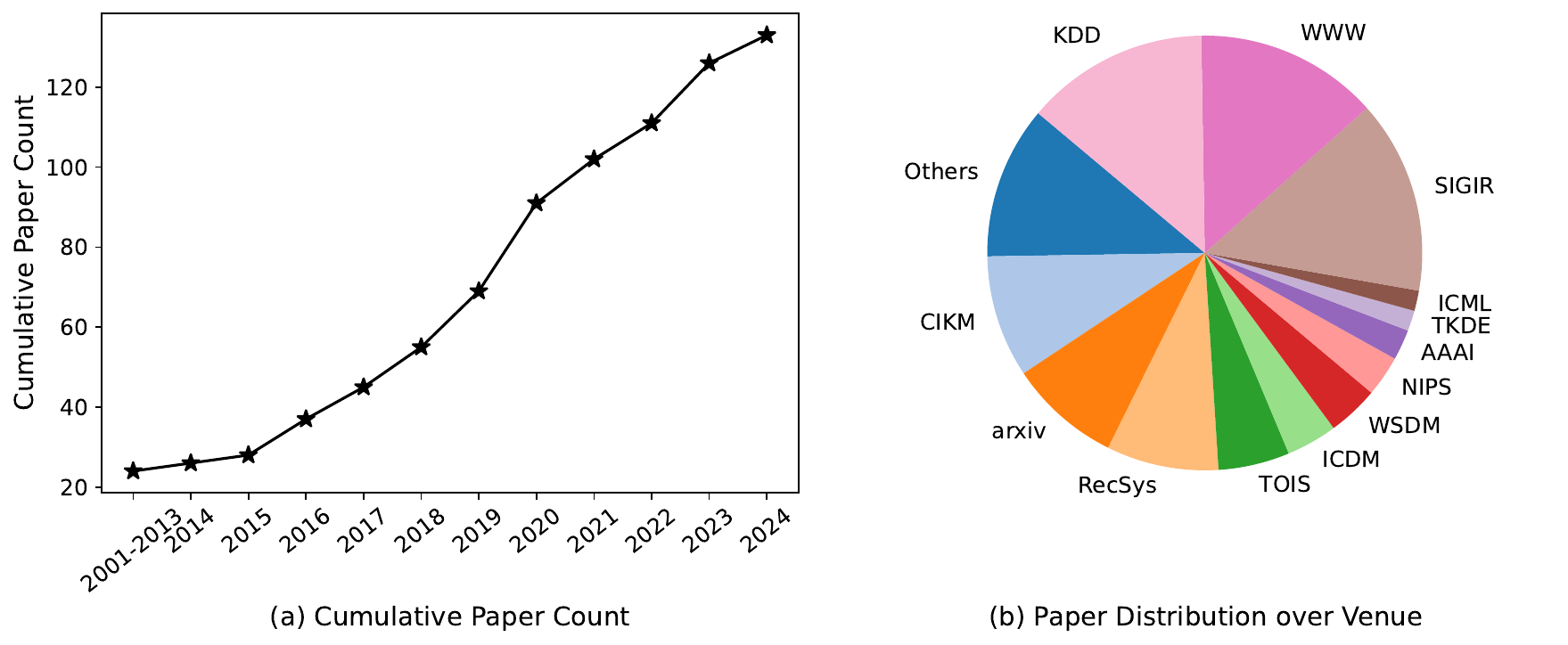}
\caption{
Distributions of the reviewed papers of retrieval methods in recommender systems over years (a) and over venues (b).
}
\label{fig:distribution}
\vspace{-2mm}
\end{figure}
We collect over 100 research papers on retrieval methods in recommender systems. In conducting our literature search, we utilize DBLP and Google Scholar as the primary search engines, using keywords such as \textit{candidate matching}, \textit{candidate retrieval}, \textit{large-scale recommendation}, and \textit{collaborative filtering} in the titles and abstracts to filter relevant papers from top-tier conferences and journals in data mining, machine learning, recommender systems, and artificial intelligence. These include, but are not limited to, KDD, ICDM, WWW, SIGIR, ICML, NIPS, AAAI, WSDM, CIKM, RecSys, TKDE, and TOIS, covering the period from 2001 to 2024, as shown in Figure~\ref{fig:distribution}. We traverse the citation graph of the identified papers and incorporate pertinent studies. In addition to reviewing published papers, we also screen preprints on arXiv, identifying those with novel and intriguing ideas to provide a more comprehensive perspective.
Based on this corpus, we summarize the efforts made to achieve precise and efficient personalized retrieval into three main areas, as illustrated in Figure~\ref{fig:taxonomy}:
\begin{figure}[!t]
\centering
\includegraphics[width=1\linewidth]{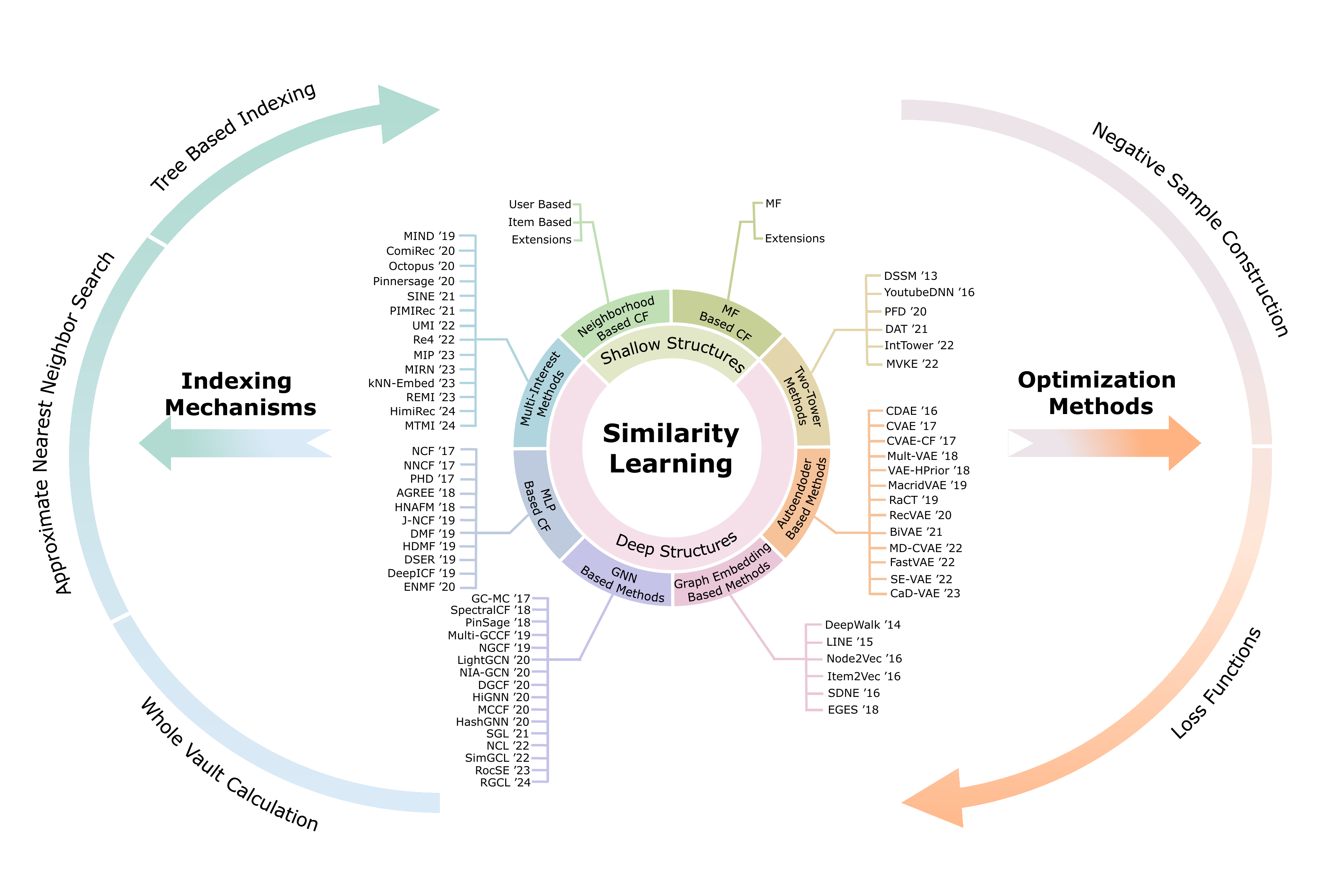}
\caption{
Taxonomy for retrieval methods in recommender systems.
}
\label{fig:taxonomy}
\vspace{-2mm}
\end{figure}
\begin{itemize}[topsep = 3pt,leftmargin =*]
    \item \textbf{Similarity Learning.}
    Many studies focus on improving the representation of users and items to enhance similarity learning. Traditional methods, based on CF and MF~\cite{sarwar2001item, koren2009matrix, manotumruksa2017deep}, use shallow structures for this purpose. Recently, there has been a growing trend towards using deep structures~\cite{covington2016deep, wang2018billion, pinsage} to better learn the representations and similarity relationships of users and items.
    \item \textbf{Indexing Mechanisms.}
    In the retrieval stage, indexing schemes are essential for effectively organizing and retrieving large-scale items. Some efforts have been directed towards improving indexing mechanisms, such as tree based methods that use a tree structure as an index~\cite{tdm, jtm, otm}. Deep Retrieval~\cite{deepretrieval} uses a matrix for indexing, enabling efficient item clustering and retrieval.
    \item \textbf{Optimization Methods.}
    Various learning strategies have been developed to optimize the training of retrieval methods, particularly in negative sample construction. Users typically interact with a small subset of items compared to the vast number available in real systems, making the lack of reliable negative data a challenge for learning from implicit feedback. This includes various negative sampling strategies~\cite{he2017neural, rendle2012bpr}, which select a subset of missing data as negative instances, and whole-data based strategies~\cite{devooght2015dynamic, hu2008collaborative, liang2016modeling} that treat all missing data as negative instances.
\end{itemize}

% contribution 和 organization
The aim of this survey is to thoroughly review the literature on retrieval methods in recommender systems and discuss the current state of retrieval in the industry, as well as future directions. This survey provides researchers and practitioners interested in recommender systems with a general understanding of the latest developments in the field of retrieval for recommendation. The key contributions of this survey are summarized as follows:
\begin{itemize}[topsep = 3pt,leftmargin =*]
    \item \textbf{First Comprehensive Review.}
    This review addresses the deficiency in summarizing retrieval-related efforts in recommender systems. 
    % We explore both traditional collaborative filtering methods and modern neural approaches using embedding-based retrieval. Additionally, we provide a detailed overview of the entire retrieval pipeline, including training data preparation, model training, and evaluation metrics, with a focus on key topics such as various negative sampling strategies.
    We categorize these efforts into three main areas: improving similarity learning between user and item through both shallow and deep structures, enhancing indexing mechanisms for efficient large-scale retrieval, and optimizing training methods of retrieval. To the best of our knowledge, this is the first comprehensive review on the retrieval methods for recommendation, which not only highlights the progress achieved but also identifies areas for future research and development.
    \item \textbf{Extensive Benchmarking Experiments.}
    We provide a comprehensive set of benchmarking experiments, covering various retrieval methods on three public datasets. As the study of retrieval methods is deeply rooted in practical applications and faces numerous real-world challenges, we summarize online performance statistics from published papers, providing valuable insights.
    % \ljh{TO BE DISCUSSED: can we move online performance statistics to the next part. Since these s are not the outcome of our conducted experiments actually.}
    \item \textbf{Case Study on Industrial Practices of Retrieval.}
    % \ljh{I suggest move the "online performance statistics" here. we first summarize xxx, and the provide a xxx.}
    Additionally, we provide a case study on the current industrial practices of the retrieval stage at a specific company, covering the entire retrieval process and online serving, along with key insights and challenges encountered.
\end{itemize}

The remaining of this article is structured as follows: Section~\ref{sec:background}  introduces the background of retrieval in recommender systems, including problem formulation, basic retrieval strategies and input data for retrieval. 
% Sections~\ref{sec:tra} and ~\ref{sec:neural} delve into early retrieval methods centered around CF and MF, and neural methods based on EBR, respectively. Section~\ref{sec:neg} discusses key aspects of model learning, such as widely-used loss functions and different negative sampling strategies used in the retrieval stage. 
Sections~\ref{sec:tra} and ~\ref{sec:neural} detail the efforts that focus on similarity learning using shallow and deep structures, respectively. Section~\ref{sec:index} discusses improvements in indexing mechanisms for efficient retrieval. Section~\ref{sec:neg} covers optimization methods of retrieval, including key aspects of model learning, such as widely-used loss functions and different negative sampling strategies. 
Section~\ref{sec:metric} presents commonly used evaluation metrics for both offline and online settings.
Section~\ref{sec:exp} summarizes commonly used datasets and conducts detailed experiments on three public datasets to assess the effectiveness of various methods. Section~\ref{sec:industry} presents a specific case study of current industrial practices in the retrieval stage at a specific company. We discuss open challenges and suggest promising future research directions in Section~\ref{sec:future}, and conclude the survey in Section~\ref{sec:concl}.
\section{BACKGROUND}\label{sec:background}
In this section, we first formulate the task of retrieval methods. Then we discuss the existing retrieval strategies from a broad perspective, including non-personalized and personalized retrieval. 
Next, we cover the common input data or information used in retrieval methods, including the user-item rating matrix and rich side information.
Finally, we introduce the multi-channel retrieval, which is widely used in real-world recommender systems.

\subsection{Problem Formulation}
We denote the universal user and item set as $\mathcal{U}$ and $\mathcal{I}$, respectively.
For a given user $u$, the objective of the retrieval method is to identify all potentially relevant items from the vast candidate pool $\mathcal{I}$. 
Generally, the retrieval method can be formulated as:
\begin{equation}\label{equ:general_formal}
\mathcal{C}(u, \mathcal{I}) = \{i \in \mathcal{I} \mid s(u, i) > \theta\},
\end{equation}
where $\mathcal{C}(u, \mathcal{I})$ is the candidate set for user $u$, $s(u, i)$ is a scoring function that measures the relevance of item $i$ to user $u$, and $\theta$ is a threshold for relevance.
Unlike the ranking phase, which deals with a smaller candidate set, the retrieval stage may involve filtering through millions of items. Consequently, efficiency becomes a critical concern for models used in this stage.
As a result, most retrieval methods employ a dual-tower architecture where user and item representations are learned separately.
Formally, given user $u \in \mathcal{U}$ and item $i \in \mathcal{I}$, the process can be abstracted as:
\begin{equation}\label{equ:formalize}
s(u, i) = f(\phi_u(u), \phi_i(i)),
\end{equation}
where $\phi_u:\mathcal{U}\xrightarrow{}\mathcal{H}$ and $\phi_i:\mathcal{I}\xrightarrow{}\mathcal{H}$ represent the mappings of user space $\mathcal{U}$ and item space $\mathcal{I}$ to a new space $\mathcal{H}$, respectively.
$f$ denotes the similarity measure between user and item, utilizing metrics like inner product or cosine similarity.
For any user-item pair $(u, i)$, $s(u, i)$ provides a score reflecting the similarity  between $u$ and $i$. This score enables the ranking of all items in the corpus $\mathcal{I}$ according to their predicted similarity scores. By ranking the items based on these scores, the retrieval method can efficiently generate a set of items, forming the retrieved set for user $u$.

\subsection{Retrieval Strategy}
Before exploring specific retrieval methods, it is important to distinguish between retrieval strategies and methods, and clarify the relationship between two of them, as they operate on different levels.
The retrieval strategy serves as the high-level concept that guides the design of various retrieval methods. Generally, as illustrated in Figure~\ref{fig:strategy} (a) and Figure~\ref{fig:strategy} (b), retrieval strategies can be broadly categorized into (1) non-personalized retrieval and (2) personalized retrieval. Non-personalized retrieval primarily aims at identifying and presenting trending topics. Though such an approach might not directly align with a user's specific interests, there is a high likelihood of user clicks due to the popularity of the content. In contrast, personalized retrieval tailors its recommendations to match user preference, significantly enhancing user engagement and retention. This category includes varied strategies like item-to-item (I2I) and user-to-item (U2I), which will be detailed further.
These strategies provide general guidance for the design of different retrieval methods. Understanding this distinction helps clarify how different approaches are applied in practice.

\subsubsection{Non-Personalized Retrieval}
Non-personalized retrieval plays a pivotal role in the retrieval stage without tailoring to individual user preference, as shown in Figure~\ref{fig:strategy} (a). One common strategy is promoting popular items, which is inspired by the ~`wisdom of the crowd'~\cite{surowiecki2005wisdom}. It is particularly effective for new users or in the absence of sufficient user interaction data~\cite{adomavicius2005toward}. 
Another strategy focuses on less mainstream items, aiming to diversify the user's preference and unearth user potential interest. Newly released item promotion~\cite{celma2008hits} is also a widely used strategy, which pushes the latest item to users, fueling exploration and maintaining a fresh and dynamic environment~\cite{fleder2009blockbuster}. 
Besides, the platform would design various operational algorithms to serve specific business goals or events, guiding users towards particular topics or items. Each of these strategies serves to enhance user engagement without personalization, necessitating the use of more advanced strategies and techniques.

\subsubsection{Personalized Retrieval}
\begin{figure}[!t]
\centering
\includegraphics[width=1\linewidth]{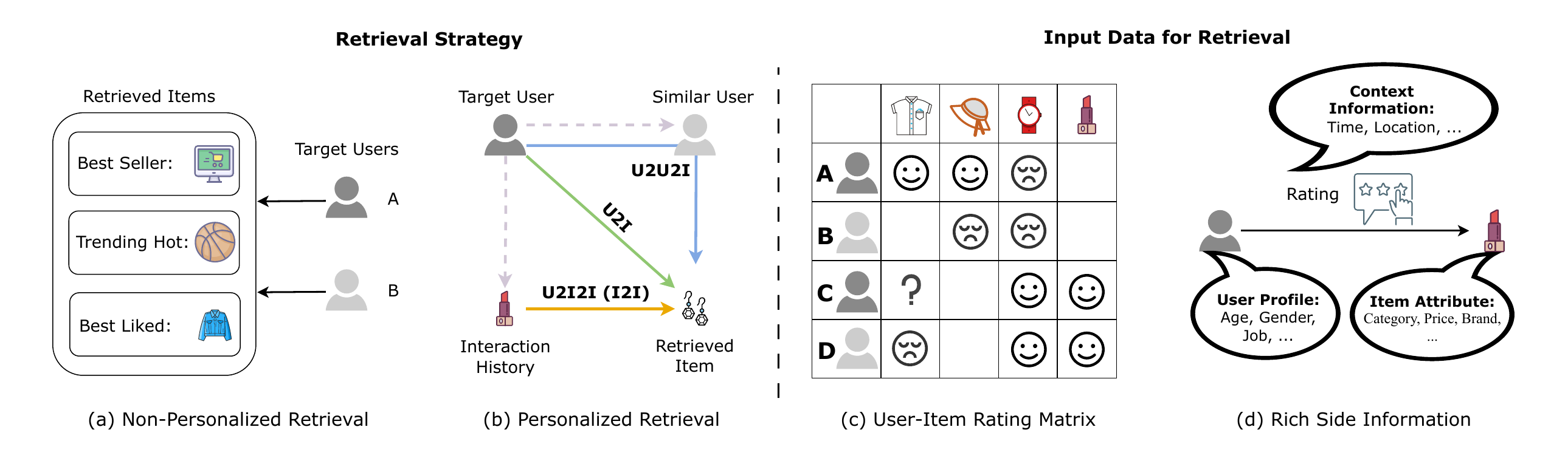}
\caption{An illustration of different retrieval strategies and input data for retrieval. Figure (a) illustrates non-personalized retrieval, a strategy that offers the same recommendation list to different users, such as ~`trending hot', without tailoring to individual user interests. Figure (b) depicts personalized retrieval, which includes three retrieval strategies: U2I, U2U2I, and U2I2I (I2I). Figure (c) shows the user-item rating matrix, which is the core data or information used in retrieval methods. Figure (d) presents information beyond the user-item rating matrix, commonly involving side information such as user profiles, item attributes, and context information.
}
\label{fig:strategy}
\vspace{-2mm}
\end{figure}

Personalized retrieval recommends items that align with user interests. While popular items may draw broad attention, they do not always match individual user preferences. Hence, personalized retrieval aims to cater to users' dynamic information needs and maintain their stickiness.
As is shown in Figure ~\ref{fig:strategy} (b), U2I, U2I2I (often shortened as I2I for simplicity), and U2U2I are three popular personalized retrieval strategies, where `U' stands for the user, `I' stands for the item, and `2' is short for `to' that denotes the process of seeking similarity or connection. For instance, U2I directly correlates the target user with items he/she might like; U2I2I identifies items similar to those the target user has previously interacted with; and U2U2I finds users with similar tastes to suggest items for the target user. 
Each path introduced above offers a unique strategy to discover and recommend items that user might be interested in.

% \ljh{CF and CB seem to be the retrieval methods instead of strategies, since they are algorithms to evaluate similarity. Suggest move this paragraph in section 3}
% \hjj{Section 3 explores specific retrieval methods. CF and CB are different ways to measure similarity, and when combined with U2I and I2I, they form various high-level retrieval strategies. I think it fits well in this section.}
Additionally, recommender systems often use collaborative filtering (CF) and content based (CB) classifications. CF operates on the principle of gauging similarity based on user interaction data, typically through a rating matrix reflecting user interactions with items. This approach assumes that users who have agreed in the past are likely to agree in the future. On the other hand, CB methods focus on item attributes or user profiles to determine similarity. It analyzes item features and recommend items similar to those a user has shown interest in before, based on the content of the items themselves.
Both CF and CB offer distinct ways to measure similarity, with CF leveraging user behavior patterns across the community, and CB concentrating on the item attributes and user profiles.
The combination of these strategies allows for a more nuanced approach to personalized retrieval, enhancing user experience by connecting individuals with content that resonates with their interests and past interactions.

\subsection{Input Data for Retrieval}
In this section, we will introduce the common input data or information used in retrieval methods, which includes the user-item rating matrix and rich side information. We categorize the side information into three aspects: user profile, item attributes, and contextual information.
\subsubsection{User-Item Rating Matrix}
The user-item rating matrix, denoted as $R$, is the fundamental input data for retrieval, capturing user preferences towards various items. 
Each entry $R_{ij}$ within the matrix indicates the preference of user $i$ for item $j$; if $R_{ij} > 0$, it suggests a positive preference. Preferences can be recorded directly through explicit ratings or indirectly through binary indicators that reflect actions such as clicks, views, or purchases. 
In Figure~\ref{fig:strategy} (c), we show the user-item rating matrix, where positive feedback is indicated by smiley faces and negative feedback by sad faces. It is important to note that the matrix is typically sparse, as the known preferences are usually limited, with many entries unspecified. 
We usually regard these unspecified (i.e., not interacted) entries as negative samples.

\subsubsection{Rich Side Information}
In addition to the user-item rating matrix, three types of side information play crucial roles in enhancing retrieval methods in recommender systems~\cite{shi2014collaborative}, as illustrated in Figure ~\ref{fig:strategy} (d). One vital source is the user profile, which might include demographics information, such as a user's gender, age, and hobbies. Item attributes also provide rich information, detailing properties of the items like their category or content. Contextual information provides insights into the circumstances under which user interactions with items occur~\cite{adomavicius2010context}. Commonly included in this category are timestamps, which track when interactions like ratings or purchases take place~\cite{xiong2010temporal, koren2009collaborative}. Additionally, other relevant details like a user's location during an app download~\cite{bohmer2011falling} or their hunger status when rating a food menu~\cite{ono2009context} can also enrich the understanding of user-item interactions, further tailoring the retrieval process to specific user conditions or preferences.

Additionally, in recent years, user-generated content (UGC) has become increasingly prevalent on UGC platforms, enriching item attributes. Tags, for instance, are short textual labels that users freely assign to items without being limited to a predefined list of categories~\cite{robu2009emergence}. These tags can describe the characteristics of an item or express users' feelings, serving as a valuable source of data. 
Reviews and comments published by users constitute another crucial source of side information that has been successfully leveraged to improve the performance~\cite{levi2012finding}.

Empirical evidence indicates that incorporating rich side information yields substantial improvements over using only the user–item rating matrix. For instance, \citet{tso2008tag} report that fusing tag information with ratings increases Recall@10 from 0.32 to 0.34 compared to pure user and item based collaborative filtering on the Last.fm~\cite{song} dataset. SocialMF~\cite{jamali2010matrix} incorporates social trust propagation into matrix factorization, reducing RMSE from 1.175 to 1.075 on the Epinions dataset~\cite{epinions} and from 0.878 to 0.821 on Flixster~\cite{flixster} relative to basic matrix factorization. With respect to contextual data, \citet{koren2009collaborative} demonstrate that adding temporal dynamics reduces RMSE from 0.913 to 0.897 on the Netflix movie-rating dataset~\cite{bennett2007netflix} compared to a static SVD++ baseline. Details on datasets appear in Section~\ref{sec:exp}, and metric definitions appear in Section~\ref{sec:metric}.

\subsection{Multi-Channel Retrieval}~\label{sec:mc}
Multi-channel retrieval is widely used in recommender systems, which employs various independent retrieval methods to individually retrieve different subsets of items~\cite{covington2016deep, grbovic2018real}. 
As is shown in Figure ~\ref{fig:cascade} (b), these retrieved item subsets are then aggregated to form a comprehensive candidate pool for downstream ranking stages. 
% This strategy balances computational efficiency with recall rate. It ensures rapid candidate retrieval through simple methods, while diverse channels designed from different perspectives aim to approximate an ideal recall rate without compromising the effectiveness of ranking.
The primary motivation is to maximize the coverage of users' diverse interests and improve the recall rate~\cite{zhang2019deep}. By using different retrieval methods, multi-channel retrieval captures a wide range of potential user preferences and enhances the overall performance of recommender systems.

The choice of retrieval channel is highly dependent on the specific applications. 
For example, the `trending list' and `interest tags' are essential for timely platforms like news streaming, while `friend favorites' is more important for social media platforms.
The number of items to be retrieved (i.e., top-$K$) can vary for different channels, and would be typically optimized through offline evaluation and online A/B testing~\cite{huang2025unleashing}.
% Each channel retrieves a set of candidate items, and for different channels, the value of top-$K$ (the number of items retrieved) can vary. The selection of top-$K$ is a hyperparameter typically optimized through offline evaluation and online A/B testing.
% \begin{figure}[!htbp]
% \centering
% \includegraphics[width=0.85\linewidth]{figures/multichannel.pdf}
% \caption{A depiction of a typical two-stage recommender system in the real world, where the retrieval stage includes multi channels.}
% \label{fig:multichannel}
% \end{figure}

\subsection{Comparison with Multi-Stage Retrieval in Ad Hoc Search}
Multi-stage cascaded pipelines are also well established in ad hoc search~\cite{clarke2016assessing, chen2017efficient}. \citet{clarke2016assessing} introduce a framework that quantifies efficiency–effectiveness trade-offs in multi-stage query systems without relying on relevance judgments.
\citet{chen2017efficient} propose a cost-aware cascade ranking model that jointly optimizes per-stage cutoffs to balance latency against retrieval quality in a static document index.

Although both ad hoc search and recommender system pipelines employ the same core principle, i.e., progressively narrowing a large corpus through increasingly expensive stages, they differ along several key dimensions:
\begin{itemize}[topsep = 3pt,leftmargin =*]
    \item \textbf{Personalization and dynamics.} Ad hoc search typically serves identical results to all users over a fixed index, whereas recommender systems have to tailor candidate sets to individual user profiles and adapt continuously as new items and interactions arrive.
    \item \textbf{Data freshness and indexing.} Ad hoc indices are updated at regular intervals, making feature extraction relatively stable; recommender pipelines often require real-time or near-real-time indexing to incorporate the latest user behavior and item metadata.
    \item \textbf{Feature representations.} Early ad hoc stages use term-based statistics (e.g., BM25~\cite{robertson1995okapi}), while the retrieval stage in recommendation leverages learned embeddings that combine user history, content attributes, and contextual signals.
\end{itemize}

\section{SIMILARITY LEARNING USING SHALLOW STRUCTURES}\label{sec:tra}
% Early efforts in retrieval models primarily focus on CF.
In Sections~\ref{sec:tra} and ~\ref{sec:neural}, we discuss efforts focused on accurate user-item similarity learning, thereby enhancing effectiveness. 
This section centers on similarity learning using shallow structures, which primarily revolves around the concept of collaborative filtering (CF).
CF, as the name suggests, leverages collective feedback, evaluations, and opinions to help users filter through vast amounts of information, which is a cornerstone technique in the retrieval stage of recommender systems. 
% As illustrated in Figure~\ref{fig:cf}(a), the interaction between users and items can typically be represented by a bipartite graph, where positive feedback is indicated by smiley faces and negative feedback by sad faces. In the context of CF, for computational convenience, this interaction graph is transformed into a user-item rating matrix in Figure~\ref{fig:cf}(b).
These approaches can be divided into two categories: (1) neighborhood based~\cite{sarwar2001item} and (2) matrix factorization (MF) based~\cite{koren2009matrix, manotumruksa2017deep} methods. In this section, we will explore these two types of traditional retrieval methods in details.
% \begin{figure}[!htbp]
% \centering
% \includegraphics[width=0.85\linewidth]{figures/cf.pdf}
% \caption{An illustration of data utilized in CF-based techniques.}
% \label{fig:cf}
% \end{figure}

\subsection{Neighborhood Based Collaborative Filtering Methods}

The neighborhood based CF method (also called memory based CF methods in some literature~\cite{su2009survey, chen2018survey}) is a fundamental subset of collaborative filtering techniques used in the retrieval stage. These methods operate by clustering users or items based on similarities in their interactions and preferences, thereby forming ~`neighborhoods'. There are primarily two types of neighborhood based CF methods: user based and item based, each with its distinct processes, advantages, and application contexts.

\subsubsection{User Based CF}
This approach focuses on finding similar users to the target user. Similarity between users is typically calculated using measures such as cosine similarity or pearson correlation coefficient based on their ratings in a user-item rating matrix~\cite{breese2013empirical, sarwar2000analysis}. The similarity between user $u$ and user $v$ can be represented as:
\begin{equation}
s(u, v) = \frac{\sum_{i \in I_{uv}} (r_{ui} - \overline{r_u})(r_{vi} - \overline{r_v})}{\sqrt{\sum_{i \in I_{uv}} (r_{ui} - \overline{r_u})^2} \sqrt{\sum_{i \in I_{uv}} (r_{vi} - \overline{r_v})^2}},
\end{equation}
where $I_{uv}$ is the set of items rated by both users $u$ and $v$, $r_{ui}$ is the rating of user $u$ for item $i$, and $\overline{r_u}$ is the average rating of user $u$.
Once similar users are identified, the system recommends items that these similar users have liked but the target user has not yet interacted with. 
User based CF can provide highly personalized recommendations since it directly leverages user behaviors and preferences. However, its scalability~\cite{karypis2001evaluation} is a significant issue due to the computational complexity of comparing each user with every other user, especially in systems with a large number of users. It also suffers from the cold-start problem for new users with insufficient interaction data.

\subsubsection{Item Based CF}
Item based CF focuses on calculating item similarity based on the patterns of users who have interacted with them~\cite{sarwar2001item, linden2003amazon}. The similarity between item $i$ and item $j$ can be calculated as:
\begin{equation}
s(i, j) = \frac{\sum_{u \in U_{ij}} (r_{ui} - \overline{r_i})(r_{uj} - \overline{r_j})}{\sqrt{\sum_{u \in U_{ij}} (r_{ui} - \overline{r_i})^2} \sqrt{\sum_{u \in U_{ij}} (r_{uj} - \overline{r_j})^2}},
\end{equation}
where $U_{ij}$ is the set of users who have rated both items $i$ and $j$, $r_{ui}$ is the rating of user $u$ for item $i$, and $\overline{r_i}$ is the average rating of item $i$.
Recommendations to a user are made by identifying items similar to those the user has already liked or interacted with, assuming that users will prefer items similar to their previous interests.
Item based CF tends to be more scalable than user based CF, as the item-item similarity matrix is typically more stable and changes less frequently than user-user similarities~\cite{sarwar2001item, bell2007scalable, takacs2007major}. It also mitigates the cold-start problem for long-tail users with limited interactions to some extent, but still struggles with completely cold-start users.
Item based CF is particularly effective in systems where item relationships are more pronounced and stable, such as movie or book recommendation systems where items have longer lifespans and well-defined features, but it is not suitable in cases where user tastes are highly diverse or when new items frequently enter the system.

Both user based CF and item based CF enhance the retrieval process by incorporating the vast amount of user-item interactions. However, the choice between them, or the decision to integrate both approaches, depends largely on the specific requirements, scalability considerations, and the nature of user interactions within the recommendation platform. Modern recommender systems often employ hybrid models that combine the strengths of user based CF and item based CF to overcome their respective limitations.

\subsubsection{Extensions to Neighborhood Based CF}
Traditional neighborhood based CF methods, like those described above, primarily utilize direct pairwise interactions between users and items to make recommendations. These are either based on user-user or item-item similarities derived directly from the user-item rating matrix. The Swing algorithm~\cite{yang2020large} introduces a quasi-local structure that goes beyond these direct interactions by examining the relationship among users and items in a more comprehensive manner.
The Swing algorithm is detailed using the concept of a swing score, which measures the relationship strength between two items based on their shared interactions with users. Specifically, the swing score between item $i$ and item $j$ is defined in ~\eqref{equ:swing}, where $U_i$ is the set of users who have clicked item $i$, and $I_u$ is the set of items that user $u$ has clicked. The parameter $\alpha$ serves as a smoothing coefficient to  manage the influence of highly interactive user-item pairs. By utilizing such quasi-local structures, Swing can uncover more complex and subtle patterns in user-item interactions, leading to a richer and more accurate collaborative filtering result. 
\begin{equation}\label{equ:swing}
s(i, j)=\sum_{u \in U_i \cap U_j} \sum_{v \in U_i \cap U_j} \frac{1}{\alpha+\left|I_u \bigcap I_v\right|}
\end{equation}

Additionally, some studies have built upon the user-item rating matrix by using side information to refine similarity calculations~\cite{shi2014collaborative}. 
~\citet{melville2002content} first propose filling missing values in the user-item rating matrix using item side information such as titles and genres. 
Another category of approaches focuses on using tags as side information~\cite{firan2007benefit, shepitsen2008personalized}. ~\citet{tso2008tag} develop a method to blend user-user or item-item similarities based on both tags and ratings within a neighborhood based CF framework.
To combat the inherent noisiness of tags, schemes to enhance tag reliability have been introduced, such as assigning different weights to different tags and integrating these with conventional neighborhood based CF approaches~\cite{liang2010connecting}.

\subsection{Matrix Factorization Based Collaborative Filtering Methods}
% netflix
\subsubsection{MF Based CF}
Matrix factorization (MF) based Collaborative Filtering (CF) methods~\cite{koren2009matrix}, also referred to as latent factor models~\cite{koren2008factorization, su2009survey}, play a pivotal role in the retrieval stage. Originating from the need to effectively map both users and items into a latent factor space, these methods estimate user-item interactions, typically ratings, by calculating the dot product of their latent factors. 
The essence of MF involves decomposing the user-item rating matrix \( R \) (of shape \( M \times N \), where \( M \) is the number of users and \( N \) is the number of items) into two lower-dimensional matrices: \( U \) (of shape \( K \times M \), representing latent user preferences) and \( V \) (of shape \( K \times N \), representing latent item attributes), aiming to approximate the original rating matrix \( R \) with the product \( U^T \times V \) as shown in \eqref{equ:mf}.
Optimization of these methods, given the sparsity of the rating matrix and computational constraints, is typically achieved through techniques such as stochastic gradient descent, complemented by regularization terms to mitigate overfitting.
We use the same notation in previous works~\cite{shi2014collaborative} and as is shown in ~\eqref{equ:mf2}, \( U \) and \( V \) represent matrices of latent factors, where \( U^* \) and \( V^* \)
denote their optimal values determined through the optimization process. Each column vector \( U_i \) in \( U \) corresponds to the latent factors associated with user \( i \), while \( V_j \) in \( V \) corresponds to the latent factors for item \( j \). The indicator function \( I_{ij} \) equals 1 if \( R_{ij} > 0 \) and 0 otherwise. The Frobenius norm of a matrix is represented as \( \| U \|_F \), and \( \lambda_U \) and \( \lambda_V \) are regularization parameters used to prevent overfitting in the model.

\begin{equation}\label{equ:mf}
R \approx U^T \times V.
\end{equation}
\begin{equation}\label{equ:mf2}
U^*, V^* = \arg \min_{U,V} \left\{ \frac{1}{2} \sum_{i=1}^M \sum_{j=1}^N I_{ij} (R_{ij} - U_i^T V_j)^2 + \frac{\lambda_U}{2} \| U \|_F^2 + \frac{\lambda_V}{2} \| V \|_F^2 \right\}
\end{equation}

% These models gained prominence during the Netflix Challenge, highlighting their capability to handle large and sparse datasets by uncovering complex, latent relationships within the data. To enhance the basic MF approach, variations like SVD++~\cite{koren2008factorization} have been introduced, which integrate neighbor-based models to incorporate both implicit and explicit feedback, thus enriching the model's accuracy and applicability. Additionally, Factorization Machines (FM)~\cite{rendle2010factorization} extend the MF concept further by incorporating a broader range of features and interactions, making them adaptable to diverse data types and sparse environments.

\subsubsection{Extentions to MF Based CF}
MF based CF gained prominence during the Netflix Challenge~\cite{takacs2008matrix}, highlighting their capability to handle large and sparse datasets by uncovering complex, latent relationships within the data.
To expand upon the foundational MF approach, advanced variations like SVD++~\cite{koren2008factorization} were developed. SVD++ not only considers the explicit ratings that users provide but also integrates implicit feedback—such as browsing or purchase history—to enhance the model's predictive accuracy and relevance. This integration helps capture a more complete picture of user behavior and preferences, greatly enriching the algorithm's applicability.

% Further extending the capabilities of traditional MF, Factorization Machines (FM)~\cite{rendle2010factorization} incorporate even more diverse data elements, such as social network connections, temporal dynamics, and other contextual information. This inclusion allows FMs to model interactions among all available features, providing a powerful framework capable of operating within highly sparse data environments and across different domains. The versatility of FMs makes them particularly effective in tasks where data interconnectivity plays a critical role.
Further extending the capabilities of traditional MF, Factorization Machine (FM)~\cite{rendle2010factorization} incorporates additional side information, which MF does not typically include. Unlike traditional MF, which focuses on learning latent representations for user and item IDs, FM is able to model interactions among more available features, providing a powerful framework capable of operating within highly sparse data environments and across different domains, where data interconnectivity plays a critical role.
Time based dynamics are considered in methods like TimeSVD++~\cite{koren2009collaborative}, which specifically incorporates temporal information to improve predictions over time.
Additionally, SLIM~\cite{ning2011slim} is a simple linear model that merges the benefits of neighborhood CF and latent factor models.
These enhancements and innovations in MF based methods underscore their vital roles in the evolution of retrieval methods, which offer a scalable solution that delivers quick and precise predictions even for large databases~\cite{bobadilla2013recommender}.
The methods discussed so far primarily use shallow structures for user-item similarity learning.
In Section~\ref{sec:neural}, we will introduce more recent works that leverage deep structures, exploring how deep learning techniques are leveraged to further improve the effectiveness of retrieval methods. 
\section{SIMILARITY LEARNING USING DEEP STRUCTURES}~\label{sec:neural}
% \begin{figure}[!htbp]
% \centering
% \includegraphics[width=1\linewidth]{figures/deepall.pdf}
% \caption{The architecture of two-tower methods.}
% \label{fig:deepall}
% \end{figure}

\begin{figure}[!htbp]
\centering
\includegraphics[width=1\linewidth]{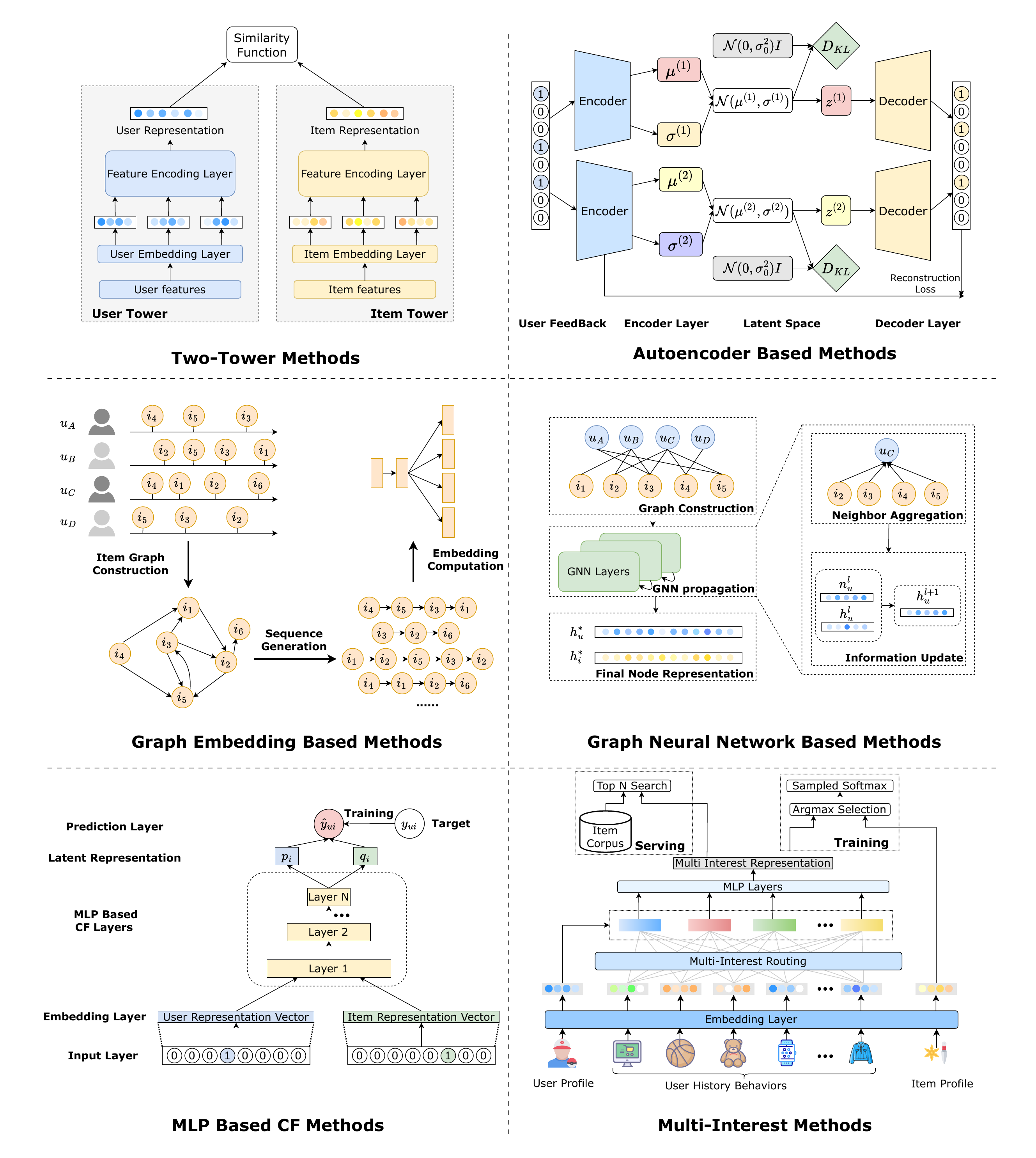}
\caption{The architecture of deep retrieval methods focused on similarity learning.}
\label{fig:deepall}
\end{figure}

Recently, with the advent of deep learning, there has been a significant shift towards more sophisticated and deep structures for user-item similarity learning. Modern recommender systems are increasingly adopting embedding based retrieval (EBR). EBR transforms the retrieval problem into a nearest neighbor search within a vector space, where both users and items are represented as points. This structure allows EBR to incorporate rich user and item features into the representation learning process, considering a broader range of features beyond mere user and item IDs.
We classify these methods into six categories based on their structure as is shown in Figure~\ref{fig:deepall} (i.e., tow-tower methods, autoencoder based methods, graph embedding based methods, graph neural network based methods, MLP based collaborative filtering methods, and multi-interest methods), and we will now discuss each category in details.

\subsection{Two-Tower Methods}
% \begin{figure}[!htbp]
% \centering
% \includegraphics[width=0.7\linewidth]{figures/twotower.pdf}
% \caption{The architecture of two-tower methods.}
% \label{fig:twotower}
% \end{figure}
The two-tower methods represent a state-of-the-art approach in the retrieval stage, offering a balance between prediction accuracy and inference efficiency~\cite{li2022inttower}. 
In this architecture, each user $u$ and item $i$ is independently mapped to a representation via two parallel sub-networks:
$\phi_u(u; \theta_u)$ and
$\phi_i(i; \theta_i)$,
where $\phi_u$ and $\phi_i$ denote the user tower and item tower, respectively, and $\theta_u,\theta_i$ are their parameters. 
Taking the user tower as an example, user features are first passed through an embedding layer, where user feature embeddings are obtained via an embedding look-up operation~\cite{zhang2016deep}. These embeddings are then processed through multiple fully connected layers to refine the representation.
The final output of the two-tower model is determined by the inner product of the user and item representations. 
\begin{equation}\label{equ:sec4.1}
s(u,i)=
\left\langle\phi_u(u; \theta_u) ,\phi_i(i; \theta_i)\right\rangle.
\end{equation}
This architecture decouples the user and item modeling, which is highly efficient and significantly reduces inference latency and computational resources.
Classic implementations of two-tower methods in industry include Microsoft's DSSM~\cite{huang2013learning}, Google's YouTubeDNN~\cite{covington2016deep}, and Airbnb's personalized embeddings~\cite{grbovic2018real}, leading to significant performance improvements.
The two-tower methods mentioned above, while efficient, often face limitations in adequately modeling user-item interactions, which can impede model performance. 
To overcome such challenges, advanced techniques have been developed to enhance the depth of interaction between the user and item side, which we will further detail in future directions in Section~\ref{sec:future}.

\subsection{Autoencoder Based Methods}
Autoencoder based methods, particularly Variational Autoencoders (VAEs)~\cite{kingma2013auto}, are increasingly utilized in the retrieval stage of recommender systems to address challenges like data sparsity and the complexity of user-item interactions. These methods excel in generating and representing complex data structures.
VAEs operate by assuming that each observed interaction vector $\mathbf{r}_u \in \{0,1\}^N$ is generated from a latent variable $\mathbf{z}$ following some probability distribution, such as standard Gaussian distribution. They consist of two main components: an encoder and a decoder. The encoder network $q_\phi(\mathbf{z}\mid \mathbf{r}_u)$ approximates the true posterior distribution of the latent variables given the input data, effectively capturing the underlying data structure:
\begin{equation}\label{equ:sec4.21}
\boldsymbol{\mu}_u = \mu(\mathbf{r}_u), 
\quad 
\log \boldsymbol{\sigma}_u^2 = \log\sigma^2(\mathbf{r}_u),
\end{equation}
\begin{equation}\label{equ:sec4.22}
\mathbf{z}_u = \boldsymbol{\mu}_u + \boldsymbol{\sigma}_u \odot \boldsymbol{\epsilon}, 
\quad \boldsymbol{\epsilon} \sim \mathcal{N}(\mathbf{0}, \mathbf{I}).
\end{equation}
The decoder then reconstructs $\mathbf{r}_u$ from the latent representation, as shown in ~\eqref{equ:sec4.23}, where $\sigma(\cdot)$ denotes the element-wise sigmoid function.  It aims to match the original input as closely as possible.
\begin{equation}\label{equ:sec4.23}
\hat{\mathbf{r}}_u = \sigma\bigl(\mathrm{Decoder}(\mathbf{z}_u)\bigr),
\end{equation}

The effectiveness of VAEs in the retrieval stage of recommender systems is showcased by their ability to learn comprehensive user preferences and item characteristics through deep neural networks.
~\citet{wu2016collaborative} introduce CDAE, which incorporates a user-specific bias into an autoencoder framework. CDAE represents a generalization of various existing CF techniques.
Mult-VAE~\cite{liang2018variational} sparks interest in VAE for CF by showcasing its superiority over other benchmarks. Simultaneously, ~\citet{lee2017augmented} explore the VAE framework for CF with an emphasis on conditional and joint VAE formulations to integrate auxiliary user data.
CVAE~\cite{li2017collaborative} and MD-CVAE~\cite{zhu2022mutually} enhance collaborative latent variable modeling by incorporating item content information.
~\citet{karamanolakis2018item} apply VAE to personalized recommendations, delving into user-dependent priors. More recent efforts by ~\citet{kim2019enhancing} evaluate VAEs enhanced by VampPrior~\cite{tomczak2018vae}. 
MacridVAE~\cite{ma2019learning} and CaD-VAE~\cite{wang2023causal} focus on modeling the latent factors behind interactions to accurately capture the distribution of latent variables.
RaCT~\cite{lobel2019towards} introduces an actor-critic reinforcement learning technique to optimize VAEs, while RecVAE~\cite{shenbin2020recvae} examines various regularization methods to refine VAE's performance in CF. BiVAE~\cite{truong2021bilateral} treats users and items symmetrically and proposed bilateral user and item based inference models to autoencode users and items under a unified framework.
FastVAE~\cite{chen2022fast} improves the efficiency of the model using an inverted multi-index technique. SE-VAE~\cite{cho2022stochastic} advances latent variable modeling by integrating a multi-expert system with stochastic expert selection, enhancing the model’s adaptability and accuracy.

\subsection{Graph Embedding Based Methods}
Graph embedding based methods have been pivotal in enhancing the retrieval process in recommender systems. These methods extend the principles of word embeddings, such as Word2Vec~\cite{mikolov2013distributed}, to user–item interaction graphs.
These methods learn a mapping that assigns each node $v$ in the graph $G=(\mathcal{V},\mathcal{E})$ to a vector $\Phi(v)$. They optimize a neighborhood‐preserving objective of the form:
\begin{equation}\label{equ:sec4.3}
\min_{\Phi} \sum_{v\in\mathcal{V}} \sum_{c\in\mathcal{N}(v)} -\log \Pr\bigl(c \,|\, \Phi(v)\bigr),
\quad
\Pr\bigl(c | \Phi(v)\bigr)
= \frac{\exp\bigl(\Phi(c)^\top \Phi(v)\bigr)}
       {\sum_{u\in\mathcal{V}} \exp\bigl(\Phi(u)^\top \Phi(v)\bigr)},
\end{equation}
where $\mathcal{N}(v)$ denotes a chosen context set for node $v$ (e.g., adjacent nodes, nodes co‐occurring in random walks, or nodes sharing side‐information). Minimizing this loss ensures that frequently co‐occurring or closely related nodes receive similar embeddings, while unrelated nodes are pushed apart.
Item2Vec~\cite{barkan2016item2vec} is a direct adaptation of Word2Vec to items, treating the user behavior sequences of items (like products or songs) as sentences to uncover similarities based on co-occurrence. This enables better recommendations by capturing contextual similarities among items. For example, Airbnb successfully applied Item2Vec~\cite{grbovic2018real} to personalize user experiences in online accommodations. DeepWalk~\cite{perozzi2014deepwalk} uses random walks on graphs to generate node sequences, learning representations that reflect the social structure of interactions. Node2Vec~\cite{grover2016node2vec} enhances this approach by balancing breadth-first and depth-first sampling, capturing diverse connectivity patterns in the graph.
Further refining this, EGES~\cite{wang2018billion} integrates side information (like user demographics or item attributes) into the embeddings, enriching the model's contextual understanding. LINE~\cite{tang2015line} preserves both direct connections and neighbor similarities, ensuring embeddings reflect both explicit and implicit relationships.
SDNE~\cite{wang2016structural} uses autoencoders to learn node representations, adeptly capturing nonlinear relationships within complex, hierarchical structures.
All these models underscore the importance of graph embeddings in understanding and leveraging complex user-item interaction patterns. 

\subsection{Graph Neural Network Based Methods}
% \begin{figure}[!htbp]
% \centering
% \includegraphics[width=0.93\linewidth]{figures/gnn.pdf}
% \caption{The framework of GNN based methods in the retrieval stage.}
% \label{fig:gnn}
% \end{figure}
Recent advancements have incorporated Graph Neural Network (GNN) techniques into the retrieval stage of recommender systems~\cite{berg2017graph, he2020lightgcn,sun2019multi, liu2020deoscillated}, known for their efficacy in handling complex relational data.
To understand the distinction between graph embedding based methods and GNN based methods, it is essential to first explain their foundational differences. Graph embedding based methods typically involve deriving item sequences from the graph and then generating embeddings from these sequences. These methods focus on capturing the structural properties and proximities within the graph to represent items in a continuous vector space.
In contrast, GNN based methods utilize graph convolution operations, which involve iterative message passing and aggregation across the graph's nodes and edges. 
In a typical message-passing paradigm, for each node $v$ at layer $k$, one computes
\begin{equation}\label{equ:sec4.4}
\begin{aligned}
m_v^{(k)} &= \mathrm{AGG}\bigl(\{\,h_u^{(k-1)} : u \in \mathcal{N}(v)\}\bigr),\\
h_v^{(k)} &= \mathrm{UPDATE}\bigl(h_v^{(k-1)},\,m_v^{(k)}\bigr),
\end{aligned}
\end{equation}
where $\mathcal{N}(v)$ is the set of neighbors of $v$, $h_v^{(k)}$ is the representation of node $v$ at layer $k$, $\mathrm{AGG}(\cdot)$ defines a pooling or aggregation function over neighbors (e.g., mean, sum, or attention), and $\mathrm{UPDATE}(\cdot)$ is a function—often a linear transformation followed by a nonlinearity—that integrates the previous representation with the aggregated message.
GNNs excel at capturing intricate user-item interactions, optimizing the retrieval process by exploiting high-order relationships. Figure~\ref{fig:deepall} illustrates the pipeline for applying GNNs in retrieval.
\begin{itemize}[topsep = 3pt,leftmargin =*]
    \item \textbf{Graph Construction.}
    Applying GNNs directly to user-item graphs~\cite{berg2017graph, he2020lightgcn, sun2020neighbor, zheng2018spectral} can face challenges due to sparse connections or large scale. Enhancements such as adding connections~\cite{sun2019multi, liu2020deoscillated} or virtual nodes~\cite{dgcf, hignn} improve graph connectivity and retrieval quality. Techniques like Multi-GCCF~\cite{sun2019multi} and DGCF~\cite{liu2020deoscillated} enhance graphs by linking two-hop neighbors, while HiGNN~\cite{hignn} clusters similar entities to enhance user-item graphs.
    For better efficiency, sampling methods like those used in Multi-GCCF~\cite{sun2019multi} and PinSage~\cite{pinsage} help process large-scale graphs by balancing the retention of meaningful information and computational demands.
    % Directly applying GNNs to the original user-item bipartite graph~\cite{berg2017graph, chen2020revisiting, he2020lightgcn, li2019hierarchical, sun2020neighbor, tan2020learning, wang2019binarized, wang2019neural, wu2020joint, zhang2019star, zheng2018spectral} can be straightforward but may face effectiveness and efficiency issues due to sparse connections or the graph's large scale. Addressing effectiveness, enhancing the original graph by adding connections~\cite{sun2019multi, liu2020deoscillated, ji2020dual} or virtual nodes~\cite{dgcf, hignn} can deepen the understanding of user-item relationships, improving the quality of the recommendations. 
    % For example, Multi-GCCF~\cite{sun2019multi} and DGCF~\cite{liu2020deoscillated} enhance graph connectivity by linking two-hop neighbors, creating user-user and item-item graphs to better represent user-item proximity. HiGNN~\cite{hignn}, on the other hand, creates user-item graphs by clustering similar entities and using cluster centers as new graph nodes, effectively revealing the hierarchical structure within user-item interactions.
    % For efficiency, strategic sampling methods are essential for processing large-scale graphs, balancing the need to retain meaningful information while reducing computational demands. Multi-GCCF~\cite{sun2019multi} and NIA-GCN~\cite{sun2020neighbor} employ random sampling to determine neighbor sets. PinSage~\cite{pinsage} introduces a random-walk sampling technique to select neighborhoods of a fixed size based on frequency of visits, allowing non-adjacent nodes to link. 
    \item \textbf{Neighbor Aggregation.}
    This step propagates information through the graph. Methods vary from mean-pooling~\cite{berg2017graph, tan2020learning} to more sophisticated methods like attention mechanisms~\cite{mccf, disenhan} that weigh neighbors based on relevance. PinSage~\cite{pinsage} uses normalized visit counts to determine neighbor importance.
    % The neighbor aggregation step, which is crucial for propagating information through the graph, varies in its approach to incorporate neighbors' information. Techniques range from mean-pooling~\cite{berg2017graph, sun2019multi, tan2020learning, zhang2019star, zhang2019inductive}, which treats all neighbors equally, to more nuanced methods like degree normalization~\cite{chen2020revisiting, he2020lightgcn, wu2020joint} and attention mechanisms that weigh neighbors based on relevance or structure~\cite{mccf, disenhan, wang2019multi}. PinSage~\cite{pinsage}, for example, uses normalized visit counts to gauge neighbor importance. Recent models like MCCF~\cite{mccf} and DisenHAN~\cite{disenhan} apply attention mechanism to differentiate neighbor importance further, highlighting the need for methods that consider the intensity of connections between nodes. Besides, NGCF~\cite{wang2019neural} enhances the relevance of item features to user interests and vice versa through an element-wise product.
    \item \textbf{Information Update.}
    Updating node representations is crucial. Some methods fully replace node representations with aggregated information~\cite{berg2017graph, he2020lightgcn}, while others combine original attributes with neighbor information through pooling or transformations~\cite{li2019hierarchical, pinsage}. LightGCN~\cite{he2020lightgcn} simplifies this process by omitting non-linearities, enhancing performance and efficiency.
    % Updating the representation of the node is crucial for information propagation. Approaches differ on whether to keep the original representation of nodes. Some methods fully replace the node's representation with the aggregated information from neighbors~\cite{berg2017graph, he2020lightgcn, dgcf, zhang2019star}, ignoring the inherent user preference and item features. Alternatively, combining the original node's attributes with the information of its neighbors, either through straightforward pooling~\cite{sun2020neighbor, wang2019neural, wu2020joint, zhang2019inductive} or more sophisticated methods like concatenation and non-linear transformations~\cite{li2019hierarchical, sun2019multi, pinsage}, allows for a richer representation. 
    % Notably, some models like LightGCN~\cite{he2020lightgcn} simplify the transformation process by omitting non-linearities, improving both performance and computational efficiency by maintaining a balance between original and aggregated information.
    \item \textbf{Final Node Representation.}
    After multiple layers of aggregation and updating, the final node representation is used for predictions. Methods like weighted pooling and concatenation integrate information from different layers, leveraging the unique characteristics captured at each depth.
    % Layer-by-layer application of neighbor aggregation and information updating results in depth-specific node representations. Usually, the representation from the last layer is used for predictions~\cite{berg2017graph, li2019hierarchical, tan2020learning, disenhan, pinsage, zhang2019star}, but recognizing the unique contributions of different layers has led to more sophisticated integration methods~\cite{wang2019neural}. These methods, including weighted pooling, which offers flexibility in emphasizing the significance of various layers, and concatenation, which preserves comprehensive information across all layers. These approach allows for a more nuanced representation by leveraging the distinct characteristics captured at each depth.
\end{itemize}

Recent studies~\cite{yu2022graph, ye2023towards, tang2024towards} have demonstrated that contrastive learning in GNN based retrieval significantly improves robustness. \citet{yu2022graph} argue that costly graph perturbations are unnecessary; instead, generating contrastive views by injecting uniform noise into the embedding space achieves superior performance and efficiency. Subsequent research~\cite{ye2023towards, tang2024towards} has further enhanced model resilience by applying structure denoising and adversarial embedding perturbations.
Although GNN based retrieval methods show remarkable results in understanding and leveraging complex user-item interaction patterns, enhancing their efficiency for large-scale retrieval remains a critical problem.

\subsection{MLP Based Collaborative Filtering Methods}
Unlike conventional collaborative filtering techniques that often rely on linear models and explicit factorization methods, neural collaborative filtering (NCF)~\cite{he2017neural} uses an MLP to model non‐linear user–item interactions.
Concretely, let $\mathbf{e}_u,\mathbf{e}_i$ be the embedding vectors for user $u$ and item $i$. NCF concatenates these embeddings as
\begin{equation}\label{equ:sec4.51}
\mathbf{h}^{(0)}_{u,i} \;=\; \bigl[\mathbf{e}_u \,\|\, \mathbf{e}_i\bigr],
\end{equation}
and then applies $L$ hidden layers:
\begin{equation}\label{equ:sec4.52}
\mathbf{h}^{(\ell)}_{u,i} \;=\; Activation\bigl(\mathbf{W}^{(\ell)}\,\mathbf{h}^{(\ell-1)}_{u,i} + \mathbf{b}^{(\ell)}\bigr), 
\quad \ell=1,\dots,L,
\end{equation}
where $\mathbf{W}^{(\ell)}$, $\mathbf{b}^{(\ell)}$. The final layer output $\mathbf{h}^{(L)}_{u,i}$ is then projected to a scalar score via another affine transformation and sigmoid activation.
% ~\citet{he2017neural} introduced neural collaborative filtering (NCF), a deep neural network approach that overcomes the limitations of linear matrix factorization (MF) methods by using a multi-layer perceptron (MLP) to capture user-item interactions through nonlinear optimization.
~\citet{bai2017neural} enhance NCF by incorporating local neighborhood information using convolutional filters, improving interaction representation between users and items.
~\citet{liu2017phd} propose PhD, a hybrid recommendation model combining user-item interactions with side information using probabilistic matrix factorization.
~\citet{cao2018attentive} develop a group recommendation method integrating an attention mechanism with MLP within the NCF framework.
~\citet{chen2018heterogeneous} utilize heterogeneous information networks (HIN) to represent diverse data types, capturing complex semantics through meta paths.
~\citet{fu2018novel} investigate both local and global co-occurrences in user-user and item-item relations to capture contextual information more comprehensively.
~\citet{zhou2019content} enrich the model's understanding by incorporating textual representations from items, enhancing both explicit and implicit feedback.
~\citet{chen2019joint} combine deep feature learning and deep interaction modeling using two MLP networks with a rating matrix.
~\citet{otunba2019deep} propose an ensemble framework integrating generalized matrix factorization (GMF) and MLP networks, leveraging the strengths of deep and shallow architectures.
~\citet{xue2019deep} focus on capturing high-order item relations in item based CF models using an NCF based approach.
~\citet{yi2019deep} develop deep matrix factorization (DMF), integrating diverse side information from users and items to enhance the retrieval process.
~\citet{chen2020efficient} propose efficient whole-data learning strategies to avoid common sampling techniques in CF retrieval methods.
Additionally, closely related methods like CML~\cite{hsieh2017collaborative} is based on metric learning, and NBPO~\cite{yu2020sampler} employs noisy-label robust learning techniques for retrieval.

\subsection{Multi-Interest Methods}
% \begin{figure}[!htbp]
% \centering
% \includegraphics[width=0.85\linewidth]{figures/multiinterest.pdf}
% \caption{A depiction of multi-interest retrieval methods.}
% \label{fig:multiinterest}
% \end{figure}
% \ljh{but how about traditional sequence methods like GRU4Rec and SDM? they model the user behavior sequence but only assign one vector for one user/item}
The deep neural retrieval methods mentioned above typically assign a single vector to represent each user, encapsulating their preferences and behaviors. However, this single vector representation often falls short in capturing the complexity and diversity of a user's multiple interests. As a result, recent works have developed models that encapsulate multiple interest vectors for each user.
Let $\mathcal{H}(u)$ be the set of items with which user $u$ has interacted, and let each item $i$ have an embedding $\mathbf{e}_i$. A general approach is to generate $K$ interest vectors for user $u$:
\begin{equation}\label{equ:sec4.6}
\bigl\{\mathbf{h}_{u,1}, \mathbf{h}_{u,2}, \dots, \mathbf{h}_{u,K}\bigr\}, 
\qquad
\mathbf{h}_{u,k} \;=\; f_k\bigl(\{\mathbf{e}_i : i \in \mathcal{H}(u)\}\bigr),
\end{equation}
where each $f_k(\cdot)$ denotes a mechanism, such as dynamic routing or self-attention, that extracts the $k^{th}$ interest representation from the user’s historical interactions. Consequently, each $\mathbf{h}_{u,k}$ corresponds to one latent interest of user $u$.
One of the pioneering models is MIND~\cite{mind}, which introduces a dynamic routing mechanism to assign each user interaction to specific interests. ComiRec~\cite{comirec} builds on this by employing a self-attentive mechanism to distill user interests from their activity patterns.
Further advancements include models like SINE~\cite{sine} and Octopus~\cite{octopus}, which establish interest pools and use attention mechanisms to activate relevant interests based on historical interactions. DGCF~\cite{dgcf} extends these concepts by integrating a dynamic routing mechanism into graph structures, focusing on the independence among interests for enhanced multi-interest learning.

The dynamic nature of user preferences is further explored in works like Pinnersage~\cite{pinnersage} and MIP~\cite{mip}, emphasizing the importance of models adapting to the evolution of user preferences over time. Models like PIMIRec~\cite{pimirec} and UMI~\cite{umi} integrate time information, interactivity, and user profiles to represent user interests more contextually and temporally.
Re4~\cite{re4} introduces backward flow regularization to enhance multi-interest learning by capturing the details of user interest evolution. MIRN~\cite{mirn} uses EM routing to segment user behavior for tailored interests. kNN-Embed~\cite{knn_embed} improves diversity in candidate generation by representing users as mixtures over learned item clusters.
HimiRec~\cite{himirec} explores hierarchical interest modeling for structured user preferences. MTMI~\cite{mtmi} advocates for a multi-tower architecture to diversify user representation, addressing adaptability and complexity in a dynamic environment. REMI~\cite{remi} enhances retrieval quality by mining hard negatives that consider user interests and employing routing regularization for improved robustness and accuracy.
\section{INDEXING MECHANISM FOR EFFICIENT RETRIEVAL} \label{sec:index}

After obtaining high-quality user and item representations in Sections~\ref{sec:tra} and ~\ref{sec:neural}, we now discuss efficient indexing mechanisms in modern retrieval systems, which are crucial for organizing and retrieving large-scale items effectively. The indexing methods can be broadly categorized into three main types: (1) no index with whole vault calculations, (2) approximate nearest neighbor search, and (3) tree based indexing.
% \ljh{can we give the illustrative figure for these three types with 1x3 block figures?}

\subsection{Whole Vault Calculation}
In the simplest form of retrieval, there is no indexing mechanism, and the system performs whole vault calculations. This approach involves computing the similarity between the user and every item in the candidate pool. 
\begin{equation}\label{equ:index1}
    s(u, i)=\left\langle\mathbf{h}_{u}, \mathbf{h}_{i}\right\rangle \quad \forall i \in \mathcal{I}.
\end{equation}
In ~\eqref{equ:index1}, $\mathbf{h}_u$ and $\mathbf{h}_i$ are the embedding vectors for user $u$ and item $i$, and $\langle \mathbf{h}_u, \mathbf{h}_i\rangle$ denotes their inner product.
While this method can be straightforward to implement, it is computationally expensive and impractical for large-scale systems with millions of items due to the sheer volume of calculations required.

\subsection{Approximate Nearest Neighbor Search}
To efficiently retrieve the top-$K$ items from a vast candidate item pool after obtaining the embeddings for users and items, approximate nearest neighbor (ANN) search techniques have been developed.
These techniques are essential for handling large datasets where full pairwise comparisons would be computationally prohibitive.
\begin{equation}\label{equ:index2}
    s(u, i)=g\left(\mathbf{h}_{u}, \mathcal{Q}\left(\mathbf{h}_{i}\right)\right).
\end{equation}
$\mathcal{Q}(\mathbf{h}_{i})$ represents the index-based, compressed or quantized version of $\mathbf{h}_{i}$, and $g(\cdot,\cdot)$ is a scoring function that approximates the similarity between $\mathbf{h}_{u}$ and the indexed representation of $\mathbf{h}_{i}$.
Various techniques and tools facilitate ANN search, such as Faiss~\cite{johnson2019billion}, Milvus~\cite{wang2021milvus}, and ScaNN~\cite{guo2020accelerating}, which are instrumental in industry applications for efficiently retrieving the top-$K$ nearest vectors. The construction of popular ANN indices, such as HNSW~\cite{malkov2018efficient} and IVFPQ in Faiss~\cite{johnson2019billion}, involves time-consuming procedures, including clustering algorithms like $K$-means. Despite these challenges, the advantages of using ANN search in practical applications, particularly in speeding up the retrieval process and reducing computational loads, make it a valuable technique in modern retrieval systems.

\subsection{Tree Based Indexing}
Tree based indexing methods have gained attention in recent years for their ability to efficiently organize and retrieve large-scale items. These methods leverage hierarchical structures to capture the diversity of user interests and represent information more effectively.
Tree based indexing methods offer several advantages over vector based approaches. They can significantly reduce the computational complexity by structuring the data hierarchically, enabling logarithmic scale computation relative to the size of the corpus during prediction. Additionally, they allow for the interaction between candidate items and users within the model, enhancing the expressive power of the retrieval method. 
These methods not only make the use of arbitrary advanced models possible, but also excel at discovering novel and effective recommendations by exploring the entire dataset and utilizing sophisticated deep models to uncover potential user interests. 

TDM~\cite{tdm} is a pioneering work in this area. It ties each item in the corpus to a leaf node via clustering, using a tree structure as an index. 
To model user preference, TDM assigns to each non‐leaf node $n$ at level $l$ a score based on its most promising child at level $l+1$. Formally, if we denote by $p^{(l)}(n \mid u)$ the preference of user $u$ for node $n$ at level $l$, then
\begin{equation}\label{equ:index3}
p^{(l)}(n \mid u)
\;=\;
\frac{1}{\alpha^{(l)}}
\;\max_{\,c \,\in\,\mathrm{Children}(n)}\;p^{(l+1)}(c \mid u)\,,
\end{equation}
where $p^{(l+1)}(c \mid u)$ is the preference for child node $c$ in the next finer level, and $\alpha^{(l)}$ is a level‐specific normalization constant. In this way, a node’s score propagates upward from the highest‐scoring child, ensuring that the top-$K$ candidates at level $l$ are drawn from among the children of the top-$K$ nodes at level $l-1$. The user’s overall interest is thus represented by the distribution of scores across the entire tree. At inference time, retrieving the top-$K$ leaf items is equivalent to performing a sequence of small classification tasks—selecting the best nodes level by level—yielding an efficient, scalable retrieval process.
JTM~\cite{jtm} improves upon TDM by synchronizing the objectives of index learning and model training, addressing inconsistencies, and enhancing the model's overall performance.
OTM~\cite{otm} further refines this approach by addressing the discrepancy between training and testing in tree based methods. It introduces the notion of bayes optimality under beam search and calibration under beam search, providing general tools for analyzing and improving tree based methods.

Furthermore, Deep Retrieval (DR)~\cite{deepretrieval} innovates by learning a retrievable structure end-to-end, overcoming data scarcity at the leaf level in large-scale recommender systems. DR enables efficient item clustering and retrieval by using a matrix for indexing. These advanced indexing mechanisms play a critical role in improving the efficiency and effectiveness of retrieval stages in recommender systems, offering advantages over early methods by reducing computational loads and enhancing retrieval performance. However, they do require substantial engineering effort to implement, underscoring the need for skilled development to fully realize their performance benefits.
\section{OPTIMIZATION METHODS OF RETRIEVAL}~\label{sec:neg}
In this section, we discuss the learning strategies and optimization methods for the retrieval stage, which involves three main aspects: data preparation, loss functions, and evaluation metrics.
In scenarios involving implicit user feedback, the lack of reliable negative data poses a significant challenge.
Consequently, data preparation, especially negative sample construction is critical. We will begin with various negative sampling strategies~\cite{he2017neural, rendle2012bpr} that select a subset of the missing data to act as negative instances, and whole-data based strategies~\cite{devooght2015dynamic, hu2008collaborative, liang2016modeling} that treat all missing data as negative instances. Then we will explore common loss functions and evaluation metrics used in retrieval tasks.

\subsection{Negative Sample Construction}
In many real-world scenarios,  it is hard to collect explicit user preference. 
In datasets with implicit feedback, observed interactions indicate a user's positive preference for an item, while all other unobserved items remain unlabeled. This means that only positive feedback is observed, and negative feedback is indistinguishable from missing values in the unobserved data. Additionally, users typically interact with a relatively small selection of items compared to the vast number of items available in real systems. Consequently, the absence of reliable negative data makes learning from implicit feedback particularly challenging.
To optimize retrieval methods with implicit feedback, two prevalent strategies have been commonly employed.
Negative sampling strategies~\cite{he2017neural, rendle2012bpr} involve selecting a subset of the missing data to serve as negative instances. While this method is efficient, it can potentially reduce the model's performance due to its selective nature of sampling.
On the other hand, whole-data based strategies~\cite{devooght2015dynamic, hu2008collaborative, liang2016modeling} treat all missing data as negative instances. It aims to utilize the complete dataset, which can enhance coverage and may improve the model's effectiveness. However, this strategy can be less efficient due to the extensive data processing involved and may lead to the issue of false negatives, where items that are not explicitly interacted with by users are incorrectly treated as irrelevant. We will discuss these strategies in further detail.

\subsubsection{Negative Sampling Strategies}\label{sec:negsample}
The negative sampling strategy extracts negative instances from unobserved data, significantly reducing the training sample size and enhancing the efficiency of the training process~\cite{he2017neural}. This strategy is prevalent in various retrieval methods, including traditional ones like BPR~\cite{rendle2012bpr} and neural based ones like NCF~\cite{he2017neural}. 
We classify various negative sampling strategies applied in retrieval stage into four types: static negative sampling~\cite{he2016fast, yu2017selection}, hard negative sampling~\cite{ding2020simplify, rendle2014improving, zhang2013optimizing}, adversarial negative sampling~\cite{he2018adversarial, jin2020sampling, wang2017irgan, chae2018cfgan, wang2019adversarial, guo2020ipgan, zhou2021pure} and graph based negative sampling~\cite{pinsage, huang2021mixgcf}.

\begin{itemize}[topsep = 3pt,leftmargin =*]
    \item \textbf{Static negative sampling.}
    When the probability of each item being sampled as a negative instance remains constant throughout training, this sampling strategy is referred to as Static Negative Sampling. Within static negative sampling, the simplest and most widely used method is Random Negative Sampling (RNS), also known as Uniform Negative Sampling. However, this strategy often selects uninformative instances that contribute minimally to model updates, leading to the heuristic strategy of Popularity-biased Negative Sampling (PNS). PNS was first introduced in Word2Vec~\cite{mikolov2013distributed}. ~\cite{caselles2018word2vec} has investigated the importance of hyperparameters when applied to recommendation. A more common strategy based on item popularity uses the frequency of an item in the training set as the weight for selecting negative examples, that is, favoring more popular items~\cite{he2016fast, yu2017selection}. This strategy can be explained by popularity bias: if a user does not interact with a highly popular item, it likely indicates disinterest. However, empirical results suggest that PNS does not consistently outperform RNS in retrieval methods~\cite{wang2020reinforced}.
    
    \item \textbf{Hard negative sampling.} It is challenging for static negative sampling methods to dynamically adapt and adjust the distribution of candidate negative samples. Consequently, they struggle to identify more informative negative samples. Hard negative sampling selects hard negatives that are misclassified or have higher prediction scores, which are more relevant for improving performance~\cite{ding2020simplify, rendle2014improving, zhang2013optimizing}. The most common method for mining hard negatives involves selecting samples closest to the user, i.e., those most similar in the embedding space. 
    Other innovative strategies include a sampling-bias-corrected algorithm for more accurate item frequency estimation from streaming data~\cite{yi2019sampling} and strategy that utilize both batch and uniformly sampled negatives to mitigate selection bias in implicit recommendations~\cite{yang2020mixed}.
    
    \item \textbf{Adversarial negative sampling.}
    Adversarial negative sampling methods typically involve a generator and a discriminator. The generator acts as a sampler, creating samples to confuse the discriminator, which must determine whether a given sample is a positive instance or one generated by the generator. The goal is still to learn a better distribution of negative samples. Some GAN based methods adaptively evolve the sampling probability by optimizing adversarial objectives~\cite{he2018adversarial, jin2020sampling, wang2017irgan, chae2018cfgan, wang2019adversarial, guo2020ipgan, zhou2021pure}. 
    However, adversarial negative sampling faces significant challenges: its complex framework, unstable performance, and lengthy training times all limit its practical applications. Additionally, the competition between the generator and discriminator may not always converge to an ideal Nash equilibrium, indicating that there is still room for exploration and improvement.
    
    \item \textbf{Graph based negative sampling.}
    Hard negative sampling and adversarial negative sampling effectively utilize the semantic information in the embedding space, while graph based negative sampling methods further integrate the structural information of samples within the graph. For instance, ~\cite{pinsage} sampled negative nodes based on their PageRank scores. ~\cite{huang2021mixgcf} proposed MixGCF, which implemented two strategies: positive mixing and hop mixing. Positive mixing enriches original negative samples with positive embeddings to infuse them with positive representations, while hop mixing enhances negative samples through GNN aggregation of neighborhood information. These strategies have achieved SOTA results in sampling methods for GNN based recommender systems.
\end{itemize}

Additionally, there are some cutting-edge efforts in the field. For instance, while negative sampling strategies primarily aim to improve the quality of negative samples, the ratio of negative sampling determines the quantity of these samples. SimpleX~\cite{mao2021simplex} has shown that even basic CF methods, when combined with the appropriate negative sampling ratio and loss function, can outperform current SOTA retrieval methods. Additionally, in scenarios where only positive and unlabeled data are accessible, sampling inevitably introduces certain biases, such as the false negative issue, a typical example of sample bias. CLRec~\cite{zhou2021contrastive} theoretically demonstrates that popularity based selection in contrastive loss equates to reducing exposure bias through inverse propensity weighting, offering a new perspective on understanding the effectiveness of contrastive learning.
% The workflow of generalized genative sampling can be summarized as follows:
% \begin{itemize}
%     \item \textbf{Positive example selection}: Firstly, the positive examples are identified from the training dataset based on user's interaction history and explicit feedback.
%     \item \textbf{Item pool generation}: In this step, a pool of unobserved items is created, which include items that have never been interacted with by the target user.
%     \item \textbf{Negative sample generation}: A set of negative items is sampled or generated based on the item pool under certain paradigm for each positive example. The number of negative samples can vary depending on training source or the specific implementation.
%     \item \textbf{Training process}: Training dataset is constructed by positive and negative samples. Then a retrieval model is trained on the dataset, aiming at learning the user interaction pattern and distinguishing positive samples from negative ones for more accurate recommendations.
% \end{itemize}

\subsubsection{Whole-Data Based Strategies}
Whole-data based strategies, also known as non-sampling strategies, treat all unobserved data as negative examples but assign them a lower weight than positive examples. This approach ensures that the entire dataset is utilized, potentially offering better coverage.
For instance, WMF~\cite{hu2008collaborative} assigns a uniform weight to all unobserved entries. EALS~\cite{he2016fast} and ExpoMF~\cite{liang2016modeling} adjust the weight of unobserved entries based on item popularity. The rationale is that popular items are more likely to be seen by users and should therefore be given higher weights as negatives. These methods have demonstrated the ability to leverage whole datasets with potentially better coverage~\cite{he2016fast, liang2016modeling, xin2018batch}, while inefficiency is the challenge.

To mitigate inefficiency, several methods have been developed, such as batch based ALS~\cite{hu2008collaborative, he2016fast, bayer2017generic} and mini-batch SGD~\cite{xin2018batch}. These techniques accelerate the learning process, making whole-data based strategies more practical. 
However, these methods are generally suitable only for traditional retrieval methods with a linear prediction layer~\cite{bayer2017generic, xin2018batch, chen2020efficient} and regression loss function. Consequently, non-sampling methods have been widely applied in traditional retrieval methods~\cite{hu2008collaborative, he2016fast, liang2016modeling} but are less common in neural retrieval methods.
Recently, the neural non-sampling retrieval method ENMF~\cite{chen2020efficient} was proposed, representing an effort to apply non-sampling strategies within neural frameworks. Individuals who are interested in this part can refer to ~\cite{chen2023revisiting} for further details.

\subsection{Loss Functions}
In this part, several commonly used loss functions in retrieval methods will be introduced. 

\subsubsection{Sampled Softmax Loss}
In recommender systems, calculating the softmax over a vast number of items is computationally expensive, making it impractical for real-time retrieval. By approximating the softmax function through negative sampling, sampled softmax loss is often used to address the computational inefficiency of the softmax function when dealing with large-scale item sets in retrieval stage. 
As is shown in ~\eqref{equ:softmax}, sampled softmax loss approximates the denominator by considering only a subset of sampled negative items.
\begin{equation}
    L_{\text {SampledSoftmax }}=-\frac{1}{|B|} \sum_{\left(u_{i}, v_{i}\right) \in B} \log \frac{\exp \left(G\left(u_{i}, v_{i}\right)\right)}{\exp \left(G\left(u_{i}, v_{i}\right)\right)+\sum_{j \in S_{i}} \exp \left(G\left(u_{i}, v_{j}\right)\right)}  
\label{equ:softmax}
\end{equation}
\begin{equation}
    G\left(u, v\right) = \mathbf{u}\cdot \mathbf{v} - \log(Q(v| u))
\label{equ:matching_score}
\end{equation}
$B$ stands for the mini-batch and $u_{i}, v_{i}$ represent the embedding of user $i$ and item $i$, respectively.
$G\left(u_{i}, v_{i}\right)$ is the scoring function that calculates the interaction score between user $u_{i}$ and item $v_{i}$. $S_i$ refers to the set of sampled negative items for user $u_i$. These are items that the user has not interacted with and are considered as negative examples during training.
Usually, we compute the matching score between user and item using dot product and adjust based on the sampling probability \(\log(Q(v|u))\) to prevent popular items from being overly penalized. This adjustment occurs only during training; for predictions, we rely solely on the dot product, which is compatible with vector databases like Faiss~\cite{johnson2019billion} and Milvus~\cite{wang2021milvus}.

Sampled softmax loss ensures that the model focuses on differentiating between the positive items and the selected negative samples, and is particularly useful in the retrieval stage of recommender systems, where the goal is to quickly and accurately identify a subset of relevant items from a large item pool. By reducing the computational overhead, it enables faster training and inference times, making it feasible to deploy in real-world scenarios with millions of items.

\subsubsection{Noise Contrastive Estimation Loss\label{nceloss}}
Noice Contrastive Estimation(NCE) loss is an efficient training objective for large-scale recommender systems, particularly in the retrieval stage. It addresses the challenge of computing probabilities over an extensive set of items by transforming the problem into a binary classification task. NCE approximates the softmax function by contrasting observed data (positive examples) against artificially generated noise (negative examples), simplifying the likelihood computation.
NCE loss can be formulated as ~\eqref{equ:NCE}, where $B$ stands for the mini-batch, and \( G(u_{i}, v_{i}) \) is the matching score between user \( u_{i} \) and item \( v_{i} \) in ~\eqref{equ:matching_score}. $\sigma$ is the sigmoid function and \( S_i \) represents the set of negative items sampled for user \( u_i \).

\begin{equation}\label{equ:NCE}
\mathcal{L}_{\text{NCE}} = -\frac{1}{|B|} \sum_{\left(u_{i}, v_{i}\right) \in B} \left[ \log \sigma(G\left(u_{i}, v_{i}\right)) + \sum_{j \in S_i} \log (1 - \sigma(G\left(u_{i}, v_{j}\right))) \right]
\end{equation}

During training, the method learns to differentiate between true user-item interactions and sampled noise. By adjusting the parameters to maximize the likelihood of observed interactions and minimize the likelihood of noise, NCE effectively handles the computational complexity of large-scale retrieval tasks.

\subsubsection{Pairwise Loss}
Pairwise loss is a type of Learning-to-Rank approach commonly used in retrieval methods. It focuses on optimizing the ranking of items for a given user by comparing pairs of items, which ensures that preferred items are ranked higher than less preferred ones.
One common method to implement pairwise loss is using Marginal Hinge loss, defined as:

\begin{equation}
L_{\text{Hinge}} = \frac{1}{|B|} \sum_{(u_i, v_{i+}, v_{i-}) \in B} \max(0, m - \mathbf{u}_i \cdot \mathbf{v}_{i+} + \mathbf{u}_i \cdot \mathbf{v}_{i-})
\end{equation}
where
\( B \) is a batch containing triplets \((u_i, v_{i+}, v_{i-})\). \( \mathbf{u}_i \) is the embedding vector for user \( u_i \) and \( \mathbf{v}_{i+} \) and \( \mathbf{v}_{i-} \) are the embedding vectors for the positive and negative items, respectively.
\( m \) is a margin parameter ensuring a minimum difference between the scores of positive and negative items. The goal is to ensure the matching degree of positive items is higher than that of negative items by at least the margin $m$.

Bayesian Personalized Ranking (BPR) loss is another pairwise loss, which maximizes the probability that a positive item is ranked higher than a negative item. It is defined as:
\begin{equation}
L_{\text{BPR}} = -\frac{1}{|B|} \sum_{(u_i, v_{i+}, v_{i-}) \in B} \log \sigma(\mathbf{u}_i \cdot (\mathbf{v}_{i+} - \mathbf{v}_{i-})), 
\end{equation}
where $\sigma$ is the sigmoid function. This formulation avoids the need for a margin parameter and focuses on maximizing the ranking order directly.

\section{EVALUATION METRICS}~\label{sec:metric}
In this section, we collect and summarize the most commonly used metrics for assessing retrieval-stage performance. We split these into (1) offline metrics, which are computed on held-out logs or public datasets, and (2) online (A/B) metrics, which are measured in a live serving environment.
\subsection{Offline Metrics}
Offline evaluation is typically performed on historical logs or public benchmarks. These metrics measure how well a retrieval method ranks known positive interactions versus held-out negatives or unobserved items. 
Recall@K, Precision@K, and F1@K are closely related metrics in the retrieval stage.
We categorize the ground truth labels as either positive or negative, within the predicted labels, this results in classifications such as True Positive (TP), True Negative (TN), False Positive (FP), and False Negative (FN).
Recall and Precision indicate the proportion of correctly predicted positive samples relative to the predicted positives and ground truth positives, respectively. The F1-score combines recall and precision into a single metric, as is shown in ~\eqref{equ:metric1}.
\begin{equation}\label{equ:metric1}
Recall =\frac{T P}{T P+F N} \quad Precision =\frac{T P}{T P+F P} \quad F 1 =\frac{2 * \text { precision } * \text { recall }}{\text { precision }+ \text { recall }}
\end{equation}

Hit Rate(HR) is another popular metric, which measures the proportion of hit items in the predicted top-$K$ list relative to the total items in the test set. A hit occurs if a user interacts with any of the top-$K$ retrieved items, which can be formulated as:
\begin{equation}\label{equ:metric2}
    HR@K = \frac{Number of Hits@K}{Test}.
\end{equation}

While Precision and Recall do not account for the order of items, Average Precision(AP) calculates the average precision score across multiple items within the list. It considers the precision at each rank and takes into account both the presence and position of relevant items, giving higher scores to relevant items appearing earlier in the list. Mean Average Precision (MAP) extends AP by averaging the AP scores across all users. AP and MAP are formulated in ~\eqref{equ:metric3}, where $\operatorname{rel}(k)$ indicates relevance and $p_{k}$ is the precision at cut-off $k$.
\begin{equation}\label{equ:metric3}
AP @ K=\frac{1}{|K|} \sum_{k=1}^{K} \frac{p_{k} \cdot \operatorname{rel}(k)}{k} \quad M A P=\frac{1}{|U|} \sum_{u \in U} A P_{u}
\end{equation}

\subsection{Online Metrics}
Online evaluation is conducted via A/B testing on live traffic. The primary metrics include click-through rate (CTR), conversion rate (CVR), revenue per mille (RPM), and gross merchandise value (GMV), which are defined in ~\eqref{equ:metric4}, where $\#$ denotes the quantity:
\begin{equation}\label{equ:metric4}
CTR =\frac{\#clicks}{\#impressions} \quad CVR =\frac{\#conversions}{\#clicks} \quad RPM =\frac{Revenue}{\#impressions}\times 1000 \quad GMV=\sum_{i \in \mathcal{S}} price_{i}  \times quantity_{i}.
\end{equation}
When measuring CTR, an `impression' is recorded when an item is rendered in a user’s viewport for at least one second, and a `click' is recorded when the user taps or clicks that item. Closely related to CTR is CVR, which measures the proportion of clicks that result in a downstream action (e.g., `Add-to-Cart' or `Purchase'). 
Since CVR reflects actual revenue-generating behavior, it typically serves as the primary business key performance indicator once engagement (CTR) reaches satisfactory levels. 
RPM indicates revenue generated per thousand impressions. GMV represents the total monetary value of all transactions originating from recommended items, where $\mathcal{S}$ denotes the set of items purchased. 

\section{DATASETS AND EXPERIMENTS}~\label{sec:exp}
\begin{table}[!t]
\centering
\caption{Statistics of the datasets.}
\label{tab:data statics}
\scalebox{1.1}{
\begin{tabular}{c|cccccc}
\hline
Dataset       & \# Users & \# Items & \# Samples &  Density \\ \hline
MovieLens-1M & $6,040$     &  $3,706$    & $1,000,209$    &   0.04468   \\
Amazon-Book     & $52,643$    & $91,599$  & $2,894,108$  & 0.00062  \\           
Yelp2018     & $31,668$    & $38,048$  & $1,561,406$  & 0.00130   \\           \hline
\end{tabular}
% \vspace{-15pt}
}
\end{table}
In this section, we first introduce commonly used large-scale public datasets for retrieval tasks and select representative retrieval methods to evaluate their performance across three datasets. Additionally, since the study of retrieval methods is deeply rooted in practical applications and faces numerous real-world challenges, designs purely based on theoretical research may not fully capture the potential benefits of innovations at the retrieval stage. Therefore, we summarize online performance statistics from published papers to gain further insights.
\subsection{Datasets}
Many real-world datasets have been released to advance research in retrieval methods, such as MovieLens~\cite{movielens}, Yelp~\cite{yelp}, AmazonBooks~\cite{amazonbooks}, Gowalla~\cite{gowalla}, and Criteo~\cite{criteo}. Additionally, a large collection of preprocessed datasets is available in RecBole~\cite{zhao2021recbole}. However, most of these datasets are relatively small, typically containing fewer than millions of interactions between tens of thousands of users and items, which differs significantly from the large-scale data in industry. Therefore, we will primarily introduce some larger, industrial-scale datasets.
\begin{itemize}[topsep = 3pt,leftmargin =*]
    \item \textbf{MovieLens~\cite{movielens}} dataset is a widely used dataset in the field of recommender systems. It was released by the GroupLens Research Project at the University of Minnesota. The dataset contains user ratings for movies, along with some additional information such as user age, occupation, zipcode and item genre. Movielens-1M and Movielens-20M are commonly used in various research studies~\cite{mao2021simplex, he2020lightgcn}.
    % This dataset contains over 6000 users and nearly 4000 movies.
    \item \textbf{Gowalla~\cite{gowalla}} dataset was collected from the Gowalla social network, which was a location based social networking website where users could check in at physical locations and share their activities with friends. 
    % This dataset has over 100000 users and over 1200000 items along with nearly 4 million interaction records.
    \item \textbf{Amazon-Book~\cite{amazonbooks}} dataset is a sub-dataset from the Amazon review dataset, which contains millions of reviews written by Amazon customers for various products available on the Amazon e-commerce platform. 
    % It contains about 15 million users, 3 million items with 50 million interactions.
    Additionally, other sub-datasets in Amazon, such as Amazon-CDs, Amazon-Movies, Amazon-Beauty, and Amazon-Electronics, are also used in some research studies~\cite{sun2020neighbor, mao2021simplex, mao2021ultragcn}.
    \item \textbf{Yelp~\cite{yelp}} dataset consists of reviews and related information collected from the Yelp platform, which is known for providing crowd-sourced reviews about businesses, particularly restaurants and other local services. 
    % In its 2022 version, nearly 20 million users, over 150000 items and around 70 million interactions are included.
    \item \textbf{CiteUlike-A~\cite{citeulike}} dataset was gathered from CiteULike and Google Scholar. CiteULike enables users to create their own collections of articles. Each article includes its abstract, title, and tags.
    \item \textbf{MillionSongData~\cite{song}} dataset is a freely available collection of audio features and metadata for a million contemporary popular music tracks. Typically, play counts are binarized and interpreted as implicit preference data.
\end{itemize}

To provide a clear understanding of the performance of various retrieval methods, we present the results of several retrieval methods on the MovieLens-1M, Amazon-Book, and Yelp2018 datasets below. 
For Amazon-Book and Yelp2018, following previous work~\cite{he2016vbpr, he2017neural}, we apply the 10-core filtering to ensure data quality, retaining only users and items with at least ten interactions. Due to its relatively smaller size, MovieLens-1M is used in its original form without core filtering. 
The statistics of these datasets are shown in Table~\ref{tab:data statics}.

\subsection{Experiment Settings}
\textbf{Evaluation Metrics.}
We adopt two commonly used metrics to evaluate the performance of various retrieval methods: HitRate@K and Recall@K. Specifically, we set K=50 on MovieLens-1M, and K=100 on Amazon-Book and Yelp2018 datasets. See Section~\ref{sec:metric} to have a deeper understanding of different metrics.

\textbf{Methods.}
We comprehensively evaluate 11 well-known retrieval methods including CF based, GNN based and autoencoder based. We also implement random and popularity based retrieval for comparison. Detailed results can be found at Table \ref{tab:main_result}.

\begin{table}[!t]
    \centering
    \caption{The best result is given in bold, and the second-best result is underlined. All the benchmark models are trained under the same setting  and the experiments have been run for three times to get the average metric results.}
    \scalebox{1.05}{
    \begin{tabular}{c|cc|cc|cc}
    \toprule
        \multirow{2}{*}{\textbf{Methods}} & \multicolumn{2}{c|}{\textbf{MovieLens-1M}} & \multicolumn{2}{c|}{\textbf{Amazon-Book}} & \multicolumn{2}{c}{\textbf{Yelp2018}} \\ \cline{2-7}
        & HitRate@50 & Recall@50 & HitRate@100 & Recall@100 & HitRate@100 & Recall@100 \\ \hline
        % \multicolumn{7}{c}{Baseline Methods}\\ \hline
        Random & 0.1819 & 0.0138 & 0.0062 & 0.0012 & 0.0119 & 0.0027  \\ 
        Pop    & 0.7703 & 0.2058 & 0.1072 & 0.0305 & 0.1561 & 0.0490  \\
        % \multicolumn{7}{c}{CF-Based Methods}\\ \hline
        ItemKNN~\cite{sarwar2001item}  & 0.9191 & 0.3896 & \textbf{0.6397} & \textbf{0.2996} & 0.5792 & 0.2488  \\
        BPR~\cite{rendle2012bpr}    & 0.9322 & 0.4198 & 0.5117 & 0.2028 & 0.5360 & 0.2148  \\
        % NeuMF  & & & - & - & - & -  \\
        % ConvNCF  & & & - & - & - & - \\
        ENMF~\cite{chen2020efficient}  & 0.9435 & 0.4419 & 0.5593 & 0.2348 & \underline{0.6129} & 0.2644 \\
        % ENMF  & 0.3732 & 0.1265 & 0.4616 & 0.1739 & 0.4899 & 0.2961   \\\hline
        % \multicolumn{7}{c}{Graph Embedding Methods}\\ \hline
        % DeepWalk    &  &  & - & - & - & - \\
        % LINE    & 0.9196 & 0.3824 & 0.4216 & 0.1541 & 0.4633 & 0.1730  \\
        % Node2Vec    &  &  & - & - & - & -  \\
        % Item2Vec    &  &  & - & - & - & -  \\\hline
        % \multicolumn{7}{c}{GNN-Based Methods}\\ \hline
        % GCMC  & 0.8999 & 0.3585  & 0.3965 & 0.1377 & 0.5347 & 0.2148 \\
        NGCF~\cite{wang2019neural}    & 0.9369 & 0.4256 & 0.5062 & 0.1980 & 0.5834 & 0.2455  \\
        DGCF~\cite{dgcf}    & 0.9399 & 0.4326 & 0.5334 & 0.2165 & 0.5878 & 0.2498  \\
        LightGCN~\cite{he2020lightgcn}    & 0.9412 & 0.4334 & 0.5574 & 0.2317 & 0.5927 & 0.2543  \\
        SGL~\cite{wu2021self}    & 0.9422 & 0.4320 & 0.5882 & 0.2527 & \textbf{0.6264} & \textbf{0.2771} \\ 
        % NCL~\cite{lin2022improving}  & 0.9498 & \underline{0.4507} & \underline{0.5992} & \underline{0.2623} & \textbf{0.6303} & \textbf{0.2834}  \\
        % \multicolumn{7}{c}{Two-Tower Methods}\\ \hline
        % YoutubeDNN    &  &  & - & - & - & -  \\
        % SimpleX    & 0.9067 & 0.3505 & 0.5639 & 0.2343 & 0.5723 & 0.2389  \\\hline
        % \multicolumn{7}{c}{Autoencoder-Based Methods}\\ \hline
        CDAE~\cite{wu2016collaborative} & 0.9482 & 0.4407 & 0.5307 & 0.2189 & 0.5807 & 0.2463  \\
        MultVAE~\cite{liang2018variational} & 0.9505 & \underline{0.4497} & 0.5369 & 0.2222 & 0.5937 & 0.2566  \\
        MultDAE~\cite{liang2018variational} & \underline{0.9508} & 0.4462 & 0.5596 & 0.2368 & 0.6085 & 0.2655  \\ 
        % NCE-PLRec & 0.9523 & 0.4563 & 0.5449 & 0.2241 & 0.6000 & 0.2601 \\
        % RaCT & 0.4726 & 0.1847 & - & - & - & -  \\
        RecVAE~\cite{shenbin2020recvae} & \textbf{0.9551} & \textbf{0.4683} & \underline{0.5627} & \underline{0.2403} & 0.6094 & \underline{0.2680}  \\
        % \multicolumn{7}{c}{Multi Interest Retrieval Methods}\\ \hline
        % MIND &  &  & - & - & - & - \\ 
        % ComiRec &  &  & - & - & - & -  \\ \hline
        \bottomrule
    \end{tabular}
    }
    \label{tab:main_result}
\end{table}

\textbf{Implementation Details.}
As with previous work~\cite{he2017neural}, for each dataset, we randomly select 80\% of each user's historical interactions for the training set and use the remaining 20\% as the test set. Additionally, we randomly select 10\% of the interactions from the training set to serve as the validation set. 
We implement these methods using the RecBole open-source framework\footnote{\url{https://github.com/RUCAIBox/RecBole}} and its provided code~\cite{zhao2021recbole}, and we adopt the optimal hyperparameters listed on the RecBole website.\footnote{\url{https://recbole.io/hyperparameters/index.html}}

\subsection{Experimental Results}\label{sec:offline_exp}
In Table~\ref{tab:main_result}, we showcase a variety of contemporary retrieval methods, highlighting key examples from categories such as CF based, GNN based, and autoencoder based approaches. The top-performing methods are indicated in bold, while the second-best results are underlined, offering multiple insights.

First, there has been a continuous evolution within each category, leading to enhanced performance through more precise and tailored modeling techniques. Second, it appears no single model type uniformly excels over others across all datasets. For example, RecVAE~\cite{shenbin2020recvae} leads the field on the Movielens-1M dataset, ItemKNN~\cite{sarwar2001item} stands out on Amazon-Book, and SGL~\cite{wu2021self} excels on Yelp2018. This diversity in leading models suggests that the field of retrieval remains wide open for innovation and is not dominated by any particular approach, indicating significant potential for breakthroughs across various categories. For an in-depth analysis of performance of various models, please refer to BARS benchmark~\cite{zhu2022bars}.

\subsection{Industrial Practices}
\begin{table}[!t]
\centering
\caption{Industrial online deployments of various retrieval methods claimed in published papers.}
\scalebox{0.8}{
\begin{threeparttable}
\begin{tabular}{c|l|l|l}
\toprule
\textbf{Model Class} & \textbf{Model} & \textbf{Compared to} & \textbf{Performance Improvement} \\ \hline
\multirow{1}{*}{Traditional Methods} & Swing~\cite{yang2020large} & Item-Based CF &+9.3\% CTR, +17.6\% CVR, +20.3\% PPM \\ \hline
\multirow{3}{*}{Graph-Based Methods} & 
EGES~\cite{wang2018billion}& Item-Based CF& +1.30\% CTR \\
\cline{2-4}
& PinSage~\cite{pinsage}&Methods Employed before& +10\% \textasciitilde 30\% Repin Rate \\
\cline{2-4}
& NIA-GCN~\cite{sun2020neighbor}& DSSM~\cite{huang2013learning} & +10.19\% CTR, +9.95\% CVR \\ \hline
\multirow{5}{*}{Two-Tower Methods} & 
YoutubeDNN\cite{yi2019sampling} & Methods Employed before & +0.37\% Engagement \\
\cline{2-4}
& \multirow{2}{*}{MNS\cite{yang2020mixed}} & Sampled Softmax instead of & \multirow{2}{*}{+1.54\% High-Quality App Install Gain} \\
&& Mixed Negative Sampling & \\
\cline{2-4}
& DAT\cite{yu2021dual}& Basic Two-Tower Method & +4.17\% CTR, +3.46\% GMV \\ 
\cline{2-4}
& MVKE\cite{xu2022mixture} & Basic Two-Tower Method & +0.79\% GMV, +0.51\% Adjust Cost \\ \hline
\multirow{7}{*}{Multi-Interest Methods} & MIND\cite{mind}  & Item-Based CF & +2.85\% CTR  \\ \cline{2-4}
& \multirow{2}{*}{PinnerSage\cite{pinnersage}} & \multirow{2}{*}{Single User Embedding} & Homefeed Engagement: +4\% Volume, +2\% Propensity \\
& & & Shopping Engagement: +20\% Volume, +8\% Propensity \\ 
\cline{2-4}
& \multirow{2}{*}{UMI\cite{umi}}  & \multirow{2}{*}{Methods Employed before}  & +0.95\% pCTR, +8.84\% GMV, +5.35\% ANP \\ 
& & & +1.37\% NAC\tnote{1}, +1.79\% Diversity\\
\cline{2-4}
& \multirow{2}{*}{HimiRec\cite{himirec}} & \multirow{2}{*}{ComiRec~\cite{comirec}} & +0.31\% User Activity, +0.76\% Click, +2.00\% Engagement \\
& & & +0.45\%  Retention, +1.26\% Diversity \\ \hline
\multirow{4}{*}{Advanced Indexing Schemes} &
TDM\cite{tdm}& Methods Employed before & +2.1\% CTR, +6.4\% RPM \\ 
\cline{2-4}
& JTM\cite{jtm}& Item-Based CF & +11.3\% CTR, +12.9\% RPM \\ 
\cline{2-4}
& \multirow{2}{*}{DR\cite{deepretrieval}}& \multirow{2}{*}{Methods Employed before} & +3.0\% Video Finish Rate, +0.87\% App View Time \\
& & & + 0.036\% Second Day Retention \\\hline
\multirow{4}{*}{Methods on Cascade System} &
ARF\cite{wang2023adaptive}& LambdaLoss~\cite{burges2010ranknet} & +1.5\% Revenue, +2.3\% Conversion \\ \cline{2-4}
& \multirow{3}{*}{FS-LR\cite{zheng2024full}}& \multirow{3}{*}{Methods Employed before} & +0.12\% Usage time, +0.50\% Real show, +0.69\% Click \\ 
& & & +0.40\% Like, +0.74\% Follow, +0.94\% Forward, \\
& & & +1.08\% Comment, +0.18\% Watch Time \\
\bottomrule
\end{tabular}
\begin{tablenotes}
\footnotesize
\item[1] PPM: Payment Per Thousand Impressions; RPM: Revenue Per Mille; ANP: Averaged Number of Payments; NAC: Number of Adding to Cart; Repin Rate: Percentage of Homefeed Recommendations saved by users in Pinterest;
Adjust Cost: The Cost paid by advertisers to the platform, Adjusted based on actual costs.
\end{tablenotes}
\end{threeparttable}
}
\label{tab:industrial}
\vspace{-2mm}
\end{table}

% \begin{table}[!t]
% \centering
% \caption{Explanation of evaluation metrics.}
% \scalebox{0.93}{
% \begin{tabular}{l}
% \toprule
% PPM: \\ Payment Per Thousand Impressions; GMV: Gross Merchandise Volume; 
% RPM: Revenue Per Mille;
% Repin Rate: Percentage of \\ Homefeed Recommendations saved by users in Pinterest;
% Adjust Cost: The Cost paid by advertisers \\ to the platform, Adjusted based on actual costs.
% \\
% \bottomrule
% \end{tabular}
% }
% \label{table:evaluation_metrics}
% \end{table}
The study of retrieval methods is deeply rooted in practical applications and faces numerous real-world challenges. Therefore, designs purely based on theoretical research may not fully capture the potential benefits of innovations at the retrieval stage. In this section, we explore the industrial applications of key retrieval methods. To guarantee the reliability and accuracy, we only include online performance statistics from published papers. The results are summarized in Table~\ref{tab:industrial}, which leads to the following observations.

Firstly, the retrieval stage often deals with large-scale item sets, making experiments on industrial online platforms particularly crucial. Examples include YouTube's YouTubeDNN~\cite{yi2019sampling}, Airbnb's personalized embeddings~\cite{grbovic2018real}, Pinterest's PinSage~\cite{pinsage}, and Taobao's EGES~\cite{wang2018billion} and TDM~\cite{tdm}, to name a few. These applications span a wide range of scenarios, including online advertising, e-commerce, and online video streaming, underscoring the pivotal role of retrieval in recommender systems. Moreover, recent industrial practices in retrieval have shown a trend toward innovative approaches, such as enhancements to dual-tower structures, precise capturing and modeling of users' multiple interests, and more cutting-edge techniques like advanced indexing schemes and studies on cascade systems, rather than relying solely on a single retrieval stage. These developments are tailored to fit the unique needs of industrial recommendation scenarios. We will explore these future directions in greater detail in Section~\ref{sec:future}. 
% In Table~\ref{table:evaluation_metrics}, we also outline some of the common online evaluation metrics mentioned in Table~\ref{tab:industrial}.
\section{CASE STUDY OF CURRENT INDUSTRIAL PRACTICES}~\label{sec:industry}
In this section, we present a detailed case study of the retrieval pipeline at China Merchants Bank (CMB), a mainstream bank company. 
We focus on CMB Life APP for credit card\footnote{\url{https://english.cmbchina.com/CreditCard/}}, which attracts millions of daily active users and generates billions of user logs each day through implicit feedback, such as click behaviors.
CMB Life APP has enriched its online service ecosystem by integrating various scenarios.
% Additionally, the app has improved interaction efficiency and customer experience through enhanced search, recommendation, and other intelligent service features. 
% It integrates various scenarios, such as mall products, meal and movie tickets, live broadcasts, and advertisements. 
In the meal-and-movie ticketing service, users receive recommendations for merchants offering personalized dining and cinema discounts. In shopping malls, consumers are presented with products redeemable through accumulated loyalty points. On homepage feeds, trending news items appear based on users' interests. In automotive marketplaces, platforms recommend car brands alongside customized down-payment and monthly-payment plans tailored to individual financial profiles.
Each scenario serves different business units with distinct performance metrics, such as click-through rates and conversion rates.
\subsection{The Overall Pipeline}
\begin{figure}[!htbp]
\centering
\includegraphics[width=1\linewidth]{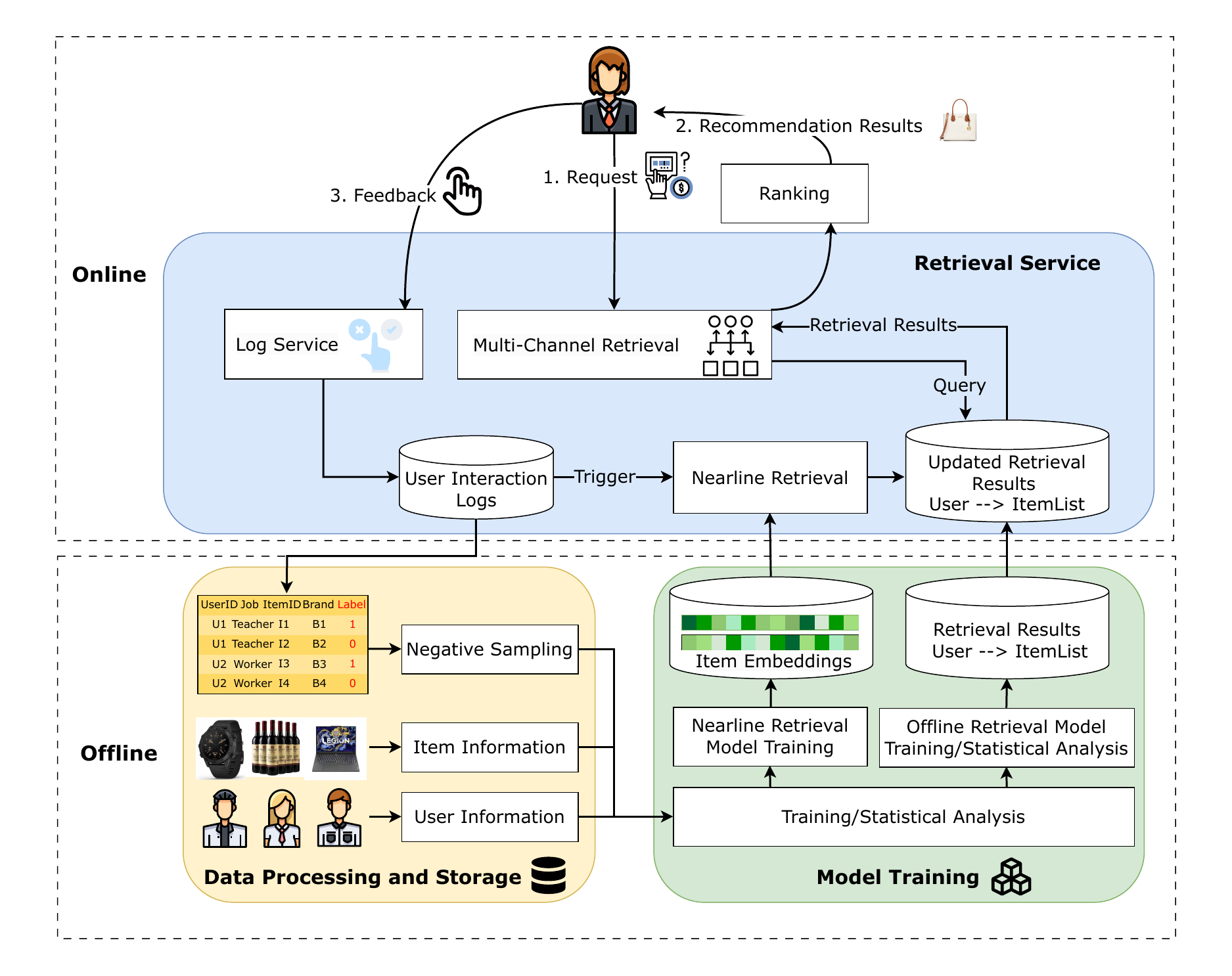}
\caption{The retrieval pipeline in CMB Life APP for credit card.}
\label{fig:industrial}
\end{figure}

We outline the overall recommendation pipeline at CMB Life APP for credit card in Figure~\ref{fig:industrial}. The pipeline is broadly divided into two main parts: offline training (with data storage), and the online recommendation service. Typically, offline processes are responsible for storing item embeddings and retrieved item list for each user. When a user makes an online request, the retrieval stage queries the $k$ most similar items from the stored retrieval results and forwards them to the subsequent stages including ranking and re-ranking, culminating in the display of recommendation results to the user. Users' interactions with these recommendation results, such as clicks or purchases, are logged as user interaction logs, which serve as data for offline training of retrieval models. This whole process forms a continuous feedback loop. We will next provide in-depth explanations of data processing and storage, training, and recommendation services, respectively.

In the data processing and storage section, we utilize three types of data sources. (1) \textit{Item information:} The item service provides detailed information about items, such as titles, prices, e.t.c. (2) \textit{User information:} This primarily includes basic user profiles.
(3) \textit{User interaction logs recorded by various business systems:} These logs capture user-item interactions such as views, clicks, purchases, and favorites. As mentioned in Section~\ref{sec:negsample}, due to the need to score a large number of items that have not been previously exposed, relying solely on ranking data with user feedback can lead to \textit{sample selection bias} (SSB)~\cite{Chen_Dong_Wang_Feng_Wang_He_2023}, which is essentially a mismatch between training and inference data. This is akin to forcing a student to take a test on material not covered in the syllabus. To address this, negative sampling is needed. In the specific context of Mobile Life, we utilize random negative sampling, whose proportion depends on the volume of data and the balance of positive and negative samples. These three data sources are combined and used for offline training or statistical analysis of retrieval models.

As for retrieval models, we categorize them into two types: offline retrieval and nearline retrieval. 
\begin{itemize}[topsep = 3pt,leftmargin =*]
    \item \textbf{Offline retrieval.} 
    Offline retrieval typically involves methods that do not require real-time updating, such as popular or new item retrieval. Once offline statistical analysis is complete or training process is finished, the retrieval results, such as user-to-item lists, are generated and saved for quick access during retrieval. 
    \item \textbf{Nearline retrieval.}
    Nearline retrieval involves methods that require periodic training and updating, which necessitates maintaining item embeddings. When a user's interaction history updates, nearline retrieval is triggered.
    To be specific, if a user recently clicked on an item, nearline retrieval select the top items from the item pool based on the dot product of item embeddings. The retrieved items are then combined with the offline user-to-item list to create an updated retrieval list.
\end{itemize}
During online recommendation service, when a user initiates a request, the system leverages multi-channel retrieval to query the user-specific retrieval pool, combining both offline and nearline results. These results are then passed to subsequent ranking models, ultimately displaying the final recommendations to the user.

\subsection{Multi-Channel Retrieval}
\begin{figure}[!htbp]
\centering
\includegraphics[width=1\linewidth]{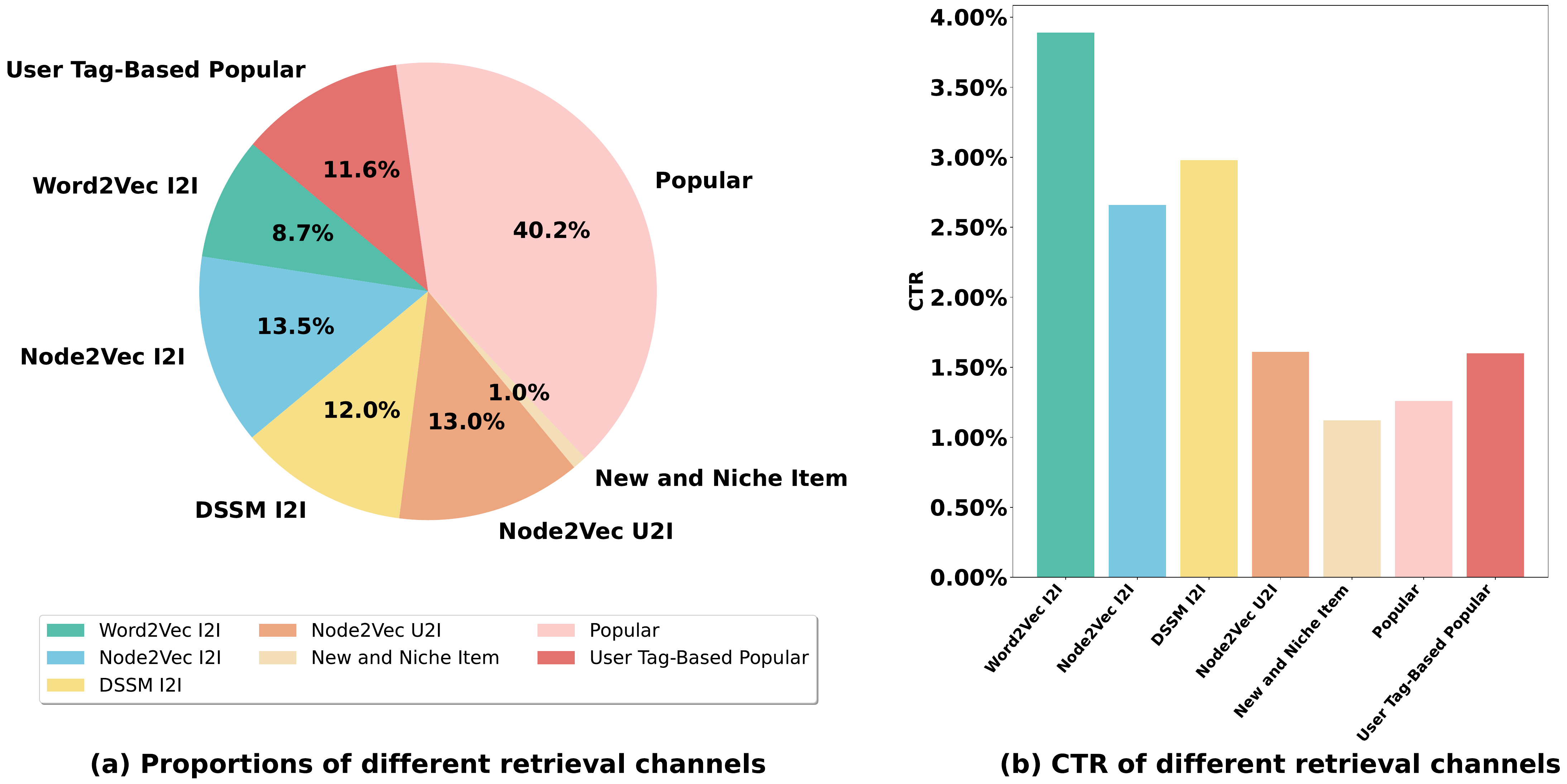}
\caption{Overview of multiple channels in the retrieval stage of CMB Life APP for credit card.}
\label{fig:proportion}
\end{figure}

As illustrated in Figure~\ref{fig:proportion}, our industrial setting features seven retrieval channels, which includes three nearline I2I retrieval methods, one offline U2I retrieval method, and three other statistical based retrieval methods. During online service, the I2I approach retrieves the top-$K$ similar items using the embedding of an item that a user clicks or purchases and places them into the user's retrieval pool; the U2I retrieval model retrieves directly through the user ID; and the statistical based retrieval methods fetch the top-$K$ items directly from the database based on predefined quotas. We will now introduce each channel.

For Word2Vec, we feed comprehensive text fields from within the scenario such as item titles, attributes, and categories to generate item embeddings, which are then stored in an embedding database. The underlying principle is that items with similar titles, attributes, and categories are more likely to appeal to users.
For Node2Vec, we create positive sample pairs from user behavior sequences by using a sliding window of length 2 to generate positive item pair item $i$ and item $j$. Simultaneously, we form negative item pairs by pairing item $i$ with a random item $k$ from the item pool, ensuring that $k\neq j$, to train the Node2Vec retrieval model. In the setting of I2I, we directly store the generated item embeddings, while in the setting of U2I, we obtain user embeddings by weighting the embeddings of the recent sequence of items interacted with by the user. We then use the user embedding to retrieve the top-$K$ items, storing them in the format of User$\xrightarrow{}$ItemList. The rationale is that items interacted with consecutively in a user’s behavior sequence are perceived as more similar by the user, making it likely that they will engage with other items close to item $i$ after visiting it.
For DSSM, we generate positive user-item pairs from user behavior sequences and create negative samples by pairing the user with a randomly sampled item from the item pool.
These data are then used to train the DSSM retrieval model, which yields item embeddings that are then stored in the embedding database. DSSM may offer more personalization than Node2Vec in this scenario, as the implicit graph structure includes both users and items.

Additionally, we have three retrieval methods based on statistical analysis. The New and Niche Item retrieval refers to storing recently least exposed items, based on the idea that new and niche items typically have lower exposure and interactions, leading to significant bias in interaction volume predictions. Exposing a certain proportion of such items can unearth overlooked items. The Popular retrieval method involves scoring items based on recent clicks, exposures, favorites, and purchases, then storing those with the highest scores. This method is especially useful for supplementing retrieval results for some users, particularly inactive or new users, by providing popular items as a fallback. User Tag Based Popular retrieval categorizes users into buckets based on fundamental attributes, which may yield more relevant results.

Figure~\ref{fig:proportion} displays the distribution and online performance of the seven retrieval channels mentioned above. 
It is important to note that this section evaluates the live production system, diverging from the offline benchmark analysis in Section~\ref{sec:offline_exp}; as the experimental contexts and objectives differ, the results from these two sections are not directly comparable. In the retrieval stage, the focus is on covering potential user interests with various approaches, making the integration of multiple retrieval results a significant topic. Currently, in the scenario of CMB Life APP for credit cards, our retrieval strategy is divided into two layers. The first layer includes multiple retrieval methods, consisting of three I2I retrieval methods, one U2I retrieval method, and a retrieval method for new and niche items. The second layer involves supplementary retrieval approaches, including popular item retrieval and user tag based popular item retrieval. This supplementary layer is only activated if the first layer's results are insufficient. For example, if we aim to retrieve 300 items and the first layer retrieves only 180, the second layer will add the top 120 to reach 300. If the first layer already retrieves 300 items, the second layer will not be triggered. As seen in Figure~\ref{fig:proportion}, supplementary retrieval accounts for about 50\% in this scenario, largely due to the presence of many inactive users.

\subsection{Key Insights and Challenges}
In this section, we discuss some insights and challenges during the efforts in the retrieval stage. 
\begin{itemize}[topsep = 3pt,leftmargin =*]
    \item 
    During the training of retrieval models, attention must be paid to the construction of negative samples and the sampling ratios.
    The proportion of negative sampling largely depends on the volume of data and the balance of positive and negative samples. Typically, the optimal sampling ratio is determined by evaluating offline metrics. Different sampling ratios can be tested to identify the best performing ratio, similar to a hyperparameter grid search. This careful tuning ensures that the model effectively distinguishes between relevant and irrelevant items, thereby improving the overall performance of retrieval.
    \item
    As retrieval methods are followed by ranking and re-ranking stage and do not directly target final business metrics like CTR and CVR, there can be discrepancies between online and offline metrics. The primary reason for this inconsistency is that the retrieval stage is just the first part of the cascade system. Subsequent stages, such as ranking and re-ranking, further filter and prioritize the retrieved items based on more complex models. Therefore, offline evaluations might not fully capture the impact of these later stages. As a result, online experiments are often necessary to assess the actual performance of various retrieval methods, as they provide a more holistic view of the end-to-end recommender system's effectiveness.
    \item 
    Multi-channel retrieval is essential for maximizing recall rate. By using different retrieval strategies, we can cover a wide range of user preferences, thereby improving overall recall. Even if channels focusing on new and niche items have a slightly lower CTR, they still contribute significantly to enhancing the overall recall rate.
    \item 
    In the context of the CMB Life APP for credit cards, the integration strategy employs a multi-channel retrieval approach with additional supplementary retrieval methods.  This involves how to adjust the volume of retrieved items from each channel. 
    Also, snake-merge is commonly used in industry, which involves alternating the selection from each retrieval channel to create a mixed list of retrieval results. However, these methods are quite rudimentary, and needs better ways to adjust the weights and integrate retrieval results from multiple retrieval channels.
\end{itemize} 
\section{CHALLENGES AND FUTURE DIRECTIONS}~\label{sec:future}
In this section, we propose several open challenges and potential future directions related to retrieval methods in recommender systems. Some of these topics are crucial yet underexplored in the field, while others present promising avenues for future research.
\begin{figure}[!htbp]
\centering
\includegraphics[width=1\linewidth]{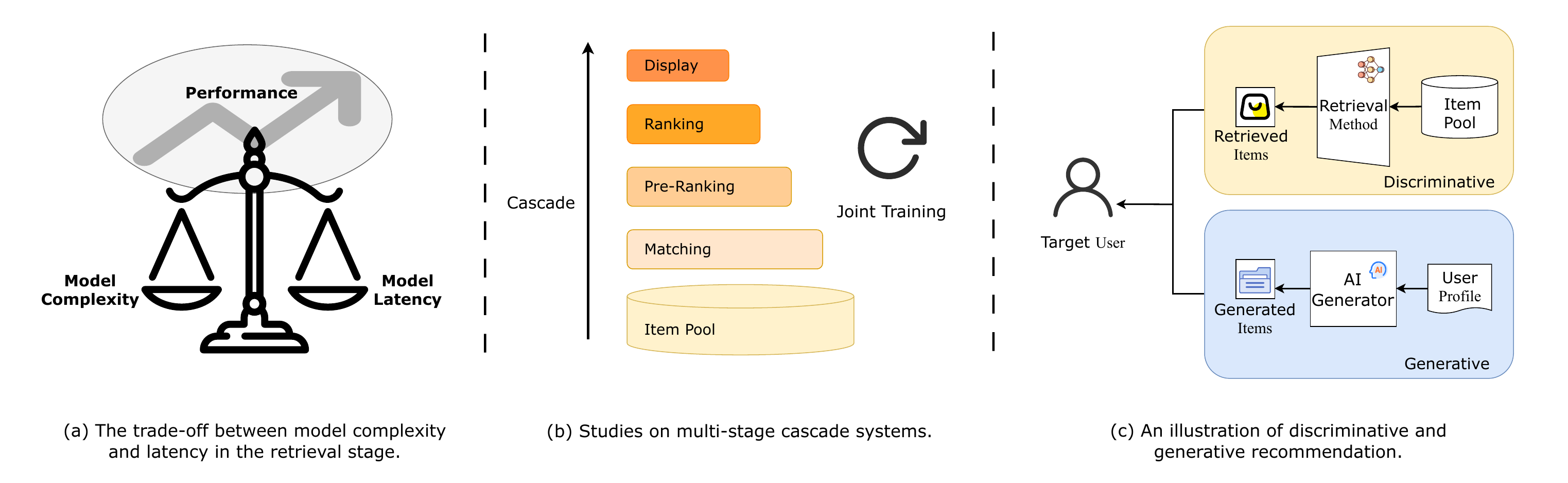}
\caption{A depiction of several underexplored challenges and possible future directions. }
\label{fig:future}
\end{figure}

\subsection{From Coarse Features to More Informative Interaction Patterns}
As mentioned in Section~\ref{sec:introduction}, retrieval differs from ranking. Due to the large-scale candidate item set and efficiency constraints at the retrieval stage, it is challenging to use complex cross-features in retrieval methods. Consequently, many earlier approaches, such as the commonly used two-tower methods~\cite{huang2013learning, covington2016deep}, separately obtain user and item representations and then compute a simple dot product or cosine similarity. However, these coarse features of users and items fail to provide comprehensive representations. Therefore, recent work has focused on balancing and making trade-offs between performance, model complexity, and real-time latency.
\begin{itemize}[topsep = 3pt,leftmargin =*]
    \item \textbf{Enhance the depth of interaction between the user and item side.}
    For instance, DAT~\cite{yu2021dual} uses an adaptive-mic mechanism to implicitly model interactions between the two towers.
    Another approach, IntTower~\cite{li2022inttower}, enhances the original two-tower methods by interacting the last hidden vector on the item side with multiple hidden vectors on the user side explicitly. Additionally, it employs positive and negative sample pairs to guide the model through self-supervised learning, capturing interactive signals implicitly.
    MVKE~\cite{xu2022mixture} generates multiple user interest representations, which are then aggregated through an attention mechanism with the item representations to refine each user’s item-related representations.
    \item \textbf{Knowledge distillation.}
    Moreover, models like PFD~\cite{xu2020privileged} have been developed where two-tower and single-tower models are trained in parallel. The rich information of the single-tower model guides the learning of the two-tower model, with only the two-tower model being used for online retrieval services after training.
    During offline training, the teacher model includes a large number of informative features, such as interaction features that the two-tower methods cannot access, like clicks within the same category by a user in the past 24 hours, and post-click information that is unavailable during online services. The teacher model does not require pre-training and is updated synchronously with the student model. Initially, both models update independently until the teacher model stabilizes, at which point distillation begins.
    Beyond recommendation, some research on knowledge distillation has also been applied to information retrieval (IR) tasks, such as ENDX~\cite{wang2022enhancing}, TRMD~\cite{choi2021improving}, VIRT~\cite{li2021virt}, ERNIE-Search~\cite{lu2022ernie}.
    \item \textbf{Advanced indexing schemes.}
    In the retrieval stage, efficient indexing schemes are essential for managing and accessing large-scale items effectively. Tree based methods~\cite{tdm}, while offering enhanced expressive power and the ability to capture diverse user interests through hierarchical representations, also present significant challenges. The primary drawback is the need for tight integration between algorithm and engineering, requiring customized indexing structures that demand substantial engineering effort. Unlike two-tower methods that can leverage open-source tools like Faiss~\cite{johnson2019billion}, tree based methods necessitate bespoke solutions, complicating deployment and optimization. The co-design approach of aligning model and index, although addressing consistency issues, adds further intricacy. 
\end{itemize}

\subsection{From Single-Stage Retrieval Optimization to Cascade System Studies}
In recent years, there has been a notable shift from single-stage retrieval models to exploring multi-stage cascade ranking systems~\cite{huang2023cooperative, wang2023adaptive, huang2024recall, zhu2023reloop2}.
For instance, Gallagher et al.~\cite{gallagher2019joint} explore deriving gradients for cascade rankers, employing end-to-end methods for optimization, while Fan et al.~\cite{fan2019mobius} advocate for unifying multiple stages into a single stage using hard negative sampling. Furthermore, Fei et al.~\cite{fei2021gemnn} suggest a strategy to share features across different stages.
Hron et al.~\cite{hron2021component} suggest using multiple channels together during the retrieval stage and learning how to merge the items retrieved from various channels.
Rankflow~\cite{qin2022rankflow} suggest training rankers on their specific data distributions to alleviate the SSB issue and better utilize the interactions between each stage.
In CoRR~\cite{huang2023cooperative}, the retriever is enhanced by deriving high-quality training signals from the ranker, while the ranker benefits from learning to discriminate hard negatives sampled by the retriever. GPRP~\cite{zheng2024full} addresses both the selection bias in each stage of the system pipeline and the underlying interests of users. The core idea is to estimate the selection bias in subsequent stages and then develop a ranking model that aligns with this bias, ensuring the top-ranked results are delivered in the final output. 
RAR~\cite{huang2024recall} stands for Recall Augmented Ranking, which optimizes the recommendation process by utilizing multi-stage information.
ARF~\cite{wang2023adaptive} introduces a novel perspective on optimizing cascade ranking systems by emphasizing the adaptability of optimization targets to data complexities and model capabilities.
These studies highlight the importance of viewing the retrieval process from a cascade system perspective, where each stage is optimized not in isolation but in concert with the others to improve the overall performance and effectiveness of the recommender system.

\subsection{From Discriminative Recommendation to Generative Recommendation}
Discriminative recommendation retrieve items from an existing corpus via scoring. However, this retrieval based approach has two main drawbacks: (1) the predefined items in the corpus may not meet the diverse information needs of all users, and (2) users typically provide feedback through passive interactions like clicks, which is inefficient and often imprecise.
Recent advancements in AI-Generated Content (AIGC) offer promising solutions to these issues. Generative AI can create personalized items that better align with specific user needs. Moreover, large language models (LLMs) with advanced language understanding and generation capabilities make it easier for users to express their preferences through natural language, reducing the effort required to tailor recommendations~\cite{lin2023can}.
Generative recommendation differ significantly from discriminative recommendation. 
Unlike discriminative recommendation, which typically rely on pointwise scoring, generative models streamline the process by directly generating recommendations. 

In rating prediction tasks, several studies have tested LLMs, many of which are based on ChatGPT~\cite{zhiyuli2023bookgpt, dai2023uncovering, liu2023chatgpt, wang2023recmind,xi2023towards}. 
Recent studies have found that the context length limit of an LLM makes it impractical to input all items. To address this, two approaches have been explored. The first is straightforward recommendation, which uses a prompt containing only user information (e.g., ID or metadata) and asks the LLM to generate recommendations directly~\cite{xu2023openp5, zhang2023chatgpt, di2023evaluating}. The second approach is selective recommendation, which includes both user information and a list of candidate items in the prompt, asking the LLM to select items from these candidates~\cite{geng2022recommendation, li2023prompt, zhang2023recommendation, li2023preliminary, dai2023uncovering}.

Some recent studies~\cite{li2023pbnr} have also instructed LLMs to determine whether a user will interact with a given item by generating ~`yes' or ~`no' responses~\cite{lin2024rella,bao2023tallrec}. Although these responses are generated by LLMs, this approach is still considered discriminative recommendation since it generates an answer or a probability score for each item. The shift towards generative recommendation highlights a new paradigm in recommender systems, where generative AI not only enhances recommendation quality but also simplifies the recommendation pipeline.

\section{CONCLUSION}~\label{sec:concl}

In conclusion, our survey is the first to thoroughly examine the retrieval stage of recommender systems, looking at both academic research and industry applications. 
We cover three main areas: improving similarity computation between users and items using both shallow and deep structures, enhancing indexing mechanisms for efficient retrieval, and optimizing training methods of retrieval.
We also conduct extensive benchmarking experiments, testing various methods on three public datasets.  Furthermore, we highlight current industrial applications through a case study on retrieval practices at a specific company, detailing the entire process and the challenges they face.
This exploration is crucial as the retrieval stage sets the groundwork for the subsequent recommendation processes, determining the efficiency and effectiveness of the entire system.

Additionally, we have identified existing gaps and proposed future directions for research, aiming to inspire and guide ongoing efforts in this field. By discussing both the theoretical frameworks and practical implementations, this survey serves as a valuable resource for researchers and practitioners alike, seeking to enhance or develop innovative retrieval methods.
Looking ahead, the field of retrieval in recommender systems holds promising potential for significant advancements. Our hope is that this survey will act as a catalyst for future research, fostering developments that will refine and revolutionize the capabilities of retrieval in recommendation, thus improving the overall user experience on recommendation platforms.

\section*{ACKNOWLEDGMENTS}
This work is supported by China Merchants
Bank Credit Card Center.
The Shanghai Jiao Tong University team is partially supported by National Natural Science Foundation of China (62322603, 62177033, 624B2096).

\bibliographystyle{ACM-Reference-Format}
\bibliography{ref}

%%% -*-BibTeX-*-
%%% Do NOT edit. File created by BibTeX with style
%%% ACM-Reference-Format-Journals [18-Jan-2012].

\begin{thebibliography}{217}

%%% ====================================================================
%%% NOTE TO THE USER: you can override these defaults by providing
%%% customized versions of any of these macros before the \bibliography
%%% command.  Each of them MUST provide its own final punctuation,
%%% except for \shownote{}, \showDOI{}, and \showURL{}.  The latter two
%%% do not use final punctuation, in order to avoid confusing it with
%%% the Web address.
%%%
%%% To suppress output of a particular field, define its macro to expand
%%% to an empty string, or better, \unskip, like this:
%%%
%%% \newcommand{\showDOI}[1]{\unskip}   % LaTeX syntax
%%%
%%% \def \showDOI #1{\unskip}           % plain TeX syntax
%%%
%%% ====================================================================

\ifx \showCODEN    \undefined \def \showCODEN     #1{\unskip}     \fi
\ifx \showDOI      \undefined \def \showDOI       #1{#1}\fi
\ifx \showISBNx    \undefined \def \showISBNx     #1{\unskip}     \fi
\ifx \showISBNxiii \undefined \def \showISBNxiii  #1{\unskip}     \fi
\ifx \showISSN     \undefined \def \showISSN      #1{\unskip}     \fi
\ifx \showLCCN     \undefined \def \showLCCN      #1{\unskip}     \fi
\ifx \shownote     \undefined \def \shownote      #1{#1}          \fi
\ifx \showarticletitle \undefined \def \showarticletitle #1{#1}   \fi
\ifx \showURL      \undefined \def \showURL       {\relax}        \fi
% The following commands are used for tagged output and should be
% invisible to TeX
\providecommand\bibfield[2]{#2}
\providecommand\bibinfo[2]{#2}
\providecommand\natexlab[1]{#1}
\providecommand\showeprint[2][]{arXiv:#2}

\bibitem[epi(2003)]%
        {epinions}
 \bibinfo{year}{2003}\natexlab{}.
\newblock \bibinfo{title}{{The Epinions Dataset.}}
\newblock \bibinfo{howpublished}{\url{https://alchemy.cs.washington.edu/data/epinions/}}.
\newblock


\bibitem[fli(2009)]%
        {flixster}
 \bibinfo{year}{2009}\natexlab{}.
\newblock \bibinfo{title}{{The Flixster Dataset.}}
\newblock \bibinfo{howpublished}{\url{https://www.flixster.com/}}.
\newblock


\bibitem[ama(2021)]%
        {amazonbooks}
 \bibinfo{year}{2021}\natexlab{}.
\newblock \bibinfo{title}{{The AmazonBooks Dataset.}}
\newblock \bibinfo{howpublished}{\url{https://jmcauley.ucsd.edu/data/amazon/amazonbooks}}.
\newblock


\bibitem[yel(2021)]%
        {yelp}
 \bibinfo{year}{2021}\natexlab{}.
\newblock \bibinfo{title}{{The Yelp Dataset.}}
\newblock \bibinfo{howpublished}{\url{https://www.yelp.com/dataset}}.
\newblock


\bibitem[cri(2022)]%
        {criteo}
 \bibinfo{year}{2022}\natexlab{}.
\newblock \bibinfo{title}{{The Criteo Dataset.}}
\newblock \bibinfo{howpublished}{\url{https://www.kaggle.com/c/criteo-display-ad-challenge/data}}.
\newblock


\bibitem[gow(2022)]%
        {gowalla}
 \bibinfo{year}{2022}\natexlab{}.
\newblock \bibinfo{title}{{The Gowalla Dataset.}}
\newblock \bibinfo{howpublished}{\url{https://snap.stanford.edu/data/loc-gowalla.html}}.
\newblock


\bibitem[mov(2022)]%
        {movielens}
 \bibinfo{year}{2022}\natexlab{}.
\newblock \bibinfo{title}{{The MovieLens Dataset.}}
\newblock \bibinfo{howpublished}{\url{https://grouplens.org/datasets/movielens}}.
\newblock


\bibitem[cit(2024)]%
        {citeulike}
 \bibinfo{year}{2024}\natexlab{}.
\newblock \bibinfo{title}{{The CiteUlike Dataset.}}
\newblock \bibinfo{howpublished}{\url{http://www.citeulike.org/}}.
\newblock


\bibitem[son(2024)]%
        {song}
 \bibinfo{year}{2024}\natexlab{}.
\newblock \bibinfo{title}{{The MillionSongData Dataset.}}
\newblock \bibinfo{howpublished}{\url{http://millionsongdataset.com/}}.
\newblock


\bibitem[Adomavicius and Tuzhilin(2005)]%
        {adomavicius2005toward}
\bibfield{author}{\bibinfo{person}{Gediminas Adomavicius} {and} \bibinfo{person}{Alexander Tuzhilin}.} \bibinfo{year}{2005}\natexlab{}.
\newblock \showarticletitle{Toward the next generation of recommender systems: A survey of the state-of-the-art and possible extensions}.
\newblock \bibinfo{journal}{\emph{IEEE transactions on knowledge and data engineering}} \bibinfo{volume}{17}, \bibinfo{number}{6} (\bibinfo{year}{2005}), \bibinfo{pages}{734--749}.
\newblock


\bibitem[Adomavicius and Tuzhilin(2010)]%
        {adomavicius2010context}
\bibfield{author}{\bibinfo{person}{Gediminas Adomavicius} {and} \bibinfo{person}{Alexander Tuzhilin}.} \bibinfo{year}{2010}\natexlab{}.
\newblock \showarticletitle{Context-aware recommender systems}.
\newblock In \bibinfo{booktitle}{\emph{Recommender systems handbook}}. \bibinfo{publisher}{Springer}, \bibinfo{pages}{217--253}.
\newblock


\bibitem[Bai et~al\mbox{.}(2017)]%
        {bai2017neural}
\bibfield{author}{\bibinfo{person}{Ting Bai}, \bibinfo{person}{Ji-Rong Wen}, \bibinfo{person}{Jun Zhang}, {and} \bibinfo{person}{Wayne~Xin Zhao}.} \bibinfo{year}{2017}\natexlab{}.
\newblock \showarticletitle{A neural collaborative filtering model with interaction-based neighborhood}. In \bibinfo{booktitle}{\emph{Proceedings of the 2017 ACM on Conference on Information and Knowledge Management}}. \bibinfo{pages}{1979--1982}.
\newblock


\bibitem[Bao et~al\mbox{.}(2023)]%
        {bao2023tallrec}
\bibfield{author}{\bibinfo{person}{Keqin Bao}, \bibinfo{person}{Jizhi Zhang}, \bibinfo{person}{Yang Zhang}, \bibinfo{person}{Wenjie Wang}, \bibinfo{person}{Fuli Feng}, {and} \bibinfo{person}{Xiangnan He}.} \bibinfo{year}{2023}\natexlab{}.
\newblock \showarticletitle{Tallrec: An effective and efficient tuning framework to align large language model with recommendation}. In \bibinfo{booktitle}{\emph{Proceedings of the 17th ACM Conference on Recommender Systems}}. \bibinfo{pages}{1007--1014}.
\newblock


\bibitem[Barkan and Koenigstein(2016)]%
        {barkan2016item2vec}
\bibfield{author}{\bibinfo{person}{Oren Barkan} {and} \bibinfo{person}{Noam Koenigstein}.} \bibinfo{year}{2016}\natexlab{}.
\newblock \showarticletitle{Item2vec: neural item embedding for collaborative filtering}. In \bibinfo{booktitle}{\emph{2016 IEEE 26th International Workshop on Machine Learning for Signal Processing (MLSP)}}. IEEE, \bibinfo{pages}{1--6}.
\newblock


\bibitem[Batmaz et~al\mbox{.}(2019)]%
        {batmaz2019review}
\bibfield{author}{\bibinfo{person}{Zeynep Batmaz}, \bibinfo{person}{Ali Yurekli}, \bibinfo{person}{Alper Bilge}, {and} \bibinfo{person}{Cihan Kaleli}.} \bibinfo{year}{2019}\natexlab{}.
\newblock \showarticletitle{A review on deep learning for recommender systems: challenges and remedies}.
\newblock \bibinfo{journal}{\emph{Artificial Intelligence Review}}  \bibinfo{volume}{52} (\bibinfo{year}{2019}), \bibinfo{pages}{1--37}.
\newblock


\bibitem[Bayer et~al\mbox{.}(2017)]%
        {bayer2017generic}
\bibfield{author}{\bibinfo{person}{Immanuel Bayer}, \bibinfo{person}{Xiangnan He}, \bibinfo{person}{Bhargav Kanagal}, {and} \bibinfo{person}{Steffen Rendle}.} \bibinfo{year}{2017}\natexlab{}.
\newblock \showarticletitle{A generic coordinate descent framework for learning from implicit feedback}. In \bibinfo{booktitle}{\emph{Proceedings of the 26th international conference on world wide web}}. \bibinfo{pages}{1341--1350}.
\newblock


\bibitem[Bell and Koren(2007)]%
        {bell2007scalable}
\bibfield{author}{\bibinfo{person}{Robert~M Bell} {and} \bibinfo{person}{Yehuda Koren}.} \bibinfo{year}{2007}\natexlab{}.
\newblock \showarticletitle{Scalable collaborative filtering with jointly derived neighborhood interpolation weights}. In \bibinfo{booktitle}{\emph{Seventh IEEE international conference on data mining (ICDM 2007)}}. IEEE, \bibinfo{pages}{43--52}.
\newblock


\bibitem[Bennett and Lanning(2007)]%
        {bennett2007netflix}
\bibfield{author}{\bibinfo{person}{James Bennett} {and} \bibinfo{person}{Stan Lanning}.} \bibinfo{year}{2007}\natexlab{}.
\newblock \showarticletitle{The netflix prize}.
\newblock  (\bibinfo{year}{2007}).
\newblock


\bibitem[Berg et~al\mbox{.}(2017)]%
        {berg2017graph}
\bibfield{author}{\bibinfo{person}{Rianne van~den Berg}, \bibinfo{person}{Thomas~N Kipf}, {and} \bibinfo{person}{Max Welling}.} \bibinfo{year}{2017}\natexlab{}.
\newblock \showarticletitle{Graph convolutional matrix completion}.
\newblock \bibinfo{journal}{\emph{arXiv preprint arXiv:1706.02263}} (\bibinfo{year}{2017}).
\newblock


\bibitem[Bobadilla et~al\mbox{.}(2013)]%
        {bobadilla2013recommender}
\bibfield{author}{\bibinfo{person}{Jes{\'u}s Bobadilla}, \bibinfo{person}{Fernando Ortega}, \bibinfo{person}{Antonio Hernando}, {and} \bibinfo{person}{Abraham Guti{\'e}rrez}.} \bibinfo{year}{2013}\natexlab{}.
\newblock \showarticletitle{Recommender systems survey}.
\newblock \bibinfo{journal}{\emph{Knowledge-based systems}}  \bibinfo{volume}{46} (\bibinfo{year}{2013}), \bibinfo{pages}{109--132}.
\newblock


\bibitem[B{\"o}hmer et~al\mbox{.}(2011)]%
        {bohmer2011falling}
\bibfield{author}{\bibinfo{person}{Matthias B{\"o}hmer}, \bibinfo{person}{Brent Hecht}, \bibinfo{person}{Johannes Sch{\"o}ning}, \bibinfo{person}{Antonio Kr{\"u}ger}, {and} \bibinfo{person}{Gernot Bauer}.} \bibinfo{year}{2011}\natexlab{}.
\newblock \showarticletitle{Falling asleep with angry birds, facebook and kindle: a large scale study on mobile application usage}. In \bibinfo{booktitle}{\emph{Proceedings of the 13th international conference on Human computer interaction with mobile devices and services}}. \bibinfo{pages}{47--56}.
\newblock


\bibitem[Breese et~al\mbox{.}(2013)]%
        {breese2013empirical}
\bibfield{author}{\bibinfo{person}{John~S Breese}, \bibinfo{person}{David Heckerman}, {and} \bibinfo{person}{Carl Kadie}.} \bibinfo{year}{2013}\natexlab{}.
\newblock \showarticletitle{Empirical analysis of predictive algorithms for collaborative filtering}.
\newblock \bibinfo{journal}{\emph{arXiv preprint arXiv:1301.7363}} (\bibinfo{year}{2013}).
\newblock


\bibitem[Burges(2010)]%
        {burges2010ranknet}
\bibfield{author}{\bibinfo{person}{Christopher~JC Burges}.} \bibinfo{year}{2010}\natexlab{}.
\newblock \showarticletitle{From ranknet to lambdarank to lambdamart: An overview}.
\newblock \bibinfo{journal}{\emph{Learning}} \bibinfo{volume}{11}, \bibinfo{number}{23-581} (\bibinfo{year}{2010}), \bibinfo{pages}{81}.
\newblock


\bibitem[Cao et~al\mbox{.}(2018)]%
        {cao2018attentive}
\bibfield{author}{\bibinfo{person}{Da Cao}, \bibinfo{person}{Xiangnan He}, \bibinfo{person}{Lianhai Miao}, \bibinfo{person}{Yahui An}, \bibinfo{person}{Chao Yang}, {and} \bibinfo{person}{Richang Hong}.} \bibinfo{year}{2018}\natexlab{}.
\newblock \showarticletitle{Attentive group recommendation}. In \bibinfo{booktitle}{\emph{The 41st International ACM SIGIR conference on research \& development in information retrieval}}. \bibinfo{pages}{645--654}.
\newblock


\bibitem[Caselles-Dupr{\'e} et~al\mbox{.}(2018)]%
        {caselles2018word2vec}
\bibfield{author}{\bibinfo{person}{Hugo Caselles-Dupr{\'e}}, \bibinfo{person}{Florian Lesaint}, {and} \bibinfo{person}{Jimena Royo-Letelier}.} \bibinfo{year}{2018}\natexlab{}.
\newblock \showarticletitle{Word2vec applied to recommendation: Hyperparameters matter}. In \bibinfo{booktitle}{\emph{Proceedings of the 12th ACM Conference on Recommender Systems}}. \bibinfo{pages}{352--356}.
\newblock


\bibitem[Celma and Cano(2008)]%
        {celma2008hits}
\bibfield{author}{\bibinfo{person}{{\`O}scar Celma} {and} \bibinfo{person}{Pedro Cano}.} \bibinfo{year}{2008}\natexlab{}.
\newblock \showarticletitle{From hits to niches? or how popular artists can bias music recommendation and discovery}. In \bibinfo{booktitle}{\emph{Proceedings of the 2nd KDD workshop on large-scale recommender systems and the netflix prize competition}}. \bibinfo{pages}{1--8}.
\newblock


\bibitem[Cen et~al\mbox{.}(2020)]%
        {comirec}
\bibfield{author}{\bibinfo{person}{Yukuo Cen}, \bibinfo{person}{Jianwei Zhang}, \bibinfo{person}{Xu Zou}, \bibinfo{person}{Chang Zhou}, \bibinfo{person}{Hongxia Yang}, {and} \bibinfo{person}{Jie Tang}.} \bibinfo{year}{2020}\natexlab{}.
\newblock \showarticletitle{Controllable multi-interest framework for recommendation}. In \bibinfo{booktitle}{\emph{Proceedings of the 26th ACM SIGKDD International Conference on Knowledge Discovery \& Data Mining}}. \bibinfo{pages}{2942--2951}.
\newblock


\bibitem[Chae et~al\mbox{.}(2018)]%
        {chae2018cfgan}
\bibfield{author}{\bibinfo{person}{Dong-Kyu Chae}, \bibinfo{person}{Jin-Soo Kang}, \bibinfo{person}{Sang-Wook Kim}, {and} \bibinfo{person}{Jung-Tae Lee}.} \bibinfo{year}{2018}\natexlab{}.
\newblock \showarticletitle{CFGAN: A generic collaborative filtering framework based on generative adversarial networks}. In \bibinfo{booktitle}{\emph{Proceedings of the 27th ACM international conference on information and knowledge management}}. \bibinfo{pages}{137--146}.
\newblock


\bibitem[Chai et~al\mbox{.}(2022)]%
        {umi}
\bibfield{author}{\bibinfo{person}{Zheng Chai}, \bibinfo{person}{Zhihong Chen}, \bibinfo{person}{Chenliang Li}, \bibinfo{person}{Rong Xiao}, \bibinfo{person}{Houyi Li}, \bibinfo{person}{Jiawei Wu}, \bibinfo{person}{Jingxu Chen}, {and} \bibinfo{person}{Haihong Tang}.} \bibinfo{year}{2022}\natexlab{}.
\newblock \showarticletitle{User-aware multi-interest learning for candidate matching in recommenders}. In \bibinfo{booktitle}{\emph{Proceedings of the 45th International ACM SIGIR Conference on Research and Development in Information Retrieval}}. \bibinfo{pages}{1326--1335}.
\newblock


\bibitem[Chen et~al\mbox{.}(2023b)]%
        {chen2023revisiting}
\bibfield{author}{\bibinfo{person}{Chong Chen}, \bibinfo{person}{Weizhi Ma}, \bibinfo{person}{Min Zhang}, \bibinfo{person}{Chenyang Wang}, \bibinfo{person}{Yiqun Liu}, {and} \bibinfo{person}{Shaoping Ma}.} \bibinfo{year}{2023}\natexlab{b}.
\newblock \showarticletitle{Revisiting negative sampling vs. non-sampling in implicit recommendation}.
\newblock \bibinfo{journal}{\emph{ACM Transactions on Information Systems}} \bibinfo{volume}{41}, \bibinfo{number}{1} (\bibinfo{year}{2023}), \bibinfo{pages}{1--25}.
\newblock


\bibitem[Chen et~al\mbox{.}(2020)]%
        {chen2020efficient}
\bibfield{author}{\bibinfo{person}{Chong Chen}, \bibinfo{person}{Min Zhang}, \bibinfo{person}{Yongfeng Zhang}, \bibinfo{person}{Yiqun Liu}, {and} \bibinfo{person}{Shaoping Ma}.} \bibinfo{year}{2020}\natexlab{}.
\newblock \showarticletitle{Efficient neural matrix factorization without sampling for recommendation}.
\newblock \bibinfo{journal}{\emph{ACM Transactions on Information Systems (TOIS)}} \bibinfo{volume}{38}, \bibinfo{number}{2} (\bibinfo{year}{2020}), \bibinfo{pages}{1--28}.
\newblock


\bibitem[Chen et~al\mbox{.}(2021)]%
        {pimirec}
\bibfield{author}{\bibinfo{person}{Gaode Chen}, \bibinfo{person}{Xinghua Zhang}, \bibinfo{person}{Yanyan Zhao}, \bibinfo{person}{Cong Xue}, {and} \bibinfo{person}{Ji Xiang}.} \bibinfo{year}{2021}\natexlab{}.
\newblock \showarticletitle{Exploring periodicity and interactivity in multi-interest framework for sequential recommendation}.
\newblock \bibinfo{journal}{\emph{arXiv preprint arXiv:2106.04415}} (\bibinfo{year}{2021}).
\newblock


\bibitem[Chen et~al\mbox{.}(2023a)]%
        {Chen_Dong_Wang_Feng_Wang_He_2023}
\bibfield{author}{\bibinfo{person}{Jiawei Chen}, \bibinfo{person}{Hande Dong}, \bibinfo{person}{Xiang Wang}, \bibinfo{person}{Fuli Feng}, \bibinfo{person}{Meng Wang}, {and} \bibinfo{person}{Xiangnan He}.} \bibinfo{year}{2023}\natexlab{a}.
\newblock \showarticletitle{Bias and Debias in Recommender System: A Survey and Future Directions}.
\newblock \bibinfo{journal}{\emph{ACM Transactions on Information Systems}} (\bibinfo{date}{Jul} \bibinfo{year}{2023}), \bibinfo{pages}{1–39}.
\newblock
\urldef\tempurl%
\url{https://doi.org/10.1145/3564284}
\showDOI{\tempurl}


\bibitem[Chen et~al\mbox{.}(2022)]%
        {chen2022fast}
\bibfield{author}{\bibinfo{person}{Jin Chen}, \bibinfo{person}{Defu Lian}, \bibinfo{person}{Binbin Jin}, \bibinfo{person}{Xu Huang}, \bibinfo{person}{Kai Zheng}, {and} \bibinfo{person}{Enhong Chen}.} \bibinfo{year}{2022}\natexlab{}.
\newblock \showarticletitle{Fast variational autoencoder with inverted multi-index for collaborative filtering}. In \bibinfo{booktitle}{\emph{Proceedings of the ACM Web Conference 2022}}. \bibinfo{pages}{1944--1954}.
\newblock


\bibitem[Chen et~al\mbox{.}(2018b)]%
        {chen2018heterogeneous}
\bibfield{author}{\bibinfo{person}{Liang Chen}, \bibinfo{person}{Yang Liu}, \bibinfo{person}{Zibin Zheng}, {and} \bibinfo{person}{Philip Yu}.} \bibinfo{year}{2018}\natexlab{b}.
\newblock \showarticletitle{Heterogeneous neural attentive factorization machine for rating prediction}. In \bibinfo{booktitle}{\emph{Proceedings of the 27th ACM international conference on information and knowledge management}}. \bibinfo{pages}{833--842}.
\newblock


\bibitem[Chen et~al\mbox{.}(2019b)]%
        {chen2019behavior}
\bibfield{author}{\bibinfo{person}{Qiwei Chen}, \bibinfo{person}{Huan Zhao}, \bibinfo{person}{Wei Li}, \bibinfo{person}{Pipei Huang}, {and} \bibinfo{person}{Wenwu Ou}.} \bibinfo{year}{2019}\natexlab{b}.
\newblock \showarticletitle{Behavior sequence transformer for e-commerce recommendation in alibaba}. In \bibinfo{booktitle}{\emph{Proceedings of the 1st International Workshop on Deep Learning Practice for High-Dimensional Sparse Data}}. \bibinfo{pages}{1--4}.
\newblock


\bibitem[Chen et~al\mbox{.}(2018a)]%
        {chen2018survey}
\bibfield{author}{\bibinfo{person}{Rui Chen}, \bibinfo{person}{Qingyi Hua}, \bibinfo{person}{Yan-Shuo Chang}, \bibinfo{person}{Bo Wang}, \bibinfo{person}{Lei Zhang}, {and} \bibinfo{person}{Xiangjie Kong}.} \bibinfo{year}{2018}\natexlab{a}.
\newblock \showarticletitle{A survey of collaborative filtering-based recommender systems: From traditional methods to hybrid methods based on social networks}.
\newblock \bibinfo{journal}{\emph{IEEE access}}  \bibinfo{volume}{6} (\bibinfo{year}{2018}), \bibinfo{pages}{64301--64320}.
\newblock


\bibitem[Chen et~al\mbox{.}(2017)]%
        {chen2017efficient}
\bibfield{author}{\bibinfo{person}{Ruey-Cheng Chen}, \bibinfo{person}{Luke Gallagher}, \bibinfo{person}{Roi Blanco}, {and} \bibinfo{person}{J~Shane Culpepper}.} \bibinfo{year}{2017}\natexlab{}.
\newblock \showarticletitle{Efficient cost-aware cascade ranking in multi-stage retrieval}. In \bibinfo{booktitle}{\emph{Proceedings of the 40th International ACM SIGIR Conference on Research and Development in Information Retrieval}}. \bibinfo{pages}{445--454}.
\newblock


\bibitem[Chen et~al\mbox{.}(2019a)]%
        {chen2019joint}
\bibfield{author}{\bibinfo{person}{Wanyu Chen}, \bibinfo{person}{Fei Cai}, \bibinfo{person}{Honghui Chen}, {and} \bibinfo{person}{Maarten~De Rijke}.} \bibinfo{year}{2019}\natexlab{a}.
\newblock \showarticletitle{Joint neural collaborative filtering for recommender systems}.
\newblock \bibinfo{journal}{\emph{ACM Transactions on Information Systems (TOIS)}} \bibinfo{volume}{37}, \bibinfo{number}{4} (\bibinfo{year}{2019}), \bibinfo{pages}{1--30}.
\newblock


\bibitem[Cho and Oh(2022)]%
        {cho2022stochastic}
\bibfield{author}{\bibinfo{person}{Yoon-Sik Cho} {and} \bibinfo{person}{Min-hwan Oh}.} \bibinfo{year}{2022}\natexlab{}.
\newblock \showarticletitle{Stochastic-expert variational autoencoder for collaborative filtering}. In \bibinfo{booktitle}{\emph{Proceedings of the ACM Web Conference 2022}}. \bibinfo{pages}{2482--2490}.
\newblock


\bibitem[Choi et~al\mbox{.}(2021)]%
        {choi2021improving}
\bibfield{author}{\bibinfo{person}{Jaekeol Choi}, \bibinfo{person}{Euna Jung}, \bibinfo{person}{Jangwon Suh}, {and} \bibinfo{person}{Wonjong Rhee}.} \bibinfo{year}{2021}\natexlab{}.
\newblock \showarticletitle{Improving bi-encoder document ranking models with two rankers and multi-teacher distillation}. In \bibinfo{booktitle}{\emph{Proceedings of the 44th international ACM SIGIR conference on research and development in information retrieval}}. \bibinfo{pages}{2192--2196}.
\newblock


\bibitem[Clarke et~al\mbox{.}(2016)]%
        {clarke2016assessing}
\bibfield{author}{\bibinfo{person}{Charles~LA Clarke}, \bibinfo{person}{J~Shane Culpepper}, {and} \bibinfo{person}{Alistair Moffat}.} \bibinfo{year}{2016}\natexlab{}.
\newblock \showarticletitle{Assessing efficiency--effectiveness tradeoffs in multi-stage retrieval systems without using relevance judgments}.
\newblock \bibinfo{journal}{\emph{Information Retrieval Journal}}  \bibinfo{volume}{19} (\bibinfo{year}{2016}), \bibinfo{pages}{351--377}.
\newblock


\bibitem[Covington et~al\mbox{.}(2016)]%
        {covington2016deep}
\bibfield{author}{\bibinfo{person}{Paul Covington}, \bibinfo{person}{Jay Adams}, {and} \bibinfo{person}{Emre Sargin}.} \bibinfo{year}{2016}\natexlab{}.
\newblock \showarticletitle{Deep neural networks for youtube recommendations}. In \bibinfo{booktitle}{\emph{Proceedings of the 10th ACM conference on recommender systems}}. \bibinfo{pages}{191--198}.
\newblock


\bibitem[Dai et~al\mbox{.}(2023)]%
        {dai2023uncovering}
\bibfield{author}{\bibinfo{person}{Sunhao Dai}, \bibinfo{person}{Ninglu Shao}, \bibinfo{person}{Haiyuan Zhao}, \bibinfo{person}{Weijie Yu}, \bibinfo{person}{Zihua Si}, \bibinfo{person}{Chen Xu}, \bibinfo{person}{Zhongxiang Sun}, \bibinfo{person}{Xiao Zhang}, {and} \bibinfo{person}{Jun Xu}.} \bibinfo{year}{2023}\natexlab{}.
\newblock \showarticletitle{Uncovering chatgpt’s capabilities in recommender systems}. In \bibinfo{booktitle}{\emph{Proceedings of the 17th ACM Conference on Recommender Systems}}. \bibinfo{pages}{1126--1132}.
\newblock


\bibitem[Devooght et~al\mbox{.}(2015)]%
        {devooght2015dynamic}
\bibfield{author}{\bibinfo{person}{Robin Devooght}, \bibinfo{person}{Nicolas Kourtellis}, {and} \bibinfo{person}{Amin Mantrach}.} \bibinfo{year}{2015}\natexlab{}.
\newblock \showarticletitle{Dynamic matrix factorization with priors on unknown values}. In \bibinfo{booktitle}{\emph{Proceedings of the 21th ACM SIGKDD international conference on knowledge discovery and data mining}}. \bibinfo{pages}{189--198}.
\newblock


\bibitem[Di~Palma et~al\mbox{.}(2023)]%
        {di2023evaluating}
\bibfield{author}{\bibinfo{person}{Dario Di~Palma}, \bibinfo{person}{Giovanni~Maria Biancofiore}, \bibinfo{person}{Vito~Walter Anelli}, \bibinfo{person}{Fedelucio Narducci}, \bibinfo{person}{Tommaso Di~Noia}, {and} \bibinfo{person}{Eugenio Di~Sciascio}.} \bibinfo{year}{2023}\natexlab{}.
\newblock \showarticletitle{Evaluating chatgpt as a recommender system: A rigorous approach}.
\newblock \bibinfo{journal}{\emph{arXiv preprint arXiv:2309.03613}} (\bibinfo{year}{2023}).
\newblock


\bibitem[Ding et~al\mbox{.}(2020)]%
        {ding2020simplify}
\bibfield{author}{\bibinfo{person}{Jingtao Ding}, \bibinfo{person}{Yuhan Quan}, \bibinfo{person}{Quanming Yao}, \bibinfo{person}{Yong Li}, {and} \bibinfo{person}{Depeng Jin}.} \bibinfo{year}{2020}\natexlab{}.
\newblock \showarticletitle{Simplify and robustify negative sampling for implicit collaborative filtering}.
\newblock \bibinfo{journal}{\emph{Advances in Neural Information Processing Systems}}  \bibinfo{volume}{33} (\bibinfo{year}{2020}), \bibinfo{pages}{1094--1105}.
\newblock


\bibitem[El-Kishky et~al\mbox{.}(2023)]%
        {knn_embed}
\bibfield{author}{\bibinfo{person}{Ahmed El-Kishky}, \bibinfo{person}{Thomas Markovich}, \bibinfo{person}{Kenny Leung}, \bibinfo{person}{Frank Portman}, \bibinfo{person}{Aria Haghighi}, {and} \bibinfo{person}{Ying Xiao}.} \bibinfo{year}{2023}\natexlab{}.
\newblock \showarticletitle{k NN-Embed: Locally Smoothed Embedding Mixtures for Multi-interest Candidate Retrieval}. In \bibinfo{booktitle}{\emph{Pacific-Asia Conference on Knowledge Discovery and Data Mining}}. Springer, \bibinfo{pages}{374--386}.
\newblock


\bibitem[Fan et~al\mbox{.}(2019)]%
        {fan2019mobius}
\bibfield{author}{\bibinfo{person}{Miao Fan}, \bibinfo{person}{Jiacheng Guo}, \bibinfo{person}{Shuai Zhu}, \bibinfo{person}{Shuo Miao}, \bibinfo{person}{Mingming Sun}, {and} \bibinfo{person}{Ping Li}.} \bibinfo{year}{2019}\natexlab{}.
\newblock \showarticletitle{MOBIUS: towards the next generation of query-ad matching in baidu's sponsored search}. In \bibinfo{booktitle}{\emph{Proceedings of the 25th ACM SIGKDD International Conference on Knowledge Discovery \& Data Mining}}. \bibinfo{pages}{2509--2517}.
\newblock


\bibitem[Fei et~al\mbox{.}(2021)]%
        {fei2021gemnn}
\bibfield{author}{\bibinfo{person}{Hongliang Fei}, \bibinfo{person}{Jingyuan Zhang}, \bibinfo{person}{Xingxuan Zhou}, \bibinfo{person}{Junhao Zhao}, \bibinfo{person}{Xinyang Qi}, {and} \bibinfo{person}{Ping Li}.} \bibinfo{year}{2021}\natexlab{}.
\newblock \showarticletitle{GemNN: gating-enhanced multi-task neural networks with feature interaction learning for CTR prediction}. In \bibinfo{booktitle}{\emph{Proceedings of the 44th international ACM SIGIR conference on research and development in information retrieval}}. \bibinfo{pages}{2166--2171}.
\newblock


\bibitem[Firan et~al\mbox{.}(2007)]%
        {firan2007benefit}
\bibfield{author}{\bibinfo{person}{Claudiu~S Firan}, \bibinfo{person}{Wolfgang Nejdl}, {and} \bibinfo{person}{Raluca Paiu}.} \bibinfo{year}{2007}\natexlab{}.
\newblock \showarticletitle{The benefit of using tag-based profiles}. In \bibinfo{booktitle}{\emph{2007 Latin American Web Conference (LA-WEB 2007)}}. IEEE, \bibinfo{pages}{32--41}.
\newblock


\bibitem[Fleder and Hosanagar(2009)]%
        {fleder2009blockbuster}
\bibfield{author}{\bibinfo{person}{Daniel Fleder} {and} \bibinfo{person}{Kartik Hosanagar}.} \bibinfo{year}{2009}\natexlab{}.
\newblock \showarticletitle{Blockbuster culture's next rise or fall: The impact of recommender systems on sales diversity}.
\newblock \bibinfo{journal}{\emph{Management science}} \bibinfo{volume}{55}, \bibinfo{number}{5} (\bibinfo{year}{2009}), \bibinfo{pages}{697--712}.
\newblock


\bibitem[Fu et~al\mbox{.}(2018)]%
        {fu2018novel}
\bibfield{author}{\bibinfo{person}{Mingsheng Fu}, \bibinfo{person}{Hong Qu}, \bibinfo{person}{Zhang Yi}, \bibinfo{person}{Li Lu}, {and} \bibinfo{person}{Yongsheng Liu}.} \bibinfo{year}{2018}\natexlab{}.
\newblock \showarticletitle{A novel deep learning-based collaborative filtering model for recommendation system}.
\newblock \bibinfo{journal}{\emph{IEEE transactions on cybernetics}} \bibinfo{volume}{49}, \bibinfo{number}{3} (\bibinfo{year}{2018}), \bibinfo{pages}{1084--1096}.
\newblock


\bibitem[Gallagher et~al\mbox{.}(2019)]%
        {gallagher2019joint}
\bibfield{author}{\bibinfo{person}{Luke Gallagher}, \bibinfo{person}{Ruey-Cheng Chen}, \bibinfo{person}{Roi Blanco}, {and} \bibinfo{person}{J~Shane Culpepper}.} \bibinfo{year}{2019}\natexlab{}.
\newblock \showarticletitle{Joint optimization of cascade ranking models}. In \bibinfo{booktitle}{\emph{Proceedings of the twelfth ACM international conference on web search and data mining}}. \bibinfo{pages}{15--23}.
\newblock


\bibitem[Gao et~al\mbox{.}(2020)]%
        {deepretrieval}
\bibfield{author}{\bibinfo{person}{Weihao Gao}, \bibinfo{person}{Xiangjun Fan}, \bibinfo{person}{Chong Wang}, \bibinfo{person}{Jiankai Sun}, \bibinfo{person}{Kai Jia}, \bibinfo{person}{Wenzhi Xiao}, \bibinfo{person}{Ruofan Ding}, \bibinfo{person}{Xingyan Bin}, \bibinfo{person}{Hui Yang}, {and} \bibinfo{person}{Xiaobing Liu}.} \bibinfo{year}{2020}\natexlab{}.
\newblock \showarticletitle{Deep retrieval: Learning a retrievable structure for large-scale recommendations}.
\newblock \bibinfo{journal}{\emph{arXiv preprint arXiv:2007.07203}} (\bibinfo{year}{2020}).
\newblock


\bibitem[Geng et~al\mbox{.}(2022)]%
        {geng2022recommendation}
\bibfield{author}{\bibinfo{person}{Shijie Geng}, \bibinfo{person}{Shuchang Liu}, \bibinfo{person}{Zuohui Fu}, \bibinfo{person}{Yingqiang Ge}, {and} \bibinfo{person}{Yongfeng Zhang}.} \bibinfo{year}{2022}\natexlab{}.
\newblock \showarticletitle{Recommendation as language processing (rlp): A unified pretrain, personalized prompt \& predict paradigm (p5)}. In \bibinfo{booktitle}{\emph{Proceedings of the 16th ACM Conference on Recommender Systems}}. \bibinfo{pages}{299--315}.
\newblock


\bibitem[Grbovic and Cheng(2018)]%
        {grbovic2018real}
\bibfield{author}{\bibinfo{person}{Mihajlo Grbovic} {and} \bibinfo{person}{Haibin Cheng}.} \bibinfo{year}{2018}\natexlab{}.
\newblock \showarticletitle{Real-time personalization using embeddings for search ranking at airbnb}. In \bibinfo{booktitle}{\emph{Proceedings of the 24th ACM SIGKDD international conference on knowledge discovery \& data mining}}. \bibinfo{pages}{311--320}.
\newblock


\bibitem[Grover and Leskovec(2016)]%
        {grover2016node2vec}
\bibfield{author}{\bibinfo{person}{Aditya Grover} {and} \bibinfo{person}{Jure Leskovec}.} \bibinfo{year}{2016}\natexlab{}.
\newblock \showarticletitle{node2vec: Scalable feature learning for networks}. In \bibinfo{booktitle}{\emph{Proceedings of the 22nd ACM SIGKDD international conference on Knowledge discovery and data mining}}. \bibinfo{pages}{855--864}.
\newblock


\bibitem[Guo et~al\mbox{.}(2020b)]%
        {guo2020ipgan}
\bibfield{author}{\bibinfo{person}{Guibing Guo}, \bibinfo{person}{Huan Zhou}, \bibinfo{person}{Bowei Chen}, \bibinfo{person}{Zhirong Liu}, \bibinfo{person}{Xiao Xu}, \bibinfo{person}{Xu Chen}, \bibinfo{person}{Zhenhua Dong}, {and} \bibinfo{person}{Xiuqiang He}.} \bibinfo{year}{2020}\natexlab{b}.
\newblock \showarticletitle{IPGAN: Generating informative item pairs by adversarial sampling}.
\newblock \bibinfo{journal}{\emph{IEEE transactions on neural networks and learning systems}} \bibinfo{volume}{33}, \bibinfo{number}{2} (\bibinfo{year}{2020}), \bibinfo{pages}{694--706}.
\newblock


\bibitem[Guo et~al\mbox{.}(2020a)]%
        {guo2020accelerating}
\bibfield{author}{\bibinfo{person}{Ruiqi Guo}, \bibinfo{person}{Philip Sun}, \bibinfo{person}{Erik Lindgren}, \bibinfo{person}{Quan Geng}, \bibinfo{person}{David Simcha}, \bibinfo{person}{Felix Chern}, {and} \bibinfo{person}{Sanjiv Kumar}.} \bibinfo{year}{2020}\natexlab{a}.
\newblock \showarticletitle{Accelerating large-scale inference with anisotropic vector quantization}. In \bibinfo{booktitle}{\emph{International Conference on Machine Learning}}. PMLR, \bibinfo{pages}{3887--3896}.
\newblock


\bibitem[He and McAuley(2016)]%
        {he2016vbpr}
\bibfield{author}{\bibinfo{person}{Ruining He} {and} \bibinfo{person}{Julian McAuley}.} \bibinfo{year}{2016}\natexlab{}.
\newblock \showarticletitle{VBPR: visual bayesian personalized ranking from implicit feedback}. In \bibinfo{booktitle}{\emph{Proceedings of the AAAI conference on artificial intelligence}}, Vol.~\bibinfo{volume}{30}.
\newblock


\bibitem[He et~al\mbox{.}(2020)]%
        {he2020lightgcn}
\bibfield{author}{\bibinfo{person}{Xiangnan He}, \bibinfo{person}{Kuan Deng}, \bibinfo{person}{Xiang Wang}, \bibinfo{person}{Yan Li}, \bibinfo{person}{Yongdong Zhang}, {and} \bibinfo{person}{Meng Wang}.} \bibinfo{year}{2020}\natexlab{}.
\newblock \showarticletitle{Lightgcn: Simplifying and powering graph convolution network for recommendation}. In \bibinfo{booktitle}{\emph{Proceedings of the 43rd International ACM SIGIR conference on research and development in Information Retrieval}}. \bibinfo{pages}{639--648}.
\newblock


\bibitem[He et~al\mbox{.}(2018)]%
        {he2018adversarial}
\bibfield{author}{\bibinfo{person}{Xiangnan He}, \bibinfo{person}{Zhankui He}, \bibinfo{person}{Xiaoyu Du}, {and} \bibinfo{person}{Tat-Seng Chua}.} \bibinfo{year}{2018}\natexlab{}.
\newblock \showarticletitle{Adversarial personalized ranking for recommendation}. In \bibinfo{booktitle}{\emph{The 41st International ACM SIGIR conference on research \& development in information retrieval}}. \bibinfo{pages}{355--364}.
\newblock


\bibitem[He et~al\mbox{.}(2017)]%
        {he2017neural}
\bibfield{author}{\bibinfo{person}{Xiangnan He}, \bibinfo{person}{Lizi Liao}, \bibinfo{person}{Hanwang Zhang}, \bibinfo{person}{Liqiang Nie}, \bibinfo{person}{Xia Hu}, {and} \bibinfo{person}{Tat-Seng Chua}.} \bibinfo{year}{2017}\natexlab{}.
\newblock \showarticletitle{Neural collaborative filtering}. In \bibinfo{booktitle}{\emph{Proceedings of the 26th international conference on world wide web}}. \bibinfo{pages}{173--182}.
\newblock


\bibitem[He et~al\mbox{.}(2016)]%
        {he2016fast}
\bibfield{author}{\bibinfo{person}{Xiangnan He}, \bibinfo{person}{Hanwang Zhang}, \bibinfo{person}{Min-Yen Kan}, {and} \bibinfo{person}{Tat-Seng Chua}.} \bibinfo{year}{2016}\natexlab{}.
\newblock \showarticletitle{Fast matrix factorization for online recommendation with implicit feedback}. In \bibinfo{booktitle}{\emph{Proceedings of the 39th International ACM SIGIR conference on Research and Development in Information Retrieval}}. \bibinfo{pages}{549--558}.
\newblock


\bibitem[He et~al\mbox{.}(2023)]%
        {he2023survey}
\bibfield{author}{\bibinfo{person}{Zhicheng He}, \bibinfo{person}{Weiwen Liu}, \bibinfo{person}{Wei Guo}, \bibinfo{person}{Jiarui Qin}, \bibinfo{person}{Yingxue Zhang}, \bibinfo{person}{Yaochen Hu}, {and} \bibinfo{person}{Ruiming Tang}.} \bibinfo{year}{2023}\natexlab{}.
\newblock \showarticletitle{A survey on user behavior modeling in recommender systems}.
\newblock \bibinfo{journal}{\emph{arXiv preprint arXiv:2302.11087}} (\bibinfo{year}{2023}).
\newblock


\bibitem[Hron et~al\mbox{.}(2021)]%
        {hron2021component}
\bibfield{author}{\bibinfo{person}{Jiri Hron}, \bibinfo{person}{Karl Krauth}, \bibinfo{person}{Michael Jordan}, {and} \bibinfo{person}{Niki Kilbertus}.} \bibinfo{year}{2021}\natexlab{}.
\newblock \showarticletitle{On component interactions in two-stage recommender systems}.
\newblock \bibinfo{journal}{\emph{Advances in neural information processing systems}}  \bibinfo{volume}{34} (\bibinfo{year}{2021}), \bibinfo{pages}{2744--2757}.
\newblock


\bibitem[Hsieh et~al\mbox{.}(2017)]%
        {hsieh2017collaborative}
\bibfield{author}{\bibinfo{person}{Cheng-Kang Hsieh}, \bibinfo{person}{Longqi Yang}, \bibinfo{person}{Yin Cui}, \bibinfo{person}{Tsung-Yi Lin}, \bibinfo{person}{Serge Belongie}, {and} \bibinfo{person}{Deborah Estrin}.} \bibinfo{year}{2017}\natexlab{}.
\newblock \showarticletitle{Collaborative metric learning}. In \bibinfo{booktitle}{\emph{Proceedings of the 26th international conference on world wide web}}. \bibinfo{pages}{193--201}.
\newblock


\bibitem[Hu et~al\mbox{.}(2008)]%
        {hu2008collaborative}
\bibfield{author}{\bibinfo{person}{Yifan Hu}, \bibinfo{person}{Yehuda Koren}, {and} \bibinfo{person}{Chris Volinsky}.} \bibinfo{year}{2008}\natexlab{}.
\newblock \showarticletitle{Collaborative filtering for implicit feedback datasets}. In \bibinfo{booktitle}{\emph{2008 Eighth IEEE international conference on data mining}}. Ieee, \bibinfo{pages}{263--272}.
\newblock


\bibitem[Huang et~al\mbox{.}(2024)]%
        {huang2024recall}
\bibfield{author}{\bibinfo{person}{Junjie Huang}, \bibinfo{person}{Guohao Cai}, \bibinfo{person}{Jieming Zhu}, \bibinfo{person}{Zhenhua Dong}, \bibinfo{person}{Ruiming Tang}, \bibinfo{person}{Weinan Zhang}, {and} \bibinfo{person}{Yong Yu}.} \bibinfo{year}{2024}\natexlab{}.
\newblock \showarticletitle{Recall-Augmented Ranking: Enhancing Click-Through Rate Prediction Accuracy with Cross-Stage Data}. In \bibinfo{booktitle}{\emph{Companion Proceedings of the ACM on Web Conference 2024}}. \bibinfo{pages}{830--833}.
\newblock


\bibitem[Huang et~al\mbox{.}(2025)]%
        {huang2025unleashing}
\bibfield{author}{\bibinfo{person}{Junjie Huang}, \bibinfo{person}{Jiarui Qin}, \bibinfo{person}{Jianghao Lin}, \bibinfo{person}{Ziming Feng}, \bibinfo{person}{Weinan Zhang}, {and} \bibinfo{person}{Yong Yu}.} \bibinfo{year}{2025}\natexlab{}.
\newblock \showarticletitle{Unleashing the Potential of Multi-Channel Fusion in Retrieval for Personalized Recommendations}. In \bibinfo{booktitle}{\emph{Proceedings of the ACM on Web Conference 2025}}. \bibinfo{pages}{483--494}.
\newblock


\bibitem[Huang et~al\mbox{.}(2013)]%
        {huang2013learning}
\bibfield{author}{\bibinfo{person}{Po-Sen Huang}, \bibinfo{person}{Xiaodong He}, \bibinfo{person}{Jianfeng Gao}, \bibinfo{person}{Li Deng}, \bibinfo{person}{Alex Acero}, {and} \bibinfo{person}{Larry Heck}.} \bibinfo{year}{2013}\natexlab{}.
\newblock \showarticletitle{Learning deep structured semantic models for web search using clickthrough data}. In \bibinfo{booktitle}{\emph{Proceedings of the 22nd ACM international conference on Information \& Knowledge Management}}. \bibinfo{pages}{2333--2338}.
\newblock


\bibitem[Huang et~al\mbox{.}(2021)]%
        {huang2021mixgcf}
\bibfield{author}{\bibinfo{person}{Tinglin Huang}, \bibinfo{person}{Yuxiao Dong}, \bibinfo{person}{Ming Ding}, \bibinfo{person}{Zhen Yang}, \bibinfo{person}{Wenzheng Feng}, \bibinfo{person}{Xinyu Wang}, {and} \bibinfo{person}{Jie Tang}.} \bibinfo{year}{2021}\natexlab{}.
\newblock \showarticletitle{Mixgcf: An improved training method for graph neural network-based recommender systems}. In \bibinfo{booktitle}{\emph{Proceedings of the 27th ACM SIGKDD Conference on Knowledge Discovery \& Data Mining}}. \bibinfo{pages}{665--674}.
\newblock


\bibitem[Huang et~al\mbox{.}(2023)]%
        {huang2023cooperative}
\bibfield{author}{\bibinfo{person}{Xu Huang}, \bibinfo{person}{Defu Lian}, \bibinfo{person}{Jin Chen}, \bibinfo{person}{Liu Zheng}, \bibinfo{person}{Xing Xie}, {and} \bibinfo{person}{Enhong Chen}.} \bibinfo{year}{2023}\natexlab{}.
\newblock \showarticletitle{Cooperative Retriever and Ranker in Deep Recommenders}. In \bibinfo{booktitle}{\emph{Proceedings of the ACM Web Conference 2023}}. \bibinfo{pages}{1150--1161}.
\newblock


\bibitem[Jamali and Ester(2010)]%
        {jamali2010matrix}
\bibfield{author}{\bibinfo{person}{Mohsen Jamali} {and} \bibinfo{person}{Martin Ester}.} \bibinfo{year}{2010}\natexlab{}.
\newblock \showarticletitle{A matrix factorization technique with trust propagation for recommendation in social networks}. In \bibinfo{booktitle}{\emph{Proceedings of the fourth ACM conference on Recommender systems}}. \bibinfo{pages}{135--142}.
\newblock


\bibitem[Jin et~al\mbox{.}(2020)]%
        {jin2020sampling}
\bibfield{author}{\bibinfo{person}{Binbin Jin}, \bibinfo{person}{Defu Lian}, \bibinfo{person}{Zheng Liu}, \bibinfo{person}{Qi Liu}, \bibinfo{person}{Jianhui Ma}, \bibinfo{person}{Xing Xie}, {and} \bibinfo{person}{Enhong Chen}.} \bibinfo{year}{2020}\natexlab{}.
\newblock \showarticletitle{Sampling-decomposable generative adversarial recommender}.
\newblock \bibinfo{journal}{\emph{Advances in Neural Information Processing Systems}}  \bibinfo{volume}{33} (\bibinfo{year}{2020}), \bibinfo{pages}{22629--22639}.
\newblock


\bibitem[Johnson et~al\mbox{.}(2019)]%
        {johnson2019billion}
\bibfield{author}{\bibinfo{person}{Jeff Johnson}, \bibinfo{person}{Matthijs Douze}, {and} \bibinfo{person}{Herv{\'e} J{\'e}gou}.} \bibinfo{year}{2019}\natexlab{}.
\newblock \showarticletitle{Billion-scale similarity search with GPUs}.
\newblock \bibinfo{journal}{\emph{IEEE Transactions on Big Data}} \bibinfo{volume}{7}, \bibinfo{number}{3} (\bibinfo{year}{2019}), \bibinfo{pages}{535--547}.
\newblock


\bibitem[Karamanolakis et~al\mbox{.}(2018)]%
        {karamanolakis2018item}
\bibfield{author}{\bibinfo{person}{Giannis Karamanolakis}, \bibinfo{person}{Kevin~Raji Cherian}, \bibinfo{person}{Ananth~Ravi Narayan}, \bibinfo{person}{Jie Yuan}, \bibinfo{person}{Da Tang}, {and} \bibinfo{person}{Tony Jebara}.} \bibinfo{year}{2018}\natexlab{}.
\newblock \showarticletitle{Item recommendation with variational autoencoders and heterogeneous priors}. In \bibinfo{booktitle}{\emph{Proceedings of the 3rd Workshop on Deep Learning for Recommender Systems}}. \bibinfo{pages}{10--14}.
\newblock


\bibitem[Karypis(2001)]%
        {karypis2001evaluation}
\bibfield{author}{\bibinfo{person}{George Karypis}.} \bibinfo{year}{2001}\natexlab{}.
\newblock \showarticletitle{Evaluation of item-based top-n recommendation algorithms}. In \bibinfo{booktitle}{\emph{Proceedings of the tenth international conference on Information and knowledge management}}. \bibinfo{pages}{247--254}.
\newblock


\bibitem[Kim and Suh(2019)]%
        {kim2019enhancing}
\bibfield{author}{\bibinfo{person}{Daeryong Kim} {and} \bibinfo{person}{Bongwon Suh}.} \bibinfo{year}{2019}\natexlab{}.
\newblock \showarticletitle{Enhancing VAEs for collaborative filtering: flexible priors \& gating mechanisms}. In \bibinfo{booktitle}{\emph{Proceedings of the 13th ACM conference on recommender systems}}. \bibinfo{pages}{403--407}.
\newblock


\bibitem[Kingma and Welling(2013)]%
        {kingma2013auto}
\bibfield{author}{\bibinfo{person}{Diederik~P Kingma} {and} \bibinfo{person}{Max Welling}.} \bibinfo{year}{2013}\natexlab{}.
\newblock \showarticletitle{Auto-encoding variational bayes}.
\newblock \bibinfo{journal}{\emph{arXiv preprint arXiv:1312.6114}} (\bibinfo{year}{2013}).
\newblock


\bibitem[Koren(2008)]%
        {koren2008factorization}
\bibfield{author}{\bibinfo{person}{Yehuda Koren}.} \bibinfo{year}{2008}\natexlab{}.
\newblock \showarticletitle{Factorization meets the neighborhood: a multifaceted collaborative filtering model}. In \bibinfo{booktitle}{\emph{Proceedings of the 14th ACM SIGKDD international conference on Knowledge discovery and data mining}}. \bibinfo{pages}{426--434}.
\newblock


\bibitem[Koren(2009)]%
        {koren2009collaborative}
\bibfield{author}{\bibinfo{person}{Yehuda Koren}.} \bibinfo{year}{2009}\natexlab{}.
\newblock \showarticletitle{Collaborative filtering with temporal dynamics}. In \bibinfo{booktitle}{\emph{Proceedings of the 15th ACM SIGKDD international conference on Knowledge discovery and data mining}}. \bibinfo{pages}{447--456}.
\newblock


\bibitem[Koren et~al\mbox{.}(2009)]%
        {koren2009matrix}
\bibfield{author}{\bibinfo{person}{Yehuda Koren}, \bibinfo{person}{Robert Bell}, {and} \bibinfo{person}{Chris Volinsky}.} \bibinfo{year}{2009}\natexlab{}.
\newblock \showarticletitle{Matrix factorization techniques for recommender systems}.
\newblock \bibinfo{journal}{\emph{Computer}} \bibinfo{volume}{42}, \bibinfo{number}{8} (\bibinfo{year}{2009}), \bibinfo{pages}{30--37}.
\newblock


\bibitem[Krumm et~al\mbox{.}(2008)]%
        {krumm2008user}
\bibfield{author}{\bibinfo{person}{John Krumm}, \bibinfo{person}{Nigel Davies}, {and} \bibinfo{person}{Chandra Narayanaswami}.} \bibinfo{year}{2008}\natexlab{}.
\newblock \showarticletitle{User-generated content}.
\newblock \bibinfo{journal}{\emph{IEEE Pervasive Computing}} \bibinfo{volume}{7}, \bibinfo{number}{4} (\bibinfo{year}{2008}), \bibinfo{pages}{10--11}.
\newblock


\bibitem[Lee et~al\mbox{.}(2017)]%
        {lee2017augmented}
\bibfield{author}{\bibinfo{person}{Wonsung Lee}, \bibinfo{person}{Kyungwoo Song}, {and} \bibinfo{person}{Il-Chul Moon}.} \bibinfo{year}{2017}\natexlab{}.
\newblock \showarticletitle{Augmented variational autoencoders for collaborative filtering with auxiliary information}. In \bibinfo{booktitle}{\emph{Proceedings of the 2017 ACM on Conference on Information and Knowledge Management}}. \bibinfo{pages}{1139--1148}.
\newblock


\bibitem[Levi et~al\mbox{.}(2012)]%
        {levi2012finding}
\bibfield{author}{\bibinfo{person}{Asher Levi}, \bibinfo{person}{Osnat Mokryn}, \bibinfo{person}{Christophe Diot}, {and} \bibinfo{person}{Nina Taft}.} \bibinfo{year}{2012}\natexlab{}.
\newblock \showarticletitle{Finding a needle in a haystack of reviews: cold start context-based hotel recommender system}. In \bibinfo{booktitle}{\emph{Proceedings of the sixth ACM conference on Recommender systems}}. \bibinfo{pages}{115--122}.
\newblock


\bibitem[Li et~al\mbox{.}(2019b)]%
        {li2019hierarchical}
\bibfield{author}{\bibinfo{person}{Chong Li}, \bibinfo{person}{Kunyang Jia}, \bibinfo{person}{Dan Shen}, \bibinfo{person}{C-J~Richard Shi}, {and} \bibinfo{person}{Hongxia Yang}.} \bibinfo{year}{2019}\natexlab{b}.
\newblock \showarticletitle{Hierarchical Representation Learning for Bipartite Graphs.}. In \bibinfo{booktitle}{\emph{IJCAI}}, Vol.~\bibinfo{volume}{19}. \bibinfo{pages}{2873--2879}.
\newblock


\bibitem[Li et~al\mbox{.}(2019c)]%
        {mind}
\bibfield{author}{\bibinfo{person}{Chao Li}, \bibinfo{person}{Zhiyuan Liu}, \bibinfo{person}{Mengmeng Wu}, \bibinfo{person}{Yuchi Xu}, \bibinfo{person}{Huan Zhao}, \bibinfo{person}{Pipei Huang}, \bibinfo{person}{Guoliang Kang}, \bibinfo{person}{Qiwei Chen}, \bibinfo{person}{Wei Li}, {and} \bibinfo{person}{Dik~Lun Lee}.} \bibinfo{year}{2019}\natexlab{c}.
\newblock \showarticletitle{Multi-interest network with dynamic routing for recommendation at Tmall}. In \bibinfo{booktitle}{\emph{Proceedings of the 28th ACM international conference on information and knowledge management}}. \bibinfo{pages}{2615--2623}.
\newblock


\bibitem[Li et~al\mbox{.}(2021)]%
        {li2021virt}
\bibfield{author}{\bibinfo{person}{Dan Li}, \bibinfo{person}{Yang Yang}, \bibinfo{person}{Hongyin Tang}, \bibinfo{person}{Jingang Wang}, \bibinfo{person}{Tong Xu}, \bibinfo{person}{Wei Wu}, {and} \bibinfo{person}{Enhong Chen}.} \bibinfo{year}{2021}\natexlab{}.
\newblock \showarticletitle{VIRT: improving representation-based models for text matching through virtual interaction}.
\newblock \bibinfo{journal}{\emph{arXiv preprint arXiv:2112.04195}} (\bibinfo{year}{2021}).
\newblock


\bibitem[Li et~al\mbox{.}(2023a)]%
        {li2023prompt}
\bibfield{author}{\bibinfo{person}{Lei Li}, \bibinfo{person}{Yongfeng Zhang}, {and} \bibinfo{person}{Li Chen}.} \bibinfo{year}{2023}\natexlab{a}.
\newblock \showarticletitle{Prompt distillation for efficient llm-based recommendation}. In \bibinfo{booktitle}{\emph{Proceedings of the 32nd ACM International Conference on Information and Knowledge Management}}. \bibinfo{pages}{1348--1357}.
\newblock


\bibitem[Li et~al\mbox{.}(2022)]%
        {li2022inttower}
\bibfield{author}{\bibinfo{person}{Xiangyang Li}, \bibinfo{person}{Bo Chen}, \bibinfo{person}{HuiFeng Guo}, \bibinfo{person}{Jingjie Li}, \bibinfo{person}{Chenxu Zhu}, \bibinfo{person}{Xiang Long}, \bibinfo{person}{Sujian Li}, \bibinfo{person}{Yichao Wang}, \bibinfo{person}{Wei Guo}, \bibinfo{person}{Longxia Mao}, {et~al\mbox{.}}} \bibinfo{year}{2022}\natexlab{}.
\newblock \showarticletitle{Inttower: the next generation of two-tower model for pre-ranking system}. In \bibinfo{booktitle}{\emph{Proceedings of the 31st ACM International Conference on Information \& Knowledge Management}}. \bibinfo{pages}{3292--3301}.
\newblock


\bibitem[Li and She(2017)]%
        {li2017collaborative}
\bibfield{author}{\bibinfo{person}{Xiaopeng Li} {and} \bibinfo{person}{James She}.} \bibinfo{year}{2017}\natexlab{}.
\newblock \showarticletitle{Collaborative variational autoencoder for recommender systems}. In \bibinfo{booktitle}{\emph{Proceedings of the 23rd ACM SIGKDD international conference on knowledge discovery and data mining}}. \bibinfo{pages}{305--314}.
\newblock


\bibitem[Li et~al\mbox{.}(2023b)]%
        {li2023pbnr}
\bibfield{author}{\bibinfo{person}{Xinyi Li}, \bibinfo{person}{Yongfeng Zhang}, {and} \bibinfo{person}{Edward~C Malthouse}.} \bibinfo{year}{2023}\natexlab{b}.
\newblock \showarticletitle{Pbnr: Prompt-based news recommender system}.
\newblock \bibinfo{journal}{\emph{arXiv preprint arXiv:2304.07862}} (\bibinfo{year}{2023}).
\newblock


\bibitem[Li et~al\mbox{.}(2023c)]%
        {li2023preliminary}
\bibfield{author}{\bibinfo{person}{Xinyi Li}, \bibinfo{person}{Yongfeng Zhang}, {and} \bibinfo{person}{Edward~C Malthouse}.} \bibinfo{year}{2023}\natexlab{c}.
\newblock \showarticletitle{A preliminary study of chatgpt on news recommendation: Personalization, provider fairness, fake news}.
\newblock \bibinfo{journal}{\emph{arXiv preprint arXiv:2306.10702}} (\bibinfo{year}{2023}).
\newblock


\bibitem[Li et~al\mbox{.}(2019a)]%
        {li2019fi}
\bibfield{author}{\bibinfo{person}{Zekun Li}, \bibinfo{person}{Zeyu Cui}, \bibinfo{person}{Shu Wu}, \bibinfo{person}{Xiaoyu Zhang}, {and} \bibinfo{person}{Liang Wang}.} \bibinfo{year}{2019}\natexlab{a}.
\newblock \showarticletitle{Fi-gnn: Modeling feature interactions via graph neural networks for ctr prediction}. In \bibinfo{booktitle}{\emph{Proceedings of the 28th ACM international conference on information and knowledge management}}. \bibinfo{pages}{539--548}.
\newblock


\bibitem[Li et~al\mbox{.}(2020)]%
        {hignn}
\bibfield{author}{\bibinfo{person}{Zhao Li}, \bibinfo{person}{Xin Shen}, \bibinfo{person}{Yuhang Jiao}, \bibinfo{person}{Xuming Pan}, \bibinfo{person}{Pengcheng Zou}, \bibinfo{person}{Xianling Meng}, \bibinfo{person}{Chengwei Yao}, {and} \bibinfo{person}{Jiajun Bu}.} \bibinfo{year}{2020}\natexlab{}.
\newblock \showarticletitle{Hierarchical bipartite graph neural networks: Towards large-scale e-commerce applications}. In \bibinfo{booktitle}{\emph{2020 IEEE 36th International Conference on Data Engineering (ICDE)}}. IEEE, \bibinfo{pages}{1677--1688}.
\newblock


\bibitem[Liang et~al\mbox{.}(2016)]%
        {liang2016modeling}
\bibfield{author}{\bibinfo{person}{Dawen Liang}, \bibinfo{person}{Laurent Charlin}, \bibinfo{person}{James McInerney}, {and} \bibinfo{person}{David~M Blei}.} \bibinfo{year}{2016}\natexlab{}.
\newblock \showarticletitle{Modeling user exposure in recommendation}. In \bibinfo{booktitle}{\emph{Proceedings of the 25th international conference on World Wide Web}}. \bibinfo{pages}{951--961}.
\newblock


\bibitem[Liang et~al\mbox{.}(2018)]%
        {liang2018variational}
\bibfield{author}{\bibinfo{person}{Dawen Liang}, \bibinfo{person}{Rahul~G Krishnan}, \bibinfo{person}{Matthew~D Hoffman}, {and} \bibinfo{person}{Tony Jebara}.} \bibinfo{year}{2018}\natexlab{}.
\newblock \showarticletitle{Variational autoencoders for collaborative filtering}. In \bibinfo{booktitle}{\emph{Proceedings of the 2018 world wide web conference}}. \bibinfo{pages}{689--698}.
\newblock


\bibitem[Liang et~al\mbox{.}(2010)]%
        {liang2010connecting}
\bibfield{author}{\bibinfo{person}{Huizhi Liang}, \bibinfo{person}{Yue Xu}, \bibinfo{person}{Yuefeng Li}, \bibinfo{person}{Richi Nayak}, {and} \bibinfo{person}{Xiaohui Tao}.} \bibinfo{year}{2010}\natexlab{}.
\newblock \showarticletitle{Connecting users and items with weighted tags for personalized item recommendations}. In \bibinfo{booktitle}{\emph{Proceedings of the 21st ACM conference on Hypertext and hypermedia}}. \bibinfo{pages}{51--60}.
\newblock


\bibitem[Lin et~al\mbox{.}(2023a)]%
        {lin2023can}
\bibfield{author}{\bibinfo{person}{Jianghao Lin}, \bibinfo{person}{Xinyi Dai}, \bibinfo{person}{Yunjia Xi}, \bibinfo{person}{Weiwen Liu}, \bibinfo{person}{Bo Chen}, \bibinfo{person}{Hao Zhang}, \bibinfo{person}{Yong Liu}, \bibinfo{person}{Chuhan Wu}, \bibinfo{person}{Xiangyang Li}, \bibinfo{person}{Chenxu Zhu}, {et~al\mbox{.}}} \bibinfo{year}{2023}\natexlab{a}.
\newblock \showarticletitle{How can recommender systems benefit from large language models: A survey}.
\newblock \bibinfo{journal}{\emph{arXiv preprint arXiv:2306.05817}} (\bibinfo{year}{2023}).
\newblock


\bibitem[Lin et~al\mbox{.}(2023b)]%
        {lin2023map}
\bibfield{author}{\bibinfo{person}{Jianghao Lin}, \bibinfo{person}{Yanru Qu}, \bibinfo{person}{Wei Guo}, \bibinfo{person}{Xinyi Dai}, \bibinfo{person}{Ruiming Tang}, \bibinfo{person}{Yong Yu}, {and} \bibinfo{person}{Weinan Zhang}.} \bibinfo{year}{2023}\natexlab{b}.
\newblock \showarticletitle{Map: A model-agnostic pretraining framework for click-through rate prediction}. In \bibinfo{booktitle}{\emph{Proceedings of the 29th ACM SIGKDD Conference on Knowledge Discovery and Data Mining}}. \bibinfo{pages}{1384--1395}.
\newblock


\bibitem[Lin et~al\mbox{.}(2024)]%
        {lin2024rella}
\bibfield{author}{\bibinfo{person}{Jianghao Lin}, \bibinfo{person}{Rong Shan}, \bibinfo{person}{Chenxu Zhu}, \bibinfo{person}{Kounianhua Du}, \bibinfo{person}{Bo Chen}, \bibinfo{person}{Shigang Quan}, \bibinfo{person}{Ruiming Tang}, \bibinfo{person}{Yong Yu}, {and} \bibinfo{person}{Weinan Zhang}.} \bibinfo{year}{2024}\natexlab{}.
\newblock \showarticletitle{Rella: Retrieval-enhanced large language models for lifelong sequential behavior comprehension in recommendation}. In \bibinfo{booktitle}{\emph{Proceedings of the ACM on Web Conference 2024}}. \bibinfo{pages}{3497--3508}.
\newblock


\bibitem[Linden et~al\mbox{.}(2003)]%
        {linden2003amazon}
\bibfield{author}{\bibinfo{person}{Greg Linden}, \bibinfo{person}{Brent Smith}, {and} \bibinfo{person}{Jeremy York}.} \bibinfo{year}{2003}\natexlab{}.
\newblock \showarticletitle{Amazon. com recommendations: Item-to-item collaborative filtering}.
\newblock \bibinfo{journal}{\emph{IEEE Internet computing}} \bibinfo{volume}{7}, \bibinfo{number}{1} (\bibinfo{year}{2003}), \bibinfo{pages}{76--80}.
\newblock


\bibitem[Liu et~al\mbox{.}(2023)]%
        {liu2023chatgpt}
\bibfield{author}{\bibinfo{person}{Junling Liu}, \bibinfo{person}{Chao Liu}, \bibinfo{person}{Peilin Zhou}, \bibinfo{person}{Renjie Lv}, \bibinfo{person}{Kang Zhou}, {and} \bibinfo{person}{Yan Zhang}.} \bibinfo{year}{2023}\natexlab{}.
\newblock \showarticletitle{Is chatgpt a good recommender? a preliminary study}.
\newblock \bibinfo{journal}{\emph{arXiv preprint arXiv:2304.10149}} (\bibinfo{year}{2023}).
\newblock


\bibitem[Liu et~al\mbox{.}(2017)]%
        {liu2017phd}
\bibfield{author}{\bibinfo{person}{Jie Liu}, \bibinfo{person}{Dong Wang}, {and} \bibinfo{person}{Yue Ding}.} \bibinfo{year}{2017}\natexlab{}.
\newblock \showarticletitle{PHD: A probabilistic model of hybrid deep collaborative filtering for recommender systems}. In \bibinfo{booktitle}{\emph{Asian Conference on machine learning}}. PMLR, \bibinfo{pages}{224--239}.
\newblock


\bibitem[Liu et~al\mbox{.}(2019)]%
        {liu2019user}
\bibfield{author}{\bibinfo{person}{Shang Liu}, \bibinfo{person}{Zhenzhong Chen}, \bibinfo{person}{Hongyi Liu}, {and} \bibinfo{person}{Xinghai Hu}.} \bibinfo{year}{2019}\natexlab{}.
\newblock \showarticletitle{User-video co-attention network for personalized micro-video recommendation}. In \bibinfo{booktitle}{\emph{The World Wide Web Conference}}. \bibinfo{pages}{3020--3026}.
\newblock


\bibitem[Liu et~al\mbox{.}(2020a)]%
        {octopus}
\bibfield{author}{\bibinfo{person}{Zheng Liu}, \bibinfo{person}{Jianxun Lian}, \bibinfo{person}{Junhan Yang}, \bibinfo{person}{Defu Lian}, {and} \bibinfo{person}{Xing Xie}.} \bibinfo{year}{2020}\natexlab{a}.
\newblock \showarticletitle{Octopus: Comprehensive and elastic user representation for the generation of recommendation candidates}. In \bibinfo{booktitle}{\emph{Proceedings of the 43rd International ACM SIGIR Conference on Research and Development in Information Retrieval}}. \bibinfo{pages}{289--298}.
\newblock


\bibitem[Liu et~al\mbox{.}(2020b)]%
        {liu2020deoscillated}
\bibfield{author}{\bibinfo{person}{Zhiwei Liu}, \bibinfo{person}{Lin Meng}, \bibinfo{person}{Fei Jiang}, \bibinfo{person}{Jiawei Zhang}, {and} \bibinfo{person}{Philip~S Yu}.} \bibinfo{year}{2020}\natexlab{b}.
\newblock \showarticletitle{Deoscillated graph collaborative filtering}.
\newblock \bibinfo{journal}{\emph{arXiv preprint arXiv:2011.02100}} (\bibinfo{year}{2020}).
\newblock


\bibitem[Lobel et~al\mbox{.}(2019)]%
        {lobel2019towards}
\bibfield{author}{\bibinfo{person}{Sam Lobel}, \bibinfo{person}{Chunyuan Li}, \bibinfo{person}{Jianfeng Gao}, {and} \bibinfo{person}{Lawrence Carin}.} \bibinfo{year}{2019}\natexlab{}.
\newblock \showarticletitle{Towards amortized ranking-critical training for collaborative filtering}.
\newblock \bibinfo{journal}{\emph{arXiv preprint arXiv:1906.04281}} (\bibinfo{year}{2019}).
\newblock


\bibitem[Lu et~al\mbox{.}(2022)]%
        {lu2022ernie}
\bibfield{author}{\bibinfo{person}{Yuxiang Lu}, \bibinfo{person}{Yiding Liu}, \bibinfo{person}{Jiaxiang Liu}, \bibinfo{person}{Yunsheng Shi}, \bibinfo{person}{Zhengjie Huang}, \bibinfo{person}{Shikun Feng~Yu Sun}, \bibinfo{person}{Hao Tian}, \bibinfo{person}{Hua Wu}, \bibinfo{person}{Shuaiqiang Wang}, \bibinfo{person}{Dawei Yin}, {et~al\mbox{.}}} \bibinfo{year}{2022}\natexlab{}.
\newblock \showarticletitle{Ernie-search: Bridging cross-encoder with dual-encoder via self on-the-fly distillation for dense passage retrieval}.
\newblock \bibinfo{journal}{\emph{arXiv preprint arXiv:2205.09153}} (\bibinfo{year}{2022}).
\newblock


\bibitem[Ma et~al\mbox{.}(2019)]%
        {ma2019learning}
\bibfield{author}{\bibinfo{person}{Jianxin Ma}, \bibinfo{person}{Chang Zhou}, \bibinfo{person}{Peng Cui}, \bibinfo{person}{Hongxia Yang}, {and} \bibinfo{person}{Wenwu Zhu}.} \bibinfo{year}{2019}\natexlab{}.
\newblock \showarticletitle{Learning disentangled representations for recommendation}.
\newblock \bibinfo{journal}{\emph{Advances in neural information processing systems}}  \bibinfo{volume}{32} (\bibinfo{year}{2019}).
\newblock


\bibitem[Ma et~al\mbox{.}(2020)]%
        {disenhan}
\bibfield{author}{\bibinfo{person}{Jianxin Ma}, \bibinfo{person}{Chang Zhou}, \bibinfo{person}{Hongxia Yang}, \bibinfo{person}{Peng Cui}, \bibinfo{person}{Xin Wang}, {and} \bibinfo{person}{Wenwu Zhu}.} \bibinfo{year}{2020}\natexlab{}.
\newblock \showarticletitle{Disentangled self-supervision in sequential recommenders}. In \bibinfo{booktitle}{\emph{Proceedings of the 26th ACM SIGKDD International Conference on Knowledge Discovery \& Data Mining}}. \bibinfo{pages}{483--491}.
\newblock


\bibitem[Malkov and Yashunin(2018)]%
        {malkov2018efficient}
\bibfield{author}{\bibinfo{person}{Yu~A Malkov} {and} \bibinfo{person}{Dmitry~A Yashunin}.} \bibinfo{year}{2018}\natexlab{}.
\newblock \showarticletitle{Efficient and robust approximate nearest neighbor search using hierarchical navigable small world graphs}.
\newblock \bibinfo{journal}{\emph{IEEE transactions on pattern analysis and machine intelligence}} \bibinfo{volume}{42}, \bibinfo{number}{4} (\bibinfo{year}{2018}), \bibinfo{pages}{824--836}.
\newblock


\bibitem[Manotumruksa et~al\mbox{.}(2017)]%
        {manotumruksa2017deep}
\bibfield{author}{\bibinfo{person}{Jarana Manotumruksa}, \bibinfo{person}{Craig Macdonald}, {and} \bibinfo{person}{Iadh Ounis}.} \bibinfo{year}{2017}\natexlab{}.
\newblock \showarticletitle{A deep recurrent collaborative filtering framework for venue recommendation}. In \bibinfo{booktitle}{\emph{Proceedings of the 2017 ACM on Conference on Information and Knowledge Management}}. \bibinfo{pages}{1429--1438}.
\newblock


\bibitem[Mao et~al\mbox{.}(2021a)]%
        {mao2021simplex}
\bibfield{author}{\bibinfo{person}{Kelong Mao}, \bibinfo{person}{Jieming Zhu}, \bibinfo{person}{Jinpeng Wang}, \bibinfo{person}{Quanyu Dai}, \bibinfo{person}{Zhenhua Dong}, \bibinfo{person}{Xi Xiao}, {and} \bibinfo{person}{Xiuqiang He}.} \bibinfo{year}{2021}\natexlab{a}.
\newblock \showarticletitle{SimpleX: A simple and strong baseline for collaborative filtering}. In \bibinfo{booktitle}{\emph{Proceedings of the 30th ACM International Conference on Information \& Knowledge Management}}. \bibinfo{pages}{1243--1252}.
\newblock


\bibitem[Mao et~al\mbox{.}(2021b)]%
        {mao2021ultragcn}
\bibfield{author}{\bibinfo{person}{Kelong Mao}, \bibinfo{person}{Jieming Zhu}, \bibinfo{person}{Xi Xiao}, \bibinfo{person}{Biao Lu}, \bibinfo{person}{Zhaowei Wang}, {and} \bibinfo{person}{Xiuqiang He}.} \bibinfo{year}{2021}\natexlab{b}.
\newblock \showarticletitle{UltraGCN: ultra simplification of graph convolutional networks for recommendation}. In \bibinfo{booktitle}{\emph{Proceedings of the 30th ACM international conference on information \& knowledge management}}. \bibinfo{pages}{1253--1262}.
\newblock


\bibitem[Melville et~al\mbox{.}(2002)]%
        {melville2002content}
\bibfield{author}{\bibinfo{person}{Prem Melville}, \bibinfo{person}{Raymond~J Mooney}, \bibinfo{person}{Ramadass Nagarajan}, {et~al\mbox{.}}} \bibinfo{year}{2002}\natexlab{}.
\newblock \showarticletitle{Content-boosted collaborative filtering for improved recommendations}.
\newblock \bibinfo{journal}{\emph{Aaai/iaai}}  \bibinfo{volume}{23} (\bibinfo{year}{2002}), \bibinfo{pages}{187--192}.
\newblock


\bibitem[Mikolov et~al\mbox{.}(2013)]%
        {mikolov2013distributed}
\bibfield{author}{\bibinfo{person}{Tomas Mikolov}, \bibinfo{person}{Ilya Sutskever}, \bibinfo{person}{Kai Chen}, \bibinfo{person}{Greg~S Corrado}, {and} \bibinfo{person}{Jeff Dean}.} \bibinfo{year}{2013}\natexlab{}.
\newblock \showarticletitle{Distributed representations of words and phrases and their compositionality}.
\newblock \bibinfo{journal}{\emph{Advances in neural information processing systems}}  \bibinfo{volume}{26} (\bibinfo{year}{2013}).
\newblock


\bibitem[Ning and Karypis(2011)]%
        {ning2011slim}
\bibfield{author}{\bibinfo{person}{Xia Ning} {and} \bibinfo{person}{George Karypis}.} \bibinfo{year}{2011}\natexlab{}.
\newblock \showarticletitle{Slim: Sparse linear methods for top-n recommender systems}. In \bibinfo{booktitle}{\emph{2011 IEEE 11th international conference on data mining}}. IEEE, \bibinfo{pages}{497--506}.
\newblock


\bibitem[Ono et~al\mbox{.}(2009)]%
        {ono2009context}
\bibfield{author}{\bibinfo{person}{Chihiro Ono}, \bibinfo{person}{Yasuhiro Takishima}, \bibinfo{person}{Yoichi Motomura}, {and} \bibinfo{person}{Hideki Asoh}.} \bibinfo{year}{2009}\natexlab{}.
\newblock \showarticletitle{Context-aware preference model based on a study of difference between real and supposed situation data}. In \bibinfo{booktitle}{\emph{User Modeling, Adaptation, and Personalization: 17th International Conference, UMAP 2009, formerly UM and AH, Trento, Italy, June 22-26, 2009. Proceedings 17}}. Springer, \bibinfo{pages}{102--113}.
\newblock


\bibitem[Otunba et~al\mbox{.}(2019)]%
        {otunba2019deep}
\bibfield{author}{\bibinfo{person}{Rasaq Otunba}, \bibinfo{person}{Raimi~A Rufai}, {and} \bibinfo{person}{Jessica Lin}.} \bibinfo{year}{2019}\natexlab{}.
\newblock \showarticletitle{Deep stacked ensemble recommender}. In \bibinfo{booktitle}{\emph{Proceedings of the 31st International Conference on Scientific and Statistical Database Management}}. \bibinfo{pages}{197--201}.
\newblock


\bibitem[Pal et~al\mbox{.}(2020)]%
        {pinnersage}
\bibfield{author}{\bibinfo{person}{Aditya Pal}, \bibinfo{person}{Chantat Eksombatchai}, \bibinfo{person}{Yitong Zhou}, \bibinfo{person}{Bo Zhao}, \bibinfo{person}{Charles Rosenberg}, {and} \bibinfo{person}{Jure Leskovec}.} \bibinfo{year}{2020}\natexlab{}.
\newblock \showarticletitle{Pinnersage: Multi-modal user embedding framework for recommendations at pinterest}. In \bibinfo{booktitle}{\emph{Proceedings of the 26th ACM SIGKDD International Conference on Knowledge Discovery \& Data Mining}}. \bibinfo{pages}{2311--2320}.
\newblock


\bibitem[Pei et~al\mbox{.}(2024)]%
        {himirec}
\bibfield{author}{\bibinfo{person}{Haolei Pei}, \bibinfo{person}{Yuanyuan Xu}, \bibinfo{person}{Yangping Zhu}, {and} \bibinfo{person}{Yuan Nie}.} \bibinfo{year}{2024}\natexlab{}.
\newblock \showarticletitle{HimiRec: Modeling Hierarchical Multi-interest for Recommendation}.
\newblock \bibinfo{journal}{\emph{arXiv preprint arXiv:2402.01253}} (\bibinfo{year}{2024}).
\newblock


\bibitem[Perozzi et~al\mbox{.}(2014)]%
        {perozzi2014deepwalk}
\bibfield{author}{\bibinfo{person}{Bryan Perozzi}, \bibinfo{person}{Rami Al-Rfou}, {and} \bibinfo{person}{Steven Skiena}.} \bibinfo{year}{2014}\natexlab{}.
\newblock \showarticletitle{Deepwalk: Online learning of social representations}. In \bibinfo{booktitle}{\emph{Proceedings of the 20th ACM SIGKDD international conference on Knowledge discovery and data mining}}. \bibinfo{pages}{701--710}.
\newblock


\bibitem[Pi et~al\mbox{.}(2019)]%
        {pi2019practice}
\bibfield{author}{\bibinfo{person}{Qi Pi}, \bibinfo{person}{Weijie Bian}, \bibinfo{person}{Guorui Zhou}, \bibinfo{person}{Xiaoqiang Zhu}, {and} \bibinfo{person}{Kun Gai}.} \bibinfo{year}{2019}\natexlab{}.
\newblock \showarticletitle{Practice on long sequential user behavior modeling for click-through rate prediction}. In \bibinfo{booktitle}{\emph{Proceedings of the 25th ACM SIGKDD International Conference on Knowledge Discovery \& Data Mining}}. \bibinfo{pages}{2671--2679}.
\newblock


\bibitem[Pi et~al\mbox{.}(2020)]%
        {pi2020search}
\bibfield{author}{\bibinfo{person}{Qi Pi}, \bibinfo{person}{Guorui Zhou}, \bibinfo{person}{Yujing Zhang}, \bibinfo{person}{Zhe Wang}, \bibinfo{person}{Lejian Ren}, \bibinfo{person}{Ying Fan}, \bibinfo{person}{Xiaoqiang Zhu}, {and} \bibinfo{person}{Kun Gai}.} \bibinfo{year}{2020}\natexlab{}.
\newblock \showarticletitle{Search-based user interest modeling with lifelong sequential behavior data for click-through rate prediction}. In \bibinfo{booktitle}{\emph{Proceedings of the 29th ACM International Conference on Information \& Knowledge Management}}. \bibinfo{pages}{2685--2692}.
\newblock


\bibitem[Qin et~al\mbox{.}(2021)]%
        {qin2021retrieval}
\bibfield{author}{\bibinfo{person}{Jiarui Qin}, \bibinfo{person}{Weinan Zhang}, \bibinfo{person}{Rong Su}, \bibinfo{person}{Zhirong Liu}, \bibinfo{person}{Weiwen Liu}, \bibinfo{person}{Ruiming Tang}, \bibinfo{person}{Xiuqiang He}, {and} \bibinfo{person}{Yong Yu}.} \bibinfo{year}{2021}\natexlab{}.
\newblock \showarticletitle{Retrieval \& interaction machine for tabular data prediction}. In \bibinfo{booktitle}{\emph{Proceedings of the 27th ACM SIGKDD Conference on Knowledge Discovery \& Data Mining}}. \bibinfo{pages}{1379--1389}.
\newblock


\bibitem[Qin et~al\mbox{.}(2022)]%
        {qin2022rankflow}
\bibfield{author}{\bibinfo{person}{Jiarui Qin}, \bibinfo{person}{Jiachen Zhu}, \bibinfo{person}{Bo Chen}, \bibinfo{person}{Zhirong Liu}, \bibinfo{person}{Weiwen Liu}, \bibinfo{person}{Ruiming Tang}, \bibinfo{person}{Rui Zhang}, \bibinfo{person}{Yong Yu}, {and} \bibinfo{person}{Weinan Zhang}.} \bibinfo{year}{2022}\natexlab{}.
\newblock \showarticletitle{RankFlow: Joint Optimization of Multi-Stage Cascade Ranking Systems as Flows}. In \bibinfo{booktitle}{\emph{Proceedings of the 45th International ACM SIGIR Conference on Research and Development in Information Retrieval}}. \bibinfo{pages}{814--824}.
\newblock


\bibitem[Qu et~al\mbox{.}(2016)]%
        {qu2016product}
\bibfield{author}{\bibinfo{person}{Yanru Qu}, \bibinfo{person}{Han Cai}, \bibinfo{person}{Kan Ren}, \bibinfo{person}{Weinan Zhang}, \bibinfo{person}{Yong Yu}, \bibinfo{person}{Ying Wen}, {and} \bibinfo{person}{Jun Wang}.} \bibinfo{year}{2016}\natexlab{}.
\newblock \showarticletitle{Product-based neural networks for user response prediction}. In \bibinfo{booktitle}{\emph{2016 IEEE 16th international conference on data mining (ICDM)}}. IEEE, \bibinfo{pages}{1149--1154}.
\newblock


\bibitem[Rendle(2010)]%
        {rendle2010factorization}
\bibfield{author}{\bibinfo{person}{Steffen Rendle}.} \bibinfo{year}{2010}\natexlab{}.
\newblock \showarticletitle{Factorization machines}. In \bibinfo{booktitle}{\emph{2010 IEEE International conference on data mining}}. IEEE, \bibinfo{pages}{995--1000}.
\newblock


\bibitem[Rendle and Freudenthaler(2014)]%
        {rendle2014improving}
\bibfield{author}{\bibinfo{person}{Steffen Rendle} {and} \bibinfo{person}{Christoph Freudenthaler}.} \bibinfo{year}{2014}\natexlab{}.
\newblock \showarticletitle{Improving pairwise learning for item recommendation from implicit feedback}. In \bibinfo{booktitle}{\emph{Proceedings of the 7th ACM international conference on Web search and data mining}}. \bibinfo{pages}{273--282}.
\newblock


\bibitem[Rendle et~al\mbox{.}(2012)]%
        {rendle2012bpr}
\bibfield{author}{\bibinfo{person}{Steffen Rendle}, \bibinfo{person}{Christoph Freudenthaler}, \bibinfo{person}{Zeno Gantner}, {and} \bibinfo{person}{Lars Schmidt-Thieme}.} \bibinfo{year}{2012}\natexlab{}.
\newblock \showarticletitle{BPR: Bayesian personalized ranking from implicit feedback}.
\newblock \bibinfo{journal}{\emph{arXiv preprint arXiv:1205.2618}} (\bibinfo{year}{2012}).
\newblock


\bibitem[Robertson et~al\mbox{.}(1995)]%
        {robertson1995okapi}
\bibfield{author}{\bibinfo{person}{Stephen~E Robertson}, \bibinfo{person}{Steve Walker}, \bibinfo{person}{Susan Jones}, \bibinfo{person}{Micheline~M Hancock-Beaulieu}, \bibinfo{person}{Mike Gatford}, {et~al\mbox{.}}} \bibinfo{year}{1995}\natexlab{}.
\newblock \showarticletitle{Okapi at TREC-3}.
\newblock \bibinfo{journal}{\emph{Nist Special Publication Sp}}  \bibinfo{volume}{109} (\bibinfo{year}{1995}), \bibinfo{pages}{109}.
\newblock


\bibitem[Robu et~al\mbox{.}(2009)]%
        {robu2009emergence}
\bibfield{author}{\bibinfo{person}{Valentin Robu}, \bibinfo{person}{Harry Halpin}, {and} \bibinfo{person}{Hana Shepherd}.} \bibinfo{year}{2009}\natexlab{}.
\newblock \showarticletitle{Emergence of consensus and shared vocabularies in collaborative tagging systems}.
\newblock \bibinfo{journal}{\emph{ACM Transactions on the Web (TWEB)}} \bibinfo{volume}{3}, \bibinfo{number}{4} (\bibinfo{year}{2009}), \bibinfo{pages}{1--34}.
\newblock


\bibitem[Sarwar et~al\mbox{.}(2000)]%
        {sarwar2000analysis}
\bibfield{author}{\bibinfo{person}{Badrul Sarwar}, \bibinfo{person}{George Karypis}, \bibinfo{person}{Joseph Konstan}, {and} \bibinfo{person}{John Riedl}.} \bibinfo{year}{2000}\natexlab{}.
\newblock \showarticletitle{Analysis of recommendation algorithms for e-commerce}. In \bibinfo{booktitle}{\emph{Proceedings of the 2nd ACM Conference on Electronic Commerce}}. \bibinfo{pages}{158--167}.
\newblock


\bibitem[Sarwar et~al\mbox{.}(2001)]%
        {sarwar2001item}
\bibfield{author}{\bibinfo{person}{Badrul Sarwar}, \bibinfo{person}{George Karypis}, \bibinfo{person}{Joseph Konstan}, {and} \bibinfo{person}{John Riedl}.} \bibinfo{year}{2001}\natexlab{}.
\newblock \showarticletitle{Item-based collaborative filtering recommendation algorithms}. In \bibinfo{booktitle}{\emph{Proceedings of the 10th international conference on World Wide Web}}. \bibinfo{pages}{285--295}.
\newblock


\bibitem[Shenbin et~al\mbox{.}(2020)]%
        {shenbin2020recvae}
\bibfield{author}{\bibinfo{person}{Ilya Shenbin}, \bibinfo{person}{Anton Alekseev}, \bibinfo{person}{Elena Tutubalina}, \bibinfo{person}{Valentin Malykh}, {and} \bibinfo{person}{Sergey~I Nikolenko}.} \bibinfo{year}{2020}\natexlab{}.
\newblock \showarticletitle{Recvae: A new variational autoencoder for top-n recommendations with implicit feedback}. In \bibinfo{booktitle}{\emph{Proceedings of the 13th international conference on web search and data mining}}. \bibinfo{pages}{528--536}.
\newblock


\bibitem[Shepitsen et~al\mbox{.}(2008)]%
        {shepitsen2008personalized}
\bibfield{author}{\bibinfo{person}{Andriy Shepitsen}, \bibinfo{person}{Jonathan Gemmell}, \bibinfo{person}{Bamshad Mobasher}, {and} \bibinfo{person}{Robin Burke}.} \bibinfo{year}{2008}\natexlab{}.
\newblock \showarticletitle{Personalized recommendation in social tagging systems using hierarchical clustering}. In \bibinfo{booktitle}{\emph{Proceedings of the 2008 ACM conference on Recommender systems}}. \bibinfo{pages}{259--266}.
\newblock


\bibitem[Shi et~al\mbox{.}(2023)]%
        {mip}
\bibfield{author}{\bibinfo{person}{Hui Shi}, \bibinfo{person}{Yupeng Gu}, \bibinfo{person}{Yitong Zhou}, \bibinfo{person}{Bo Zhao}, \bibinfo{person}{Sicun Gao}, {and} \bibinfo{person}{Jishen Zhao}.} \bibinfo{year}{2023}\natexlab{}.
\newblock \showarticletitle{Everyone’s preference changes differently: A weighted multi-interest model for retrieval}. In \bibinfo{booktitle}{\emph{International Conference on Machine Learning}}. PMLR, \bibinfo{pages}{31228--31242}.
\newblock


\bibitem[Shi et~al\mbox{.}(2014)]%
        {shi2014collaborative}
\bibfield{author}{\bibinfo{person}{Yue Shi}, \bibinfo{person}{Martha Larson}, {and} \bibinfo{person}{Alan Hanjalic}.} \bibinfo{year}{2014}\natexlab{}.
\newblock \showarticletitle{Collaborative filtering beyond the user-item matrix: A survey of the state of the art and future challenges}.
\newblock \bibinfo{journal}{\emph{ACM Computing Surveys (CSUR)}} \bibinfo{volume}{47}, \bibinfo{number}{1} (\bibinfo{year}{2014}), \bibinfo{pages}{1--45}.
\newblock


\bibitem[Su and Khoshgoftaar(2009)]%
        {su2009survey}
\bibfield{author}{\bibinfo{person}{Xiaoyuan Su} {and} \bibinfo{person}{Taghi~M Khoshgoftaar}.} \bibinfo{year}{2009}\natexlab{}.
\newblock \showarticletitle{A survey of collaborative filtering techniques}.
\newblock \bibinfo{journal}{\emph{Advances in artificial intelligence}}  \bibinfo{volume}{2009} (\bibinfo{year}{2009}).
\newblock


\bibitem[Sun et~al\mbox{.}(2020)]%
        {sun2020neighbor}
\bibfield{author}{\bibinfo{person}{Jianing Sun}, \bibinfo{person}{Yingxue Zhang}, \bibinfo{person}{Wei Guo}, \bibinfo{person}{Huifeng Guo}, \bibinfo{person}{Ruiming Tang}, \bibinfo{person}{Xiuqiang He}, \bibinfo{person}{Chen Ma}, {and} \bibinfo{person}{Mark Coates}.} \bibinfo{year}{2020}\natexlab{}.
\newblock \showarticletitle{Neighbor interaction aware graph convolution networks for recommendation}. In \bibinfo{booktitle}{\emph{Proceedings of the 43rd international ACM SIGIR conference on research and development in information retrieval}}. \bibinfo{pages}{1289--1298}.
\newblock


\bibitem[Sun et~al\mbox{.}(2019)]%
        {sun2019multi}
\bibfield{author}{\bibinfo{person}{Jianing Sun}, \bibinfo{person}{Yingxue Zhang}, \bibinfo{person}{Chen Ma}, \bibinfo{person}{Mark Coates}, \bibinfo{person}{Huifeng Guo}, \bibinfo{person}{Ruiming Tang}, {and} \bibinfo{person}{Xiuqiang He}.} \bibinfo{year}{2019}\natexlab{}.
\newblock \showarticletitle{Multi-graph convolution collaborative filtering}. In \bibinfo{booktitle}{\emph{2019 IEEE international conference on data mining (ICDM)}}. IEEE, \bibinfo{pages}{1306--1311}.
\newblock


\bibitem[Surowiecki(2005)]%
        {surowiecki2005wisdom}
\bibfield{author}{\bibinfo{person}{James Surowiecki}.} \bibinfo{year}{2005}\natexlab{}.
\newblock \bibinfo{booktitle}{\emph{The wisdom of crowds}}.
\newblock \bibinfo{publisher}{Anchor}.
\newblock


\bibitem[Tak{\'a}cs et~al\mbox{.}(2007)]%
        {takacs2007major}
\bibfield{author}{\bibinfo{person}{G{\'a}bor Tak{\'a}cs}, \bibinfo{person}{Istv{\'a}n Pil{\'a}szy}, \bibinfo{person}{Botty{\'a}n N{\'e}meth}, {and} \bibinfo{person}{Domonkos Tikk}.} \bibinfo{year}{2007}\natexlab{}.
\newblock \showarticletitle{Major components of the gravity recommendation system}.
\newblock \bibinfo{journal}{\emph{Acm Sigkdd Explorations Newsletter}} \bibinfo{volume}{9}, \bibinfo{number}{2} (\bibinfo{year}{2007}), \bibinfo{pages}{80--83}.
\newblock


\bibitem[Tak{\'a}cs et~al\mbox{.}(2008)]%
        {takacs2008matrix}
\bibfield{author}{\bibinfo{person}{G{\'a}bor Tak{\'a}cs}, \bibinfo{person}{Istv{\'a}n Pil{\'a}szy}, \bibinfo{person}{Botty{\'a}n N{\'e}meth}, {and} \bibinfo{person}{Domonkos Tikk}.} \bibinfo{year}{2008}\natexlab{}.
\newblock \showarticletitle{Matrix factorization and neighbor based algorithms for the netflix prize problem}. In \bibinfo{booktitle}{\emph{Proceedings of the 2008 ACM conference on Recommender systems}}. \bibinfo{pages}{267--274}.
\newblock


\bibitem[Tan et~al\mbox{.}(2020)]%
        {tan2020learning}
\bibfield{author}{\bibinfo{person}{Qiaoyu Tan}, \bibinfo{person}{Ninghao Liu}, \bibinfo{person}{Xing Zhao}, \bibinfo{person}{Hongxia Yang}, \bibinfo{person}{Jingren Zhou}, {and} \bibinfo{person}{Xia Hu}.} \bibinfo{year}{2020}\natexlab{}.
\newblock \showarticletitle{Learning to hash with graph neural networks for recommender systems}. In \bibinfo{booktitle}{\emph{Proceedings of The Web Conference 2020}}. \bibinfo{pages}{1988--1998}.
\newblock


\bibitem[Tan et~al\mbox{.}(2021)]%
        {sine}
\bibfield{author}{\bibinfo{person}{Qiaoyu Tan}, \bibinfo{person}{Jianwei Zhang}, \bibinfo{person}{Jiangchao Yao}, \bibinfo{person}{Ninghao Liu}, \bibinfo{person}{Jingren Zhou}, \bibinfo{person}{Hongxia Yang}, {and} \bibinfo{person}{Xia Hu}.} \bibinfo{year}{2021}\natexlab{}.
\newblock \showarticletitle{Sparse-interest network for sequential recommendation}. In \bibinfo{booktitle}{\emph{Proceedings of the 14th ACM international conference on web search and data mining}}. \bibinfo{pages}{598--606}.
\newblock


\bibitem[Tang et~al\mbox{.}(2024)]%
        {tang2024towards}
\bibfield{author}{\bibinfo{person}{Jiakai Tang}, \bibinfo{person}{Sunhao Dai}, \bibinfo{person}{Zexu Sun}, \bibinfo{person}{Xu Chen}, \bibinfo{person}{Jun Xu}, \bibinfo{person}{Wenhui Yu}, \bibinfo{person}{Lantao Hu}, \bibinfo{person}{Peng Jiang}, {and} \bibinfo{person}{Han Li}.} \bibinfo{year}{2024}\natexlab{}.
\newblock \showarticletitle{Towards Robust Recommendation via Decision Boundary-aware Graph Contrastive Learning}. In \bibinfo{booktitle}{\emph{Proceedings of the 30th ACM SIGKDD Conference on Knowledge Discovery and Data Mining}}. \bibinfo{pages}{2854--2865}.
\newblock


\bibitem[Tang et~al\mbox{.}(2015)]%
        {tang2015line}
\bibfield{author}{\bibinfo{person}{Jian Tang}, \bibinfo{person}{Meng Qu}, \bibinfo{person}{Mingzhe Wang}, \bibinfo{person}{Ming Zhang}, \bibinfo{person}{Jun Yan}, {and} \bibinfo{person}{Qiaozhu Mei}.} \bibinfo{year}{2015}\natexlab{}.
\newblock \showarticletitle{Line: Large-scale information network embedding}. In \bibinfo{booktitle}{\emph{Proceedings of the 24th international conference on world wide web}}. \bibinfo{pages}{1067--1077}.
\newblock


\bibitem[Tomczak and Welling(2018)]%
        {tomczak2018vae}
\bibfield{author}{\bibinfo{person}{Jakub Tomczak} {and} \bibinfo{person}{Max Welling}.} \bibinfo{year}{2018}\natexlab{}.
\newblock \showarticletitle{VAE with a VampPrior}. In \bibinfo{booktitle}{\emph{International conference on artificial intelligence and statistics}}. PMLR, \bibinfo{pages}{1214--1223}.
\newblock


\bibitem[Truong et~al\mbox{.}(2021)]%
        {truong2021bilateral}
\bibfield{author}{\bibinfo{person}{Quoc-Tuan Truong}, \bibinfo{person}{Aghiles Salah}, {and} \bibinfo{person}{Hady~W Lauw}.} \bibinfo{year}{2021}\natexlab{}.
\newblock \showarticletitle{Bilateral variational autoencoder for collaborative filtering}. In \bibinfo{booktitle}{\emph{Proceedings of the 14th ACM International Conference on Web Search and Data Mining}}. \bibinfo{pages}{292--300}.
\newblock


\bibitem[Tso-Sutter et~al\mbox{.}(2008)]%
        {tso2008tag}
\bibfield{author}{\bibinfo{person}{Karen~HL Tso-Sutter}, \bibinfo{person}{Leandro~Balby Marinho}, {and} \bibinfo{person}{Lars Schmidt-Thieme}.} \bibinfo{year}{2008}\natexlab{}.
\newblock \showarticletitle{Tag-aware recommender systems by fusion of collaborative filtering algorithms}. In \bibinfo{booktitle}{\emph{Proceedings of the 2008 ACM symposium on Applied computing}}. \bibinfo{pages}{1995--1999}.
\newblock


\bibitem[Wang et~al\mbox{.}(2016)]%
        {wang2016structural}
\bibfield{author}{\bibinfo{person}{Daixin Wang}, \bibinfo{person}{Peng Cui}, {and} \bibinfo{person}{Wenwu Zhu}.} \bibinfo{year}{2016}\natexlab{}.
\newblock \showarticletitle{Structural deep network embedding}. In \bibinfo{booktitle}{\emph{Proceedings of the 22nd ACM SIGKDD international conference on Knowledge discovery and data mining}}. \bibinfo{pages}{1225--1234}.
\newblock


\bibitem[Wang et~al\mbox{.}(2023c)]%
        {wang2023flip}
\bibfield{author}{\bibinfo{person}{Hangyu Wang}, \bibinfo{person}{Jianghao Lin}, \bibinfo{person}{Xiangyang Li}, \bibinfo{person}{Bo Chen}, \bibinfo{person}{Chenxu Zhu}, \bibinfo{person}{Ruiming Tang}, \bibinfo{person}{Weinan Zhang}, {and} \bibinfo{person}{Yong Yu}.} \bibinfo{year}{2023}\natexlab{c}.
\newblock \showarticletitle{FLIP: Towards Fine-grained Alignment between ID-based Models and Pretrained Language Models for CTR Prediction}.
\newblock \bibinfo{journal}{\emph{arXiv e-prints}} (\bibinfo{year}{2023}), \bibinfo{pages}{arXiv--2310}.
\newblock


\bibitem[Wang et~al\mbox{.}(2019b)]%
        {wang2019adversarial}
\bibfield{author}{\bibinfo{person}{Haoyu Wang}, \bibinfo{person}{Nan Shao}, {and} \bibinfo{person}{Defu Lian}.} \bibinfo{year}{2019}\natexlab{b}.
\newblock \showarticletitle{Adversarial binary collaborative filtering for implicit feedback}. In \bibinfo{booktitle}{\emph{Proceedings of the AAAI Conference on Artificial Intelligence}}, Vol.~\bibinfo{volume}{33}. \bibinfo{pages}{5248--5255}.
\newblock


\bibitem[Wang et~al\mbox{.}(2018b)]%
        {wang2018dkn}
\bibfield{author}{\bibinfo{person}{Hongwei Wang}, \bibinfo{person}{Fuzheng Zhang}, \bibinfo{person}{Xing Xie}, {and} \bibinfo{person}{Minyi Guo}.} \bibinfo{year}{2018}\natexlab{b}.
\newblock \showarticletitle{DKN: Deep knowledge-aware network for news recommendation}. In \bibinfo{booktitle}{\emph{Proceedings of the 2018 world wide web conference}}. \bibinfo{pages}{1835--1844}.
\newblock


\bibitem[Wang et~al\mbox{.}(2018a)]%
        {wang2018billion}
\bibfield{author}{\bibinfo{person}{Jizhe Wang}, \bibinfo{person}{Pipei Huang}, \bibinfo{person}{Huan Zhao}, \bibinfo{person}{Zhibo Zhang}, \bibinfo{person}{Binqiang Zhao}, {and} \bibinfo{person}{Dik~Lun Lee}.} \bibinfo{year}{2018}\natexlab{a}.
\newblock \showarticletitle{Billion-scale commodity embedding for e-commerce recommendation in alibaba}. In \bibinfo{booktitle}{\emph{Proceedings of the 24th ACM SIGKDD international conference on knowledge discovery \& data mining}}. \bibinfo{pages}{839--848}.
\newblock


\bibitem[Wang et~al\mbox{.}(2021)]%
        {wang2021milvus}
\bibfield{author}{\bibinfo{person}{Jianguo Wang}, \bibinfo{person}{Xiaomeng Yi}, \bibinfo{person}{Rentong Guo}, \bibinfo{person}{Hai Jin}, \bibinfo{person}{Peng Xu}, \bibinfo{person}{Shengjun Li}, \bibinfo{person}{Xiangyu Wang}, \bibinfo{person}{Xiangzhou Guo}, \bibinfo{person}{Chengming Li}, \bibinfo{person}{Xiaohai Xu}, {et~al\mbox{.}}} \bibinfo{year}{2021}\natexlab{}.
\newblock \showarticletitle{Milvus: A purpose-built vector data management system}. In \bibinfo{booktitle}{\emph{Proceedings of the 2021 International Conference on Management of Data}}. \bibinfo{pages}{2614--2627}.
\newblock


\bibitem[Wang et~al\mbox{.}(2017)]%
        {wang2017irgan}
\bibfield{author}{\bibinfo{person}{Jun Wang}, \bibinfo{person}{Lantao Yu}, \bibinfo{person}{Weinan Zhang}, \bibinfo{person}{Yu Gong}, \bibinfo{person}{Yinghui Xu}, \bibinfo{person}{Benyou Wang}, \bibinfo{person}{Peng Zhang}, {and} \bibinfo{person}{Dell Zhang}.} \bibinfo{year}{2017}\natexlab{}.
\newblock \showarticletitle{Irgan: A minimax game for unifying generative and discriminative information retrieval models}. In \bibinfo{booktitle}{\emph{Proceedings of the 40th International ACM SIGIR conference on Research and Development in Information Retrieval}}. \bibinfo{pages}{515--524}.
\newblock


\bibitem[Wang et~al\mbox{.}(2011)]%
        {wang2011cascade}
\bibfield{author}{\bibinfo{person}{Lidan Wang}, \bibinfo{person}{Jimmy Lin}, {and} \bibinfo{person}{Donald Metzler}.} \bibinfo{year}{2011}\natexlab{}.
\newblock \showarticletitle{A cascade ranking model for efficient ranked retrieval}. In \bibinfo{booktitle}{\emph{Proceedings of the 34th international ACM SIGIR conference on Research and development in Information Retrieval}}. \bibinfo{pages}{105--114}.
\newblock


\bibitem[Wang et~al\mbox{.}(2023a)]%
        {wang2023causal}
\bibfield{author}{\bibinfo{person}{Siyu Wang}, \bibinfo{person}{Xiaocong Chen}, \bibinfo{person}{Quan~Z Sheng}, \bibinfo{person}{Yihong Zhang}, {and} \bibinfo{person}{Lina Yao}.} \bibinfo{year}{2023}\natexlab{a}.
\newblock \showarticletitle{Causal disentangled variational auto-encoder for preference understanding in recommendation}. In \bibinfo{booktitle}{\emph{Proceedings of the 46th International ACM SIGIR Conference on Research and Development in Information Retrieval}}. \bibinfo{pages}{1874--1878}.
\newblock


\bibitem[Wang et~al\mbox{.}(2019a)]%
        {wang2019neural}
\bibfield{author}{\bibinfo{person}{Xiang Wang}, \bibinfo{person}{Xiangnan He}, \bibinfo{person}{Meng Wang}, \bibinfo{person}{Fuli Feng}, {and} \bibinfo{person}{Tat-Seng Chua}.} \bibinfo{year}{2019}\natexlab{a}.
\newblock \showarticletitle{Neural graph collaborative filtering}. In \bibinfo{booktitle}{\emph{Proceedings of the 42nd international ACM SIGIR conference on Research and development in Information Retrieval}}. \bibinfo{pages}{165--174}.
\newblock


\bibitem[Wang et~al\mbox{.}(2020a)]%
        {dgcf}
\bibfield{author}{\bibinfo{person}{Xiang Wang}, \bibinfo{person}{Hongye Jin}, \bibinfo{person}{An Zhang}, \bibinfo{person}{Xiangnan He}, \bibinfo{person}{Tong Xu}, {and} \bibinfo{person}{Tat-Seng Chua}.} \bibinfo{year}{2020}\natexlab{a}.
\newblock \showarticletitle{Disentangled graph collaborative filtering}. In \bibinfo{booktitle}{\emph{Proceedings of the 43rd international ACM SIGIR conference on research and development in information retrieval}}. \bibinfo{pages}{1001--1010}.
\newblock


\bibitem[Wang et~al\mbox{.}(2020b)]%
        {mccf}
\bibfield{author}{\bibinfo{person}{Xiao Wang}, \bibinfo{person}{Ruijia Wang}, \bibinfo{person}{Chuan Shi}, \bibinfo{person}{Guojie Song}, {and} \bibinfo{person}{Qingyong Li}.} \bibinfo{year}{2020}\natexlab{b}.
\newblock \showarticletitle{Multi-component graph convolutional collaborative filtering}. In \bibinfo{booktitle}{\emph{Proceedings of the AAAI conference on artificial intelligence}}, Vol.~\bibinfo{volume}{34}. \bibinfo{pages}{6267--6274}.
\newblock


\bibitem[Wang et~al\mbox{.}(2020c)]%
        {wang2020reinforced}
\bibfield{author}{\bibinfo{person}{Xiang Wang}, \bibinfo{person}{Yaokun Xu}, \bibinfo{person}{Xiangnan He}, \bibinfo{person}{Yixin Cao}, \bibinfo{person}{Meng Wang}, {and} \bibinfo{person}{Tat-Seng Chua}.} \bibinfo{year}{2020}\natexlab{c}.
\newblock \showarticletitle{Reinforced negative sampling over knowledge graph for recommendation}. In \bibinfo{booktitle}{\emph{Proceedings of the web conference 2020}}. \bibinfo{pages}{99--109}.
\newblock


\bibitem[Wang et~al\mbox{.}(2022)]%
        {wang2022enhancing}
\bibfield{author}{\bibinfo{person}{Yanmeng Wang}, \bibinfo{person}{Jun Bai}, \bibinfo{person}{Ye Wang}, \bibinfo{person}{Jianfei Zhang}, \bibinfo{person}{Wenge Rong}, \bibinfo{person}{Zongcheng Ji}, \bibinfo{person}{Shaojun Wang}, {and} \bibinfo{person}{Jing Xiao}.} \bibinfo{year}{2022}\natexlab{}.
\newblock \showarticletitle{Enhancing dual-encoders with question and answer cross-embeddings for answer retrieval}.
\newblock \bibinfo{journal}{\emph{arXiv preprint arXiv:2206.02978}} (\bibinfo{year}{2022}).
\newblock


\bibitem[Wang et~al\mbox{.}(2023b)]%
        {wang2023recmind}
\bibfield{author}{\bibinfo{person}{Yancheng Wang}, \bibinfo{person}{Ziyan Jiang}, \bibinfo{person}{Zheng Chen}, \bibinfo{person}{Fan Yang}, \bibinfo{person}{Yingxue Zhou}, \bibinfo{person}{Eunah Cho}, \bibinfo{person}{Xing Fan}, \bibinfo{person}{Xiaojiang Huang}, \bibinfo{person}{Yanbin Lu}, {and} \bibinfo{person}{Yingzhen Yang}.} \bibinfo{year}{2023}\natexlab{b}.
\newblock \showarticletitle{Recmind: Large language model powered agent for recommendation}.
\newblock \bibinfo{journal}{\emph{arXiv preprint arXiv:2308.14296}} (\bibinfo{year}{2023}).
\newblock


\bibitem[Wang et~al\mbox{.}(2023d)]%
        {wang2023adaptive}
\bibfield{author}{\bibinfo{person}{Yunli Wang}, \bibinfo{person}{Zhiqiang Wang}, \bibinfo{person}{Jian Yang}, \bibinfo{person}{Shiyang Wen}, \bibinfo{person}{Dongying Kong}, \bibinfo{person}{Han Li}, {and} \bibinfo{person}{Kun Gai}.} \bibinfo{year}{2023}\natexlab{d}.
\newblock \showarticletitle{Adaptive Neural Ranking Framework: Toward Maximized Business Goal for Cascade Ranking Systems}.
\newblock \bibinfo{journal}{\emph{arXiv preprint arXiv:2310.10462}} (\bibinfo{year}{2023}).
\newblock


\bibitem[Wang et~al\mbox{.}(2020d)]%
        {wang2020cold}
\bibfield{author}{\bibinfo{person}{Zhe Wang}, \bibinfo{person}{Liqin Zhao}, \bibinfo{person}{Biye Jiang}, \bibinfo{person}{Guorui Zhou}, \bibinfo{person}{Xiaoqiang Zhu}, {and} \bibinfo{person}{Kun Gai}.} \bibinfo{year}{2020}\natexlab{d}.
\newblock \showarticletitle{Cold: Towards the next generation of pre-ranking system}.
\newblock \bibinfo{journal}{\emph{arXiv preprint arXiv:2007.16122}} (\bibinfo{year}{2020}).
\newblock


\bibitem[Wu et~al\mbox{.}(2021)]%
        {wu2021self}
\bibfield{author}{\bibinfo{person}{Jiancan Wu}, \bibinfo{person}{Xiang Wang}, \bibinfo{person}{Fuli Feng}, \bibinfo{person}{Xiangnan He}, \bibinfo{person}{Liang Chen}, \bibinfo{person}{Jianxun Lian}, {and} \bibinfo{person}{Xing Xie}.} \bibinfo{year}{2021}\natexlab{}.
\newblock \showarticletitle{Self-supervised graph learning for recommendation}. In \bibinfo{booktitle}{\emph{Proceedings of the 44th international ACM SIGIR conference on research and development in information retrieval}}. \bibinfo{pages}{726--735}.
\newblock


\bibitem[Wu et~al\mbox{.}(2022)]%
        {wu2022survey}
\bibfield{author}{\bibinfo{person}{Le Wu}, \bibinfo{person}{Xiangnan He}, \bibinfo{person}{Xiang Wang}, \bibinfo{person}{Kun Zhang}, {and} \bibinfo{person}{Meng Wang}.} \bibinfo{year}{2022}\natexlab{}.
\newblock \showarticletitle{A survey on accuracy-oriented neural recommendation: From collaborative filtering to information-rich recommendation}.
\newblock \bibinfo{journal}{\emph{IEEE Transactions on Knowledge and Data Engineering}} \bibinfo{volume}{35}, \bibinfo{number}{5} (\bibinfo{year}{2022}), \bibinfo{pages}{4425--4445}.
\newblock


\bibitem[Wu et~al\mbox{.}(2016)]%
        {wu2016collaborative}
\bibfield{author}{\bibinfo{person}{Yao Wu}, \bibinfo{person}{Christopher DuBois}, \bibinfo{person}{Alice~X Zheng}, {and} \bibinfo{person}{Martin Ester}.} \bibinfo{year}{2016}\natexlab{}.
\newblock \showarticletitle{Collaborative denoising auto-encoders for top-n recommender systems}. In \bibinfo{booktitle}{\emph{Proceedings of the ninth ACM international conference on web search and data mining}}. \bibinfo{pages}{153--162}.
\newblock


\bibitem[Xi et~al\mbox{.}(2023a)]%
        {xi2023bird}
\bibfield{author}{\bibinfo{person}{Yunjia Xi}, \bibinfo{person}{Jianghao Lin}, \bibinfo{person}{Weiwen Liu}, \bibinfo{person}{Xinyi Dai}, \bibinfo{person}{Weinan Zhang}, \bibinfo{person}{Rui Zhang}, \bibinfo{person}{Ruiming Tang}, {and} \bibinfo{person}{Yong Yu}.} \bibinfo{year}{2023}\natexlab{a}.
\newblock \showarticletitle{A bird's-eye view of reranking: from list level to page level}. In \bibinfo{booktitle}{\emph{Proceedings of the Sixteenth ACM International Conference on Web Search and Data Mining}}. \bibinfo{pages}{1075--1083}.
\newblock


\bibitem[Xi et~al\mbox{.}(2023b)]%
        {xi2023towards}
\bibfield{author}{\bibinfo{person}{Yunjia Xi}, \bibinfo{person}{Weiwen Liu}, \bibinfo{person}{Jianghao Lin}, \bibinfo{person}{Jieming Zhu}, \bibinfo{person}{Bo Chen}, \bibinfo{person}{Ruiming Tang}, \bibinfo{person}{Weinan Zhang}, \bibinfo{person}{Rui Zhang}, {and} \bibinfo{person}{Yong Yu}.} \bibinfo{year}{2023}\natexlab{b}.
\newblock \showarticletitle{Towards open-world recommendation with knowledge augmentation from large language models}.
\newblock \bibinfo{journal}{\emph{arXiv preprint arXiv:2306.10933}} (\bibinfo{year}{2023}).
\newblock


\bibitem[Xiao et~al\mbox{.}(2017)]%
        {xiao2017attentional}
\bibfield{author}{\bibinfo{person}{Jun Xiao}, \bibinfo{person}{Hao Ye}, \bibinfo{person}{Xiangnan He}, \bibinfo{person}{Hanwang Zhang}, \bibinfo{person}{Fei Wu}, {and} \bibinfo{person}{Tat-Seng Chua}.} \bibinfo{year}{2017}\natexlab{}.
\newblock \showarticletitle{Attentional factorization machines: Learning the weight of feature interactions via attention networks}.
\newblock \bibinfo{journal}{\emph{arXiv preprint arXiv:1708.04617}} (\bibinfo{year}{2017}).
\newblock


\bibitem[Xie et~al\mbox{.}(2023)]%
        {remi}
\bibfield{author}{\bibinfo{person}{Yueqi Xie}, \bibinfo{person}{Jingqi Gao}, \bibinfo{person}{Peilin Zhou}, \bibinfo{person}{Qichen Ye}, \bibinfo{person}{Yining Hua}, \bibinfo{person}{Jae~Boum Kim}, \bibinfo{person}{Fangzhao Wu}, {and} \bibinfo{person}{Sunghun Kim}.} \bibinfo{year}{2023}\natexlab{}.
\newblock \showarticletitle{Rethinking multi-interest learning for candidate matching in recommender systems}. In \bibinfo{booktitle}{\emph{Proceedings of the 17th ACM Conference on Recommender Systems}}. \bibinfo{pages}{283--293}.
\newblock


\bibitem[Xin et~al\mbox{.}(2018)]%
        {xin2018batch}
\bibfield{author}{\bibinfo{person}{Xin Xin}, \bibinfo{person}{Fajie Yuan}, \bibinfo{person}{Xiangnan He}, {and} \bibinfo{person}{Joemon~M Jose}.} \bibinfo{year}{2018}\natexlab{}.
\newblock \showarticletitle{Batch is not heavy: Learning word representations from all samples}. In \bibinfo{booktitle}{\emph{Proceedings of the 56th Annual Meeting of the Association for Computational Linguistics (Volume 1: Long Papers)}}. \bibinfo{pages}{1853--1862}.
\newblock


\bibitem[Xiong et~al\mbox{.}(2010)]%
        {xiong2010temporal}
\bibfield{author}{\bibinfo{person}{Liang Xiong}, \bibinfo{person}{Xi Chen}, \bibinfo{person}{Tzu-Kuo Huang}, \bibinfo{person}{Jeff Schneider}, {and} \bibinfo{person}{Jaime~G Carbonell}.} \bibinfo{year}{2010}\natexlab{}.
\newblock \showarticletitle{Temporal collaborative filtering with bayesian probabilistic tensor factorization}. In \bibinfo{booktitle}{\emph{Proceedings of the 2010 SIAM international conference on data mining}}. SIAM, \bibinfo{pages}{211--222}.
\newblock


\bibitem[Xiong and Yu(2024)]%
        {mtmi}
\bibfield{author}{\bibinfo{person}{Tianyu Xiong} {and} \bibinfo{person}{Xiaohan Yu}.} \bibinfo{year}{2024}\natexlab{}.
\newblock \showarticletitle{Multi-Tower Multi-Interest Recommendation with User Representation Repel}.
\newblock \bibinfo{journal}{\emph{arXiv preprint arXiv:2403.05122}} (\bibinfo{year}{2024}).
\newblock


\bibitem[Xu et~al\mbox{.}(2020)]%
        {xu2020privileged}
\bibfield{author}{\bibinfo{person}{Chen Xu}, \bibinfo{person}{Quan Li}, \bibinfo{person}{Junfeng Ge}, \bibinfo{person}{Jinyang Gao}, \bibinfo{person}{Xiaoyong Yang}, \bibinfo{person}{Changhua Pei}, \bibinfo{person}{Fei Sun}, \bibinfo{person}{Jian Wu}, \bibinfo{person}{Hanxiao Sun}, {and} \bibinfo{person}{Wenwu Ou}.} \bibinfo{year}{2020}\natexlab{}.
\newblock \showarticletitle{Privileged features distillation at taobao recommendations}. In \bibinfo{booktitle}{\emph{Proceedings of the 26th ACM SIGKDD International Conference on Knowledge Discovery \& Data Mining}}. \bibinfo{pages}{2590--2598}.
\newblock


\bibitem[Xu et~al\mbox{.}(2018)]%
        {xu2018deep}
\bibfield{author}{\bibinfo{person}{Jun Xu}, \bibinfo{person}{Xiangnan He}, {and} \bibinfo{person}{Hang Li}.} \bibinfo{year}{2018}\natexlab{}.
\newblock \showarticletitle{Deep learning for matching in search and recommendation}. In \bibinfo{booktitle}{\emph{The 41st International ACM SIGIR Conference on Research \& Development in Information Retrieval}}. \bibinfo{pages}{1365--1368}.
\newblock


\bibitem[Xu et~al\mbox{.}(2023)]%
        {xu2023openp5}
\bibfield{author}{\bibinfo{person}{Shuyuan Xu}, \bibinfo{person}{Wenyue Hua}, {and} \bibinfo{person}{Yongfeng Zhang}.} \bibinfo{year}{2023}\natexlab{}.
\newblock \showarticletitle{Openp5: Benchmarking foundation models for recommendation}.
\newblock \bibinfo{journal}{\emph{arXiv preprint arXiv:2306.11134}} (\bibinfo{year}{2023}).
\newblock


\bibitem[Xu et~al\mbox{.}(2022)]%
        {xu2022mixture}
\bibfield{author}{\bibinfo{person}{Zhenhui Xu}, \bibinfo{person}{Meng Zhao}, \bibinfo{person}{Liqun Liu}, \bibinfo{person}{Lei Xiao}, \bibinfo{person}{Xiaopeng Zhang}, {and} \bibinfo{person}{Bifeng Zhang}.} \bibinfo{year}{2022}\natexlab{}.
\newblock \showarticletitle{Mixture of virtual-kernel experts for multi-objective user profile modeling}. In \bibinfo{booktitle}{\emph{Proceedings of the 28th ACM SIGKDD Conference on Knowledge Discovery and Data Mining}}. \bibinfo{pages}{4257--4267}.
\newblock


\bibitem[Xue et~al\mbox{.}(2019)]%
        {xue2019deep}
\bibfield{author}{\bibinfo{person}{Feng Xue}, \bibinfo{person}{Xiangnan He}, \bibinfo{person}{Xiang Wang}, \bibinfo{person}{Jiandong Xu}, \bibinfo{person}{Kai Liu}, {and} \bibinfo{person}{Richang Hong}.} \bibinfo{year}{2019}\natexlab{}.
\newblock \showarticletitle{Deep item-based collaborative filtering for top-n recommendation}.
\newblock \bibinfo{journal}{\emph{ACM Transactions on Information Systems (TOIS)}} \bibinfo{volume}{37}, \bibinfo{number}{3} (\bibinfo{year}{2019}), \bibinfo{pages}{1--25}.
\newblock


\bibitem[Yang et~al\mbox{.}(2020a)]%
        {yang2020mixed}
\bibfield{author}{\bibinfo{person}{Ji Yang}, \bibinfo{person}{Xinyang Yi}, \bibinfo{person}{Derek Zhiyuan~Cheng}, \bibinfo{person}{Lichan Hong}, \bibinfo{person}{Yang Li}, \bibinfo{person}{Simon Xiaoming~Wang}, \bibinfo{person}{Taibai Xu}, {and} \bibinfo{person}{Ed~H Chi}.} \bibinfo{year}{2020}\natexlab{a}.
\newblock \showarticletitle{Mixed negative sampling for learning two-tower neural networks in recommendations}. In \bibinfo{booktitle}{\emph{Companion proceedings of the web conference 2020}}. \bibinfo{pages}{441--447}.
\newblock


\bibitem[Yang et~al\mbox{.}(2020b)]%
        {yang2020large}
\bibfield{author}{\bibinfo{person}{Xiaoyong Yang}, \bibinfo{person}{Yadong Zhu}, \bibinfo{person}{Yi Zhang}, \bibinfo{person}{Xiaobo Wang}, {and} \bibinfo{person}{Quan Yuan}.} \bibinfo{year}{2020}\natexlab{b}.
\newblock \showarticletitle{Large scale product graph construction for recommendation in e-commerce}.
\newblock \bibinfo{journal}{\emph{arXiv preprint arXiv:2010.05525}} (\bibinfo{year}{2020}).
\newblock


\bibitem[Ye et~al\mbox{.}(2023)]%
        {ye2023towards}
\bibfield{author}{\bibinfo{person}{Haibo Ye}, \bibinfo{person}{Xinjie Li}, \bibinfo{person}{Yuan Yao}, {and} \bibinfo{person}{Hanghang Tong}.} \bibinfo{year}{2023}\natexlab{}.
\newblock \showarticletitle{Towards robust neural graph collaborative filtering via structure denoising and embedding perturbation}.
\newblock \bibinfo{journal}{\emph{ACM Transactions on Information Systems}} \bibinfo{volume}{41}, \bibinfo{number}{3} (\bibinfo{year}{2023}), \bibinfo{pages}{1--28}.
\newblock


\bibitem[Yi et~al\mbox{.}(2019a)]%
        {yi2019deep}
\bibfield{author}{\bibinfo{person}{Baolin Yi}, \bibinfo{person}{Xiaoxuan Shen}, \bibinfo{person}{Hai Liu}, \bibinfo{person}{Zhaoli Zhang}, \bibinfo{person}{Wei Zhang}, \bibinfo{person}{Sannyuya Liu}, {and} \bibinfo{person}{Naixue Xiong}.} \bibinfo{year}{2019}\natexlab{a}.
\newblock \showarticletitle{Deep matrix factorization with implicit feedback embedding for recommendation system}.
\newblock \bibinfo{journal}{\emph{IEEE Transactions on Industrial Informatics}} \bibinfo{volume}{15}, \bibinfo{number}{8} (\bibinfo{year}{2019}), \bibinfo{pages}{4591--4601}.
\newblock


\bibitem[Yi et~al\mbox{.}(2019b)]%
        {yi2019sampling}
\bibfield{author}{\bibinfo{person}{Xinyang Yi}, \bibinfo{person}{Ji Yang}, \bibinfo{person}{Lichan Hong}, \bibinfo{person}{Derek~Zhiyuan Cheng}, \bibinfo{person}{Lukasz Heldt}, \bibinfo{person}{Aditee Kumthekar}, \bibinfo{person}{Zhe Zhao}, \bibinfo{person}{Li Wei}, {and} \bibinfo{person}{Ed Chi}.} \bibinfo{year}{2019}\natexlab{b}.
\newblock \showarticletitle{Sampling-bias-corrected neural modeling for large corpus item recommendations}. In \bibinfo{booktitle}{\emph{Proceedings of the 13th ACM Conference on Recommender Systems}}. \bibinfo{pages}{269--277}.
\newblock


\bibitem[Ying et~al\mbox{.}(2018)]%
        {pinsage}
\bibfield{author}{\bibinfo{person}{Rex Ying}, \bibinfo{person}{Ruining He}, \bibinfo{person}{Kaifeng Chen}, \bibinfo{person}{Pong Eksombatchai}, \bibinfo{person}{William~L Hamilton}, {and} \bibinfo{person}{Jure Leskovec}.} \bibinfo{year}{2018}\natexlab{}.
\newblock \showarticletitle{Graph convolutional neural networks for web-scale recommender systems}. In \bibinfo{booktitle}{\emph{Proceedings of the 24th ACM SIGKDD international conference on knowledge discovery \& data mining}}. \bibinfo{pages}{974--983}.
\newblock


\bibitem[Yu et~al\mbox{.}(2017)]%
        {yu2017selection}
\bibfield{author}{\bibinfo{person}{Hsiang-Fu Yu}, \bibinfo{person}{Mikhail Bilenko}, {and} \bibinfo{person}{Chih-Jen Lin}.} \bibinfo{year}{2017}\natexlab{}.
\newblock \showarticletitle{Selection of negative samples for one-class matrix factorization}. In \bibinfo{booktitle}{\emph{Proceedings of the 2017 SIAM International Conference on Data Mining}}. SIAM, \bibinfo{pages}{363--371}.
\newblock


\bibitem[Yu et~al\mbox{.}(2022)]%
        {yu2022graph}
\bibfield{author}{\bibinfo{person}{Junliang Yu}, \bibinfo{person}{Hongzhi Yin}, \bibinfo{person}{Xin Xia}, \bibinfo{person}{Tong Chen}, \bibinfo{person}{Lizhen Cui}, {and} \bibinfo{person}{Quoc Viet~Hung Nguyen}.} \bibinfo{year}{2022}\natexlab{}.
\newblock \showarticletitle{Are graph augmentations necessary? simple graph contrastive learning for recommendation}. In \bibinfo{booktitle}{\emph{Proceedings of the 45th international ACM SIGIR conference on research and development in information retrieval}}. \bibinfo{pages}{1294--1303}.
\newblock


\bibitem[Yu and Qin(2020)]%
        {yu2020sampler}
\bibfield{author}{\bibinfo{person}{Wenhui Yu} {and} \bibinfo{person}{Zheng Qin}.} \bibinfo{year}{2020}\natexlab{}.
\newblock \showarticletitle{Sampler design for implicit feedback data by noisy-label robust learning}. In \bibinfo{booktitle}{\emph{Proceedings of the 43rd international ACM SIGIR conference on research and development in information retrieval}}. \bibinfo{pages}{861--870}.
\newblock


\bibitem[Yu et~al\mbox{.}(2021)]%
        {yu2021dual}
\bibfield{author}{\bibinfo{person}{Yantao Yu}, \bibinfo{person}{Weipeng Wang}, \bibinfo{person}{Zhoutian Feng}, {and} \bibinfo{person}{Daiyue Xue}.} \bibinfo{year}{2021}\natexlab{}.
\newblock \showarticletitle{A dual augmented two-tower model for online large-scale recommendation}.
\newblock \bibinfo{journal}{\emph{DLP-KDD}} (\bibinfo{year}{2021}).
\newblock


\bibitem[Zhang et~al\mbox{.}(2023a)]%
        {zhang2023chatgpt}
\bibfield{author}{\bibinfo{person}{Jizhi Zhang}, \bibinfo{person}{Keqin Bao}, \bibinfo{person}{Yang Zhang}, \bibinfo{person}{Wenjie Wang}, \bibinfo{person}{Fuli Feng}, {and} \bibinfo{person}{Xiangnan He}.} \bibinfo{year}{2023}\natexlab{a}.
\newblock \showarticletitle{Is chatgpt fair for recommendation? evaluating fairness in large language model recommendation}. In \bibinfo{booktitle}{\emph{Proceedings of the 17th ACM Conference on Recommender Systems}}. \bibinfo{pages}{993--999}.
\newblock


\bibitem[Zhang et~al\mbox{.}(2023c)]%
        {zhang2023recommendation}
\bibfield{author}{\bibinfo{person}{Junjie Zhang}, \bibinfo{person}{Ruobing Xie}, \bibinfo{person}{Yupeng Hou}, \bibinfo{person}{Wayne~Xin Zhao}, \bibinfo{person}{Leyu Lin}, {and} \bibinfo{person}{Ji-Rong Wen}.} \bibinfo{year}{2023}\natexlab{c}.
\newblock \showarticletitle{Recommendation as instruction following: A large language model empowered recommendation approach}.
\newblock \bibinfo{journal}{\emph{arXiv preprint arXiv:2305.07001}} (\bibinfo{year}{2023}).
\newblock


\bibitem[Zhang et~al\mbox{.}(2022)]%
        {re4}
\bibfield{author}{\bibinfo{person}{Shengyu Zhang}, \bibinfo{person}{Lingxiao Yang}, \bibinfo{person}{Dong Yao}, \bibinfo{person}{Yujie Lu}, \bibinfo{person}{Fuli Feng}, \bibinfo{person}{Zhou Zhao}, \bibinfo{person}{Tat-Seng Chua}, {and} \bibinfo{person}{Fei Wu}.} \bibinfo{year}{2022}\natexlab{}.
\newblock \showarticletitle{Re4: Learning to re-contrast, re-attend, re-construct for multi-interest recommendation}. In \bibinfo{booktitle}{\emph{Proceedings of the ACM Web Conference 2022}}. \bibinfo{pages}{2216--2226}.
\newblock


\bibitem[Zhang et~al\mbox{.}(2019)]%
        {zhang2019deep}
\bibfield{author}{\bibinfo{person}{Shuai Zhang}, \bibinfo{person}{Lina Yao}, \bibinfo{person}{Aixin Sun}, {and} \bibinfo{person}{Yi Tay}.} \bibinfo{year}{2019}\natexlab{}.
\newblock \showarticletitle{Deep learning based recommender system: A survey and new perspectives}.
\newblock \bibinfo{journal}{\emph{ACM computing surveys (CSUR)}} \bibinfo{volume}{52}, \bibinfo{number}{1} (\bibinfo{year}{2019}), \bibinfo{pages}{1--38}.
\newblock


\bibitem[Zhang et~al\mbox{.}(2013)]%
        {zhang2013optimizing}
\bibfield{author}{\bibinfo{person}{Weinan Zhang}, \bibinfo{person}{Tianqi Chen}, \bibinfo{person}{Jun Wang}, {and} \bibinfo{person}{Yong Yu}.} \bibinfo{year}{2013}\natexlab{}.
\newblock \showarticletitle{Optimizing top-n collaborative filtering via dynamic negative item sampling}. In \bibinfo{booktitle}{\emph{Proceedings of the 36th international ACM SIGIR conference on Research and development in information retrieval}}. \bibinfo{pages}{785--788}.
\newblock


\bibitem[Zhang et~al\mbox{.}(2016)]%
        {zhang2016deep}
\bibfield{author}{\bibinfo{person}{Weinan Zhang}, \bibinfo{person}{Tianming Du}, {and} \bibinfo{person}{Jun Wang}.} \bibinfo{year}{2016}\natexlab{}.
\newblock \showarticletitle{Deep Learning over Multi-field Categorical Data: --A Case Study on User Response Prediction}. In \bibinfo{booktitle}{\emph{Advances in Information Retrieval: 38th European Conference on IR Research, ECIR 2016, Padua, Italy, March 20--23, 2016. Proceedings 38}}. Springer, \bibinfo{pages}{45--57}.
\newblock


\bibitem[Zhang et~al\mbox{.}(2021)]%
        {zhang2021deep}
\bibfield{author}{\bibinfo{person}{Weinan Zhang}, \bibinfo{person}{Jiarui Qin}, \bibinfo{person}{Wei Guo}, \bibinfo{person}{Ruiming Tang}, {and} \bibinfo{person}{Xiuqiang He}.} \bibinfo{year}{2021}\natexlab{}.
\newblock \showarticletitle{Deep learning for click-through rate estimation}.
\newblock \bibinfo{journal}{\emph{arXiv preprint arXiv:2104.10584}} (\bibinfo{year}{2021}).
\newblock


\bibitem[Zhang et~al\mbox{.}(2023b)]%
        {mirn}
\bibfield{author}{\bibinfo{person}{Xiliang Zhang}, \bibinfo{person}{Jin Liu}, \bibinfo{person}{Siwei Chang}, \bibinfo{person}{Peizhu Gong}, \bibinfo{person}{Zhongdai Wu}, {and} \bibinfo{person}{Bing Han}.} \bibinfo{year}{2023}\natexlab{b}.
\newblock \showarticletitle{MIRN: A multi-interest retrieval network with sequence-to-interest EM routing}.
\newblock \bibinfo{journal}{\emph{Plos one}} \bibinfo{volume}{18}, \bibinfo{number}{2} (\bibinfo{year}{2023}), \bibinfo{pages}{e0281275}.
\newblock


\bibitem[Zhao et~al\mbox{.}(2021)]%
        {zhao2021recbole}
\bibfield{author}{\bibinfo{person}{Wayne~Xin Zhao}, \bibinfo{person}{Shanlei Mu}, \bibinfo{person}{Yupeng Hou}, \bibinfo{person}{Zihan Lin}, \bibinfo{person}{Yushuo Chen}, \bibinfo{person}{Xingyu Pan}, \bibinfo{person}{Kaiyuan Li}, \bibinfo{person}{Yujie Lu}, \bibinfo{person}{Hui Wang}, \bibinfo{person}{Changxin Tian}, {et~al\mbox{.}}} \bibinfo{year}{2021}\natexlab{}.
\newblock \showarticletitle{Recbole: Towards a unified, comprehensive and efficient framework for recommendation algorithms}. In \bibinfo{booktitle}{\emph{proceedings of the 30th acm international conference on information \& knowledge management}}. \bibinfo{pages}{4653--4664}.
\newblock


\bibitem[Zheng et~al\mbox{.}(2024)]%
        {zheng2024full}
\bibfield{author}{\bibinfo{person}{Kai Zheng}, \bibinfo{person}{Haijun Zhao}, \bibinfo{person}{Rui Huang}, \bibinfo{person}{Beichuan Zhang}, \bibinfo{person}{Na Mou}, \bibinfo{person}{Yanan Niu}, \bibinfo{person}{Yang Song}, \bibinfo{person}{Hongning Wang}, {and} \bibinfo{person}{Kun Gai}.} \bibinfo{year}{2024}\natexlab{}.
\newblock \showarticletitle{Full Stage Learning to Rank: A Unified Framework for Multi-Stage Systems}.
\newblock \bibinfo{journal}{\emph{arXiv preprint arXiv:2405.04844}} (\bibinfo{year}{2024}).
\newblock


\bibitem[Zheng et~al\mbox{.}(2018)]%
        {zheng2018spectral}
\bibfield{author}{\bibinfo{person}{Lei Zheng}, \bibinfo{person}{Chun-Ta Lu}, \bibinfo{person}{Fei Jiang}, \bibinfo{person}{Jiawei Zhang}, {and} \bibinfo{person}{Philip~S Yu}.} \bibinfo{year}{2018}\natexlab{}.
\newblock \showarticletitle{Spectral collaborative filtering}. In \bibinfo{booktitle}{\emph{Proceedings of the 12th ACM conference on recommender systems}}. \bibinfo{pages}{311--319}.
\newblock


\bibitem[Zhiyuli et~al\mbox{.}(2023)]%
        {zhiyuli2023bookgpt}
\bibfield{author}{\bibinfo{person}{Aakas Zhiyuli}, \bibinfo{person}{Yanfang Chen}, \bibinfo{person}{Xuan Zhang}, {and} \bibinfo{person}{Xun Liang}.} \bibinfo{year}{2023}\natexlab{}.
\newblock \showarticletitle{Bookgpt: A general framework for book recommendation empowered by large language model}.
\newblock \bibinfo{journal}{\emph{arXiv preprint arXiv:2305.15673}} (\bibinfo{year}{2023}).
\newblock


\bibitem[Zhou et~al\mbox{.}(2021a)]%
        {zhou2021contrastive}
\bibfield{author}{\bibinfo{person}{Chang Zhou}, \bibinfo{person}{Jianxin Ma}, \bibinfo{person}{Jianwei Zhang}, \bibinfo{person}{Jingren Zhou}, {and} \bibinfo{person}{Hongxia Yang}.} \bibinfo{year}{2021}\natexlab{a}.
\newblock \showarticletitle{Contrastive learning for debiased candidate generation in large-scale recommender systems}. In \bibinfo{booktitle}{\emph{Proceedings of the 27th ACM SIGKDD Conference on Knowledge Discovery \& Data Mining}}. \bibinfo{pages}{3985--3995}.
\newblock


\bibitem[Zhou et~al\mbox{.}(2019)]%
        {zhou2019content}
\bibfield{author}{\bibinfo{person}{Jianing Zhou}, \bibinfo{person}{Junhao Wen}, \bibinfo{person}{Shun Li}, {and} \bibinfo{person}{Wei Zhou}.} \bibinfo{year}{2019}\natexlab{}.
\newblock \showarticletitle{From content text encoding perspective: A hybrid deep matrix factorization approach for recommender system}. In \bibinfo{booktitle}{\emph{2019 International Joint Conference on Neural Networks (IJCNN)}}. IEEE, \bibinfo{pages}{1--8}.
\newblock


\bibitem[Zhou et~al\mbox{.}(2021b)]%
        {zhou2021pure}
\bibfield{author}{\bibinfo{person}{Yao Zhou}, \bibinfo{person}{Jianpeng Xu}, \bibinfo{person}{Jun Wu}, \bibinfo{person}{Zeinab Taghavi}, \bibinfo{person}{Evren Korpeoglu}, \bibinfo{person}{Kannan Achan}, {and} \bibinfo{person}{Jingrui He}.} \bibinfo{year}{2021}\natexlab{b}.
\newblock \showarticletitle{Pure: Positive-unlabeled recommendation with generative adversarial network}. In \bibinfo{booktitle}{\emph{Proceedings of the 27th ACM SIGKDD Conference on Knowledge Discovery \& Data Mining}}. \bibinfo{pages}{2409--2419}.
\newblock


\bibitem[Zhu et~al\mbox{.}(2019)]%
        {jtm}
\bibfield{author}{\bibinfo{person}{Han Zhu}, \bibinfo{person}{Daqing Chang}, \bibinfo{person}{Ziru Xu}, \bibinfo{person}{Pengye Zhang}, \bibinfo{person}{Xiang Li}, \bibinfo{person}{Jie He}, \bibinfo{person}{Han Li}, \bibinfo{person}{Jian Xu}, {and} \bibinfo{person}{Kun Gai}.} \bibinfo{year}{2019}\natexlab{}.
\newblock \showarticletitle{Joint optimization of tree-based index and deep model for recommender systems}.
\newblock \bibinfo{journal}{\emph{Advances in Neural Information Processing Systems}}  \bibinfo{volume}{32} (\bibinfo{year}{2019}).
\newblock


\bibitem[Zhu et~al\mbox{.}(2018)]%
        {tdm}
\bibfield{author}{\bibinfo{person}{Han Zhu}, \bibinfo{person}{Xiang Li}, \bibinfo{person}{Pengye Zhang}, \bibinfo{person}{Guozheng Li}, \bibinfo{person}{Jie He}, \bibinfo{person}{Han Li}, {and} \bibinfo{person}{Kun Gai}.} \bibinfo{year}{2018}\natexlab{}.
\newblock \showarticletitle{Learning tree-based deep model for recommender systems}. In \bibinfo{booktitle}{\emph{Proceedings of the 24th ACM SIGKDD international conference on knowledge discovery \& data mining}}. \bibinfo{pages}{1079--1088}.
\newblock


\bibitem[Zhu et~al\mbox{.}(2023)]%
        {zhu2023reloop2}
\bibfield{author}{\bibinfo{person}{Jieming Zhu}, \bibinfo{person}{Guohao Cai}, \bibinfo{person}{Junjie Huang}, \bibinfo{person}{Zhenhua Dong}, \bibinfo{person}{Ruiming Tang}, {and} \bibinfo{person}{Weinan Zhang}.} \bibinfo{year}{2023}\natexlab{}.
\newblock \showarticletitle{ReLoop2: Building Self-Adaptive Recommendation Models via Responsive Error Compensation Loop}. In \bibinfo{booktitle}{\emph{Proceedings of the 29th ACM SIGKDD Conference on Knowledge Discovery and Data Mining}}. \bibinfo{pages}{5728--5738}.
\newblock


\bibitem[Zhu et~al\mbox{.}(2022)]%
        {zhu2022bars}
\bibfield{author}{\bibinfo{person}{Jieming Zhu}, \bibinfo{person}{Quanyu Dai}, \bibinfo{person}{Liangcai Su}, \bibinfo{person}{Rong Ma}, \bibinfo{person}{Jinyang Liu}, \bibinfo{person}{Guohao Cai}, \bibinfo{person}{Xi Xiao}, {and} \bibinfo{person}{Rui Zhang}.} \bibinfo{year}{2022}\natexlab{}.
\newblock \showarticletitle{Bars: Towards open benchmarking for recommender systems}. In \bibinfo{booktitle}{\emph{Proceedings of the 45th International ACM SIGIR Conference on Research and Development in Information Retrieval}}. \bibinfo{pages}{2912--2923}.
\newblock


\bibitem[Zhu and Chen(2022)]%
        {zhu2022mutually}
\bibfield{author}{\bibinfo{person}{Yaochen Zhu} {and} \bibinfo{person}{Zhenzhong Chen}.} \bibinfo{year}{2022}\natexlab{}.
\newblock \showarticletitle{Mutually-regularized dual collaborative variational auto-encoder for recommendation systems}. In \bibinfo{booktitle}{\emph{Proceedings of The ACM Web Conference 2022}}. \bibinfo{pages}{2379--2387}.
\newblock


\bibitem[Zhuo et~al\mbox{.}(2020)]%
        {otm}
\bibfield{author}{\bibinfo{person}{Jingwei Zhuo}, \bibinfo{person}{Ziru Xu}, \bibinfo{person}{Wei Dai}, \bibinfo{person}{Han Zhu}, \bibinfo{person}{Han Li}, \bibinfo{person}{Jian Xu}, {and} \bibinfo{person}{Kun Gai}.} \bibinfo{year}{2020}\natexlab{}.
\newblock \showarticletitle{Learning optimal tree models under beam search}. In \bibinfo{booktitle}{\emph{International Conference on Machine Learning}}. PMLR, \bibinfo{pages}{11650--11659}.
\newblock


\end{thebibliography}

\end{document}